\documentclass[a4paper,11pt,openany]{article}

\usepackage{jcappub} 
\usepackage[T1]{fontenc} 
\usepackage{graphicx}
\usepackage{txfonts}

\usepackage{amssymb}
\usepackage{amsmath}
\usepackage{graphicx}
\usepackage{natbib}
\usepackage{multirow}
\usepackage{multicol}
\usepackage{txfonts}
\usepackage{microtype}
\usepackage{color}
\usepackage[colorlinks=true,citecolor=blue,linkcolor=blue]{hyperref}
\usepackage{epstopdf}
\usepackage[flushleft]{threeparttable}
\usepackage{booktabs}
\usepackage{longtable,lscape}
\usepackage{subfigure}

\title{\boldmath \centering Do gamma-ray burst measurements provide a useful test of cosmological models?}

\author[a]{Narayan Khadka,\note{Corresponding author.}}
\author[b,c,d]{Orlando Luongo,}
\author[d,e]{Marco Muccino,}
\author[a]{Bharat Ratra}

\affiliation[a]{Department of Physics, Kansas State University, 116 Cardwell Hall, Manhattan, KS 66502, USA.}
\affiliation[b]{Scuola di Scienze e Tecnologie, Università di Camerino, Via Madonna delle Carceri 9, 62032 Camerino, Italy.}
\affiliation[c]{Dipartimento di Matematica, Universit\`a di Pisa, Largo B. Pontecorvo 5, 56127 Pisa, Italy.}
\affiliation[d]{NNLOT, Al-Farabi Kazakh National University, Al-Farabi av. 71, 050040 Almaty, Kazakhstan.}
\affiliation[e]{Istituto Nazionale di Fisica Nucleare (INFN), Laboratori Nazionali di Frascati, Via Enrico Fermi 54, 00044 Frascati, Italy.}

\emailAdd{nkhadka@phys.ksu.edu}
\emailAdd{orlando.luongo@unicam.it}
\emailAdd{muccino@lnf.infn.it}
\emailAdd{ratra@phys.ksu.edu}

\abstract{{We study eight different gamma-ray burst (GRB) data sets to examine whether current GRB measurements --- that probe a largely unexplored part of cosmological redshift ($z$) space --- can be used to reliably constrain cosmological model parameters.
}
{We use three Amati-correlation samples and five Combo-correlation samples to simultaneously derive correlation and cosmological model parameter constraints. The intrinsic dispersion of each GRB data set is taken as a goodness measurement. We examine the consistency between the cosmological bounds from GRBs with those determined from better-established cosmological probes, such as baryonic acoustic oscillation (BAO) and Hubble parameter $H(z)$ measurements.
}
{We use the Markov chain Monte Carlo method implemented in \textsc{MontePython} to find best-fit correlation and cosmological parameters, in six different cosmological models, for the eight GRB samples, alone or in conjunction with BAO and $H(z)$ data.
}
{For the Amati correlation case, we compile a data set of 118 bursts, the A118 sample, which is the largest --- about half of the total Amati-correlation GRBs --- current collection of GRBs suitable for constraining cosmological parameters. This updated GRB compilation has the smallest intrinsic dispersion of the three Amati-correlation GRB data sets we examined. We are unable to define a collection of reliable bursts for current Combo-correlation GRB data.
}
{Cosmological constraints determined from the A118 sample are consistent with --- but significantly weaker than --- those from BAO and $H(z)$ data. They also are consistent with the spatially-flat $\Lambda$CDM model, in which dark energy is the cosmological constant $\Lambda$, as well as with dynamical dark energy models and non-spatially-flat models. Since GRBs probe a largely unexplored region of $z$, it is well worth acquiring more and better-quality burst data which will give a more definitive answer to the question of the title.}}

\begin{document}

\maketitle

\section{Introduction}
\label{sec:intro}
Observational data indicate that the cosmological expansion is currently accelerating. They also indicate that in the recent past the expansion was decelerated. The standard spatially-flat $\Lambda$CDM model \citep{Peebles1984} is the simplest model consistent with these observations \citep{Farooqetal2017, Scolnicetal2018, PlanckCollaboration2020, eBOSSCollaboration2020}.\footnote{For recent reviews of the observational status of the flat $\Lambda$CDM model, see \cite{DiValentinoetal2021b} and \cite{PerivolaropoulosSkara2021}.} In this model, dark energy --- in the form of a cosmological constant, $\Lambda$ --- dominates the current cosmological energy budget and powers the currently-accelerating cosmological expansion. In this model, above a redshift $z$ of about 3/4, non-relativistic cold dark matter (CDM) and baryonic matter contributes more than $\Lambda$ does to the energy budget and powered the then-decelerating cosmological expansion. While the observations are consistent with dark energy being time- and space-independent, they do not rule out slowly-evolving and weakly spatially-inhomogeneous dynamical dark energy, nor do they rule out mildly curved spatial hypersurfaces.

Significant constraints on cosmological parameters come from cosmic microwave background (CMB) anisotropy data --- that primarily probe the $z \sim 1100$ part of redshift space --- as well as from baryon acoustic oscillation (BAO) observations --- the highest of which reach to $z \sim 2.3$ --- and other lower-redshift Type Ia supernova (SNIa) and Hubble parameter [$H(z)$] measurements. Observational data in the intermediate redshift range, between $z \sim 2.3$ and $\sim 1100$, are not as constraining as the lower and higher redshift data, but hold significant promise.

Intermediate redshift observations include those of HII starburst galaxies that reach to $z \sim 2.4$ \citep{ManiaRatra2012, Chavezetal2014, GonzalezMoran2019, GonzalezMoranetal2021, Caoetal2020, Caoetal2021a, Johnsonetal2021}, quasar angular sizes that reach to $z \sim 2.7$ \citep{Caoetal2017, Ryanetal2019, Caoetal2020, Caoetal2021b, Zhengetal2021, Lianetal2021},  quasar X-ray and UV fluxes that reach to $z \sim 7.5$ \citep{RisalitiLusso2015, RisalitiLusso2019, KhadkaRatra2020a, KhadkaRatra2020b, KhadkaRatra2021, Yangetal2020, Lussoetal2020, Lietal2021, Lianetal2021}, as well as gamma-ray bursts (GRBs), that have now been detected to $z=9.4$ \citep{Cucchiara2011}, and the main subject of this paper.

Observed correlations between GRB photometric and spectroscopic properties that can be related to an intrinsic burst physical property would allow GRBs to be used as valuable standard candles that reach to high $z$ and probe a largely unexplored region of cosmological redshift space \citep[see e.g.][and references therein]{Schaefer2007, Wangetal2007, Amati2008, CapozzielloIzzo2008, Dainotti2008, Izzo2009, AmatiDellaValle2013, Wei2014, Izzo2015, 2019ApJS..245....1T}, similar to how SNeIa are used as standard candles \citep{Phillips1993} at lower $z < 2.3$. This however is still a challenge for GRBs.

After it was established that GRBs were at cosmological distances, many attempts have been made to use burst correlations to constrain cosmological parameters. The first GRB Hubble diagram of a small sample of $9$ bursts, obtained by \cite{2003ApJ...583L..67S} from the $L_{\rm iso}$--$V$ correlation\footnote{This correlation \citep{2000astro.ph..4176F} relates the burst isotropic luminosity $L_{\rm iso}$ and the variability $V$ of the light curve.}, led to a current non-relativistic matter energy density parameter limit of $\Omega_{m0} < 0.35$ at the 1$\sigma$ confidence level (for the flat $\Lambda$CDM model). Soon after, using the Ghirlanda correlation\footnote{This correlation \citep{Ghirlanda2004} relates the burst jet-corrected $\gamma$-ray energy $E_{\gamma}$ and the rest-frame peak energy $E_{\rm p}$ of the photon energy spectrum $\nu F_\nu$ (where $\nu$ is the frequency). Jets are usually assumed to have a double cone structure, thus the correction is given by $(1-\cos\theta)$, where $\theta$ is the jet half opening angle. This correlation has a limited sample due to the difficulty in determining $\theta$.}, \cite{2004ApJ...612L.101D}, with a sample of $12$ bursts, found $\Omega_{m0}=0.35\pm0.15$ for the flat $\Lambda$CDM model, and \cite{2004ApJ...613L..13G}, with $14$ GRBs as well as SNeIa, found $\Omega_{m0}=0.37\pm0.10$ and a cosmological constant energy density parameter $\Omega_\Lambda=0.87\pm0.23$, in the non-flat $\Lambda$CDM model, and $\Omega_{m0}=0.29\pm0.04$ in the flat model. Similar constraints were obtained by \cite{2005ApJ...633..611L}, using the $E_{\rm p}$--$E_{\rm iso}$--$t_{\rm b}$ correlation\footnote{This correlation relates the isotropic $\gamma$-ray energy $E_{\rm iso}$, $E_{\rm p}$, and the rest-frame break time $t_{\rm b}$ of the optical afterglow light curve.}: $0.13 < \Omega_{m0} < 0.49$ and $0.50 < \Omega_\Lambda < 0.85$ at 1$\sigma$ confidence level in the flat $\Lambda$CDM model. 

\cite{SamushiaRatra2010} considered cosmological parameter constraints in the $\Lambda$CDM model and in dynamical dark energy models from two different GRB data sets and found different constraints from the two data sets and also found that the GRB constraints were relatively broad. Similarly, \cite{LiuWei2015} showed that then-current GRB data could not significantly constrain cosmological parameters. In addition, \cite{Linetal2016} showed that most GRB correlations (including the well-known Amati correlation\footnote{This correlation \citep{Amati2002} relates the above-defined $E_{\rm iso}$ and $E_{\rm p}$ and for this reason is also referred to as the $E_{\rm p}$--$E_{\rm iso}$ correlation.}, but not the Ghirlanda correlation) have large scatter and/or their parameters differ somewhat significantly between low- and high-$z$ GRB data sets\footnote{For similar conclusions on the $E_{\rm p}$--$E_{\rm iso}$ correlation, see \cite{Li2007, Huangetal2020} and references therein.} from the calibration of the Ghirlanda correlation, by using an SNeIa distance-redshift relation [through the Pad\'e approximant (3,2)], \cite{Linetal2016} obtained $\Omega_{m0}= 0.302\pm0.142$ within the flat $\Lambda$CDM model.\footnote{Similar conclusions have been reached by \cite{Tangetal2021}, who confirmed that only the Ghirlanda correlation has no redshift dependence, and determined $\Omega_{m0} = 0.307^{+0.065}_{-0.073}$ in the flat $\Lambda$CDM model from SNeIa calibrated GRB data. As discussed in Sec.\ 5.1 below, in our analysis of the complete Amati-correlation set of 220 GRBs we do not find significant evidence for redshift evolution of the correlation parameters, in agreement with the conclusions of Ref.\ \citep{Wangetal2011}.} 

More recently, based on a cosmographic approach, an updated $E_{\rm p}$--$E_{\rm iso}$ correlation with $162$ GRBs has been used to get cosmological constraints. In \cite{Demianskietal2017a} GRBs were calibrated with SNeIa, resulting in $\Omega_{m0}=0.25_{-0.12}^{+0.29}$ within the flat $\Lambda$CDM model, whereas in \cite{Demianskietal2017b} a cosmographic expansion, up to the fifth order, involving SNeIa is used to calibrate the $E_{\rm p}$--$E_{\rm iso}$ correlation for GRBs, which are then used in conjunction with $H(z)$ and BAO measurements to constrain cosmographic parameters, resulting in a 1$\sigma$ deviation from the $\Lambda$CDM cosmological model. We will see that a number of the 162 GRBs used in these, and related, later, analyses are probably not appropriate for cosmological purposes, and this is probably reflected in these GRB data constraints being somewhat inconsistent with those from SNeIa, BAO, and $H(z)$ data, as well as with those from CMB anisotropy data.

Other recent works (involving GRB data only or in conjuction with other probes) also report inconsistencies with the $\Lambda$CDM model. \cite{Demianskietal2019} again used the $E_{\rm p}$--$E_{\rm iso}$ correlation, now also modeling the potential evolution of GRB observables, and found that calibrated GRB, SNIa, and $H(z)$ data favor a dynamical dark energy model described by a scalar field with an exponential potential energy density. \cite{orlando3} considered $E_{\rm p}$--$E_{\rm iso}$, Ghirlanda, Yonetoku\footnote{This correlation relates the peak luminosity $L_{\rm p}$, computed from the observed peak flux $F_{\rm p}$ within the time interval of $1$~s around the most intense peak of the burst light curve and in the rest frame $30$--$10^4$~keV energy band, and $E_{\rm p}$.} \citep{Yonetoku2004} and Combo\footnote{This is an hybrid correlation \citep{Izzo2015} linking the $\gamma$-ray observable $E_{\rm p}$ and X-ray afterglow observables inferred from the rest-frame $0.3$--$10$~keV light curve, i.e., the plateau luminosity $L_0$, the rest-frame duration $\tau$, and the late power-law decay index $\alpha$.} GRB correlations, calibrated them in a model-independent way via $H(z)$ data, and concluded that a joint analysis of SNeIa, BAO, and calibrated GRB data sets performed by using cosmographic methods, such as Taylor expansions, auxiliary variables and Padé approximations, did not favor the flat $\Lambda$CDM model but instead favored a mildly evolving dark energy density model. Similarly, \cite{LM2020} considered $E_{\rm p}$--$E_{\rm iso}$ and Combo correlations, calibrated them via $H(z)$ actual and machine-learned data, and again, based on a joint analysis with SNeIa and BAO, found indications against a genuine cosmological constant of the $\Lambda$CDM model. In the same direction, \cite{Rezaeietal2020} used different combinations of SNIa, quasar, and GRB data sets for testing the $\Lambda$CDM model and dynamical dark energy parametrizations and concluded that the GRB and quasar data sets were inconsistent with the flat $\Lambda$CDM model (in agreement with what was found by \cite{Lussoetal2019} for similar data). \cite{Kumaretal2021} considered strong gravitational lensing data in conjunction with SNeIa and GRBs and found that the best-fit value of the spatial curvature parameter $\Omega_{k0}$ favored a closed universe, although a flat universe can be accommodated at the 68\% confidence level.

On the other hand, some recent efforts have shown that the $E_{\rm p}$--$E_{\rm iso}$ and Combo correlations calibrated using better-established cosmological data --- such as SNIa or $H(z)$ measurements --- provide cosmological constraints that are consistent with the flat $\Lambda$CDM model. In \cite{2019MNRAS.486L..46A} an updated $E_{\rm p}$--$E_{\rm iso}$ correlation with $193$ GRBs and a calibration based on an interpolation of the $H(z)$ data set have been considered, leading to $\Omega_{m0}=0.397_{-0.039}^{+0.040}$ in a flat $\Lambda$CDM cosmology, though the value of the mass density is higher than the one established by \cite{PlanckCollaboration2020}. \cite{Montieletal2021} calibrated the $E_{\rm p}$--$E_{\rm iso}$ correlation with the latest $H(z)$ data set and included CMB, BAO and SNIa data in a search for cosmological parameter constraints within the standard cosmological model, as well as in dynamical dark energy parametrizations, finding no evidence in favour of the alternatives to the $\Lambda$CDM model. Finally, by using the Combo correlation with $174$ GRBs calibrated in a semi-model independent way, \cite{2021ApJ...908..181M}  found: a) for a flat $\Lambda$CDM model $\Omega_{m0}=0.32^{+0.05}_{-0.05}$ and $\Omega_{m0}=0.22^{+0.04}_{-0.03}$ for the two values of the Hubble constant $H_0$ of \cite{PlanckCollaboration2020} and \cite{2019ApJ...876...85R}, respectively, and b) for a non-flat $\Lambda$CDM model $\Omega_{ m0}=0.34^{+0.08}_{-0.07}$ and $\Omega_\Lambda=0.91^{+0.22}_{-0.35}$ for the $H_0$ of \cite{PlanckCollaboration2020}, and $\Omega_{m0}=0.24^{+0.06}_{-0.05}$ and $\Omega_\Lambda=1.01^{+0.15}_{-0.25}$ for the $H_0$ of \cite{2019ApJ...876...85R}.

Again, by examining an uncalibrated $E_{\rm p}$--$E_{\rm iso}$ correlation built up from a sample of bright $Fermi$-LAT GRBs \citep{2019ApJ...887...13F} and another GRB sample with lower average fluence GRBs \citep{2016A&A...585A..68W} --- that has a lower intrinsic dispersion and is the basis for the most reliable Amati-correlation GRB sample we favor here ---  \cite{KhadkaRatra2020c} obtained cosmological parameter constraints in a number of cosmological models. They found that current GRB data are not able to restrictively constrain cosmological parameters, and that cosmological parameter constraints from the more-reliable GRBs are consistent with those resulting from better-established cosmological probes.
In \cite{Caoetal2021a}, a joint $H(z)$+BAO+quasar+HII starburst galaxy+GRB fit determined $\Omega_{m0}=0.313\pm0.013$ in the flat $\Lambda$CDM model, a dark energy  constituent consistent with a cosmological constant and zero spatial curvature, though mild dark energy dynamics or a little spatial curvature are not ruled out.

Taken together, the GRB data cosmological results summarized in the previous paragraphs are mutually inconsistent. We show in this paper these results can be understood as a consequence of three things: $(i)$ if GRBs are calibrated using better-established SNIa, $H(z)$, or other data, when the uncalibrated GRB data cosmological constraints are inconsistent with the calibrating data constraints the resulting calibrated GRB data constraints might or not be consistent with the calibrating data constraints\footnote{We recommend that prior to calibrating GRB data one should first verify whether the uncalibrated GRB data cosmological constraints are consistent with the calibrating data constraints. If they are not, it is then not meaningful to calibrate the GRB data. (We have verified in the flat $\Lambda$CDM model that the $H(z)$-calibrated complete Amati-correlated 220 GRB data set and the uncalibrated 220 GRB data in conjunction with $H(z)$ data provide almost identical cosmological constraints. In both cases the results are dominated by the $H(z)$ data. Even though the cosmological constraints from the uncalibrated 220 GRB data are not consistent with those from well-established cosmological probes,  because of the $H(z)$ data domination the constraints from the $H(z)$ calibrated GRB data are consistent with those from well-established cosmological probes. Other similar examples are discussed in the text, although not all authors noted that their GRB and calibrating data constraints were inconsistent.)}; $(ii)$ only about half the uncalibrated Amati-correlation GRBs that are currently used for cosmological purposes are reliable enough for this purpose; and, $(iii)$ the current uncalibrated Combo correlation GRBs cannot be reliably used for cosmological purposes. 

We note that the procedure of making use of GRB correlations when constraining cosmological model parameters is affected by the so-called circularity problem \citep{Kodama2008}, caused by having to compute the GRB correlations in an assumed background cosmological model and not in a model-independent way \citep{Dainotti2008, SamushiaRatra2010, Bernardini2012, AmatiDellaValle2013, Wei2014, Wangetal2015, Izzo2015, Demianskietal2017a, Demianskietal2017b}. This is largely due to the lack of low-$z$ GRBs that could act as distance-scale anchor GRBs once tied to primary distance indicators, such as Cepheids, SNeIa, tip of the red-giant branch, and so on. One way out of this problem is to simultaneously fit for the parameters that characterize the GRB correlations and for the cosmological model parameters, in a number of different cosmological models. In this paper we focus on one to four parameter cosmological models, spatially-flat or non-flat, with constant or dynamical dark energy. We here focus on spatially-flat or non-flat $\Lambda$CDM, XCDM and $\phi$CDM models (see Sec.\,2 for details) to make a more direct comparison with recent results obtained in Ref.\,\cite{KhadkaRatra2020c}, where the Amati correlation and the GRB sample we favor here have been considered. 
More specifically, limiting our analyses to these models, our strategy consists of employing correlations, inserting the luminosity distance from the considered cosmological models, and checking if the resulting GRB correlations are found to be close to identical among the different cosmological models. 
The fact that correlations are the same in all the studied cosmological models may indicate that such correlations are reliable instruments for the standardization of GRBs.
On the other hand, it is worth to stress that our analysis method does not address the circularity problem, since it is unable to produce distance GRB moduli independently from any cosmological model. Thus, we are not yet in the position to discriminate among possible cosmologies.
Finally, to assess the validity of such a procedure for more complicated models, further analyses are required.

In addition, all GRB correlations are characterized by large intrinsic dispersions, possibly caused by unknown large systematic errors\footnote{Possibly including those associated with detector sensitivity, and the differences in estimated spectral parameters determined from measurements taken with different detectors or from different models.} \citep{Schaefer2007,2008MNRAS.391..411B,2009AIPC.1133..350N,2011MNRAS.418L.109G} in comparison to the case of better-established probes, such as BAO, $H(z)$, and SNIa, where many error sources have been better modeled. On the other hand, the influence of possible selection bias and evolution effects are currently debated \citep{Butler_2007,2008MNRAS.387..319G,2008MNRAS.391..639N,2009A&A...508..173A,2010PASJ...62.1495Y} and one could conclude that the large intrinsic dispersions of GRB correlations could be a consequence of as yet undiscovered GRB intrinsic properties and/or an as yet unidentified sub-class within the population of GRBs, as was the case with SN populations. 

In this paper, we use two of the above correlations, i.e. $E_{\rm p}$--$E_{\rm iso}$ and Combo, to simultaneously determine correlation and cosmological model parameters and determine which GRB data sets provide reliable cosmological constraints and what those constraints are. In particular, we use the Markov chain Monte Carlo (MCMC) method with uncalibrated GRB measurements to determine these constraints. We consider eight different GRB compilations and examine the constraints in six different cosmological models, including models with dynamical dark energy density and non-spatially-flat models. We demonstrate that for currently-available GRB data reasonable results can be found for the $E_{\rm p}$--$E_{\rm iso}$ correlation case, from a compilation of 118 bursts --- roughly half of the currently available $E_{\rm p}$--$E_{\rm iso}$ correlation bursts --- that has the smallest intrinsic dispersion (of the three data sets we studied), while Combo correlation  results fail to be predictive. Moreover, we emphasize that only relatively weak cosmological limits result from these 118 more-reliable $E_{\rm p}$--$E_{\rm iso}$ bursts.

This paper is structured as follows. In Sec.\ 2 we summarize the cosmological models we use. In Sec.\ 3 we describe the data sets we analyze. In Sec.\ 4 we summarize our analysis methods. In Sec.\ 5 we present our results. We conclude in Sec.\ 6. The GRB data sets we use are tabulated in the Appendix.

\section{Cosmological models}
To analyze cosmological constraints that follow from the GRB data, we utilize six cosmological models, three pairs of spatially-flat and non-flat models\footnote{For recent discussions of observational constraints on spatial curvature, see \cite{Farooqetal2015, Chenetal2016, Ranaetal2017, Oobaetal2018a, Oobaetal2018b, Yuetal2018, ParkRatra2019a, ParkRatra2019b, Wei2018, DESCollaboration2019, Lietal2020, Handley2019, EfstathiouGratton2020, DiValentinoetal2021a, VelasquezToribioFabris2020, Vagnozzietal2020, Vagnozzietal2021, KiDSCollaboration2020, ArjonaNesseris2021, Dhawanetal2021}, and references therein.} based on three different dark energy scenarios. These models are used to compute the luminosity distances of cosmological events at known redshifts, as a function of the cosmological parameters of the cosmological model. For this we require the expansion rate or Hubble parameter $H(z)$ as a function of the redshift $z$ for these models. 

In the $\Lambda$CDM model dark energy is non-dynamical and the expansion rate is
\begin{equation}
\label{eq:friedLCDM}
    H(z) = H_0\sqrt{\Omega_{m0}(1+z)^3 + \Omega_{k0}(1+z)^2 + \Omega_{\Lambda}},
\end{equation}
where $\Omega_{\Lambda}=1-\Omega_{m0} - \Omega_{k0}$, $\Omega_{m0}$ and $\Omega_{k0}$ are the current values of the energy density parameters of the cosmological constant, the non-relativistic matter and the spatial curvature, respectively. For the GRB-only data analyses, in the non-flat $\Lambda$CDM model $\Omega_{k0}$ and $\Omega_{m0}$ are the free parameters while in the flat $\Lambda$CDM model $\Omega_{k0}$ vanishes and $\Omega_{m0}$ is the free parameter.

In the XCDM dynamical dark energy parametrization the Hubble rate is
\begin{equation}
\label{eq:XCDM}
    H(z) = H_0\sqrt{\Omega_{m0}(1+z)^3 + \Omega_{k0}(1+z)^2 + \Omega_{X0}(1+z)^{3(1+\omega_X)}},
\end{equation}
where $\Omega_{X0} = 1-\Omega_{m0} - \Omega_{k0}$ is the current value of the $X$-fluid dark energy density parameter and $\omega_X$ its equation of state parameter (the ratio of the pressure and energy density). For the GRB-only data analyses, in the non-flat XCDM case $\Omega_{m0}$, $\Omega_{k0}$, and $\omega_X$ are picked to be the free parameters while in the spatially-flat XCDM case $\Omega_{k0}$ vanishes and $\Omega_{m0}$ and $\omega_X$ are the free parameters. When $\omega_X = -1$ the $\Lambda$CDM model is recovered.

In the $\phi$CDM model dynamical dark energy is modeled as a scalar field $\phi$ \citep{PeeblesRatra1988, RatraPeebles1988, Pavlovetal2013}.\footnote{For recent discussions of observational constraints on the $\phi$CDM model, see \cite{Avsajanishvilietal2015, SolaPeracaulaetal2018, SolaPercaulaetal2019, Zhaietal2017, Oobaetal2018c, Oobaetal2019, ParkRatra2018, ParkRatra2019c, ParkRatra2020, Sangwanetal2018, Singhetal2019, UrenaLopezRoy2020, SinhaBanerjee2021, Khadkaetal2021b}, and references therein.} The scalar field dynamical dark energy density parameter $\Omega_{\phi}(z, \alpha)$ is determined by the scalar field potential energy density, assumed to be an inverse power of $\phi$,
\begin{equation}
\label{eq:phiCDMV}
    V(\phi) = \frac{1}{2}\kappa m_{p}^2 \phi^{-\alpha},
\end{equation}
where $m_{p}$ is the Planck mass, $\alpha$ is a positive parameter,  and $\kappa$ is a constant whose value is determined using the shooting method to guarantee that the current energy budget equation $\Omega_{m0} + \Omega_{k0} + \Omega_{\phi}(z = 0, \alpha) = 1$ is satisfied.

With this potential energy density, the equations of motion of the spatially homogeneous background cosmological model are
\begin{align}
\label{field}
   & \ddot{\phi}  + 3\frac{\dot{a}}{a}\dot\phi - \frac{1}{2}\alpha \kappa m_{p}^2 \phi^{-\alpha - 1} = 0, \\
\label{friedpCDM}
   & \left(\frac{\dot{a}}{a}\right)^2 = \frac{8 \pi}{3 m_{p}^2}\left(\rho_m + \rho_{\phi}\right) - \frac{k}{a^2}.
\end{align}
Here an overdot denotes a derivative with respect to time, $k$ is positive, zero, and negative for closed, flat, and open spatial geometries (corresponding to $\Omega_{k0} < 0, =0, {\rm and} >0$), $\rho_m$ is the non-relativistic matter energy density, and $\rho_{\phi}$ is the scalar field energy density given by
\begin{equation}
    \rho_{\phi} = \frac{m^2_p}{32\pi}\left[\dot{\phi}^2 + \kappa m^2_p \phi^{-\alpha}\right].
\end{equation}
$\rho_{\phi}$ is computed by numerically solving Eqs.\ (2.4) and (2.5)\footnote{To solve these coupled differential equations one needs initial conditions. See Refs.\ \cite{PeeblesRatra1988, Omerthesis} for the initial conditions we use. The $\phi$CDM model has an attractor solution, \cite{RatraPeebles1988, Pavlovetal2013} and for reasonable initial conditions the solutions are drawn to this attractor and so there is no dependence on the chosen initial conditions.} and $\Omega_{\phi}(z, \alpha)$ is then computed from 
\begin{equation}
    \Omega_{\phi}(z, \alpha) = \frac{8 \rho_{\phi}}{3 m^2_p H^2_0}.
\end{equation}

The Hubble parameter in the $\phi$CDM model is
\begin{equation}
    H(z) = H_0\sqrt{\Omega_{m0}(1+z)^3 + \Omega_{k0}(1+z)^2 + \Omega_{\phi}\left(z, \alpha\right)}.
\end{equation}
For GRB-only data analyses, in the non-flat $\phi$CDM model $\Omega_{m0}$, $\Omega_{k0}$, and $\alpha$ are picked to be the free parameters while in the flat $\phi$CDM model $\Omega_{k0}$ vanishes and $\Omega_{m0}$ and $\alpha$ are the free parameters. When $\alpha = 0$ the $\phi$CDM model is the $\Lambda$CDM model. 

For analyses involving BAO and $H(z)$ data, instead of using $\Omega_{m0}$ as a free parameter, we use the CDM and baryonic matter (physical) energy density parameters, $\Omega_c h^2$ and $\Omega_b h^2$, as free parameters. Here $h$ is the Hubble constant in units of 100 km s$^{-1}$ Mpc$^{-1}$ and $\Omega_{m0} = \Omega_c + \Omega_b$.

\section{Data}

We consider eight different GRB data sets, as well as BAO and $H(z)$ data. The GRB data sets are summarized in Table \ref{tab:samples}, which lists the assumed correlation and the number of GRBs, as well as their redshift range. The GRB data are listed in the Appendix, in Tables \ref{tab:A118}--\ref{tab:C79}, and described in what follows. The BAO and $H(z)$ measurements are discussed at the end of this Section. 

\begin{table}
	\centering
	\caption{Summary of the GRB data sets.}
	\label{tab:samples}
	\begin{threeparttable}
	\begin{tabular}{l|ccc}
	\hline
	Correlation & Data set & Number & Redshift range \\
	\hline
	Amati & A118 & $118$ & $[0.3399, 8.2]$\\
		  & A102 & $102$ & $[0.0331, 6.32]$\\
		  & A220 & $220$ & $[0.0331, 8.2]$\\
		\hline
	Combo & C60  & $60$  & $[0.145, 8.2]$\\
	      & C174 & $174$ & $[0.117, 9.4]$\\
	      & C101 & $101$ & $[0.3285, 8.2]$\\
	      & C51  & $51$  & $[0.3399, 8.2]$\\
	      & C79  & $79$  & $[0.145, 8.2]$\\
	      \hline
	\end{tabular}
    \end{threeparttable}
\end{table}

The Amati correlation GRB data are given in Tables \ref{tab:A118} and \ref{tab:A102}, where for each source of the sample the GRB name, redshift, rest-frame spectral peak energy $E_{\rm p}$, and measured bolometric fluence $S_{\rm bolo}$, computed in the standard rest-frame energy band $1$--$10^4$~keV, are listed. For details on $E_{\rm p}$ and $S_{\rm bolo}$, see Section 4. We consider three Amati correlation data compilations.
\begin{itemize}
\item[]{\bf A118 sample.} The data set, listed in Table \ref{tab:A118} is composed of $118$ long GRBs \citep{2019ApJ...887...13F}, $93$ bursts, in their Table 5, updated from those considered in \cite{2016A&A...585A..68W} (with GRB~020127 removed because its redshift is not secure) as well as $25$ long GRBs, in their Table 2 (with the short GRB~090510 excluded because the Amati correlation does not hold for short GRBs), with \textit{Fermi}-GBM/LAT data and well-constrained spectral properties.\footnote{Only GRB~080916C has non-zero $z$ error \citep{2019ApJ...887...13F}. As noticed in \cite{KhadkaRatra2020c}, including or excluding this $z$ error results in no noticeable difference, thus we ignore it.} This is the updated version of the data used in \cite{KhadkaRatra2020c}. 
\item[]{\bf A102 sample.} These $102$ long GRBs, listed in Table \ref{tab:A102}, are compiled from those listed in \cite{Demianskietal2017a} and \cite{2019MNRAS.486L..46A}, which have not already been included in the A118 sample. 
\item[]{\bf A220 sample.} This sample, in Tables \ref{tab:A118} and \ref{tab:A102} combined, with 220 long GRBs, is the union of A118 and A102 samples.
\end{itemize}

Spectral parameters for the $25$ GRBs with \textit{Fermi}-GBM/LAT \citep{2019ApJ...887...13F} in the A118 compilation have been inferred from more refined fits performed by using multiple component models, instead of the Band model \citep{Band1993}. For the other $93$ sources in the A118 compilation, taken from \cite{2016A&A...585A..68W}, there is no available information on the model used to infer the spectral parameters. We list updated spectral parameters for all A102 GRBs, collecting them from cited papers in the first instance and, when no papers were found, from Gamma-ray Coordination Network circulars. All A102 spectral parameters were inferred from simple Band model fits of GRB spectra. Consequently, the sources of the A220 sample have spectral parameters inferred from mixed procedures and samples.

The Combo correlation data are given in Tables \ref{tab:C17460}--\ref{tab:C79}, where for each source of the sample, GRB name, redshift, $\log E_{\rm p}$, the logarithms of the plateau flux $F_0$ and rest-frame duration $\tau$, and late power-law decay index $\alpha$,\footnote{Not to be confused with the $\phi$CDM model power-law potential energy density exponent $\alpha$ in Eq.\ (\ref{eq:phiCDMV}).} are listed.
For details on $E_{\rm p}$ and the other parameters, see Section 4. 

We consider two initial Combo correlation data compilations.
\begin{itemize}
\item[]{\bf C60 sample.} This data set of $60$ long GRBs, listed in Table \ref{tab:C17460}, is the one introduced in \cite{Izzo2015}. 
\item[] {\bf C174 sample.} This is an updated data set, listed in Table \ref{tab:C17460}, composed of $174$ long GRBs \citep{2021ApJ...908..181M}. 
\end{itemize}

The values of $F_0$, $\tau$, and $\alpha$ used in the C60 and C174 samples here remain the same as in \cite{2021ApJ...908..181M}, since these quantities come from the fit of the luminosity light curves of the X-ray afterglow \citep{Izzo2015}. On the other hand, $E_{\rm p}$ values in Table \ref{tab:C17460} have been updated, using those from the A118 and our updated A102 data sets, and come from mixed methodologies involving both simple Band model and multiple component models (see the above discussion). 

A further difference between the above Amati and Combo data sets is in the number of sources. This follow from the fact that, with respect to the Amati case, the Combo correlations requires a further condition to be fulfilled by GRBs: the presence of a complete X-ray light curve. For this reason a direct comparison of the above Amati and Combo data sets cannot be immediately performed using the A118, A102, A220, and C60 and C174 data sets. In light of this, for comparative studies of Amati and Combo correlations, from C174 we extract three subsamples.
\begin{itemize}
\item[] {\bf C101 sample.} A data set of $101$ Combo GRBs, listed in Table \ref{tab:C101}, which are common between the A220 and C174 samples.
\item[] {\bf C51 sample.} A data set of $51$ Combo GRBs, listed in Table \ref{tab:C51}, which are common between the A118 and C174 samples.
\item[] {\bf C79 sample.} A data set of 79 Combo GRBs, listed in Table \ref{tab:C79}, the union of the C60 and C51 samples.
\end{itemize}

In all Combo data sets, the observables $F_0$, $\tau$, and $\alpha$, are inferred by using the same technique, i.e., by fitting with the same model of the X-ray afterglow light curves computed in the $0.3$--$10$ keV rest-frame energy band \citep{Izzo2015}; moreover, these light curve data come from the same detector, i.e., the \textit{Swift}-XRT \citep{Evans}. 
In this sense, limiting the attention to X-ray observables only, all the Combo data sets would be homogeneous and free from any systematics coming from the use of different detectors or fitting techniques/models.
On the other hand, the values of $E_{\rm p}$ do depend upon the Amati samples they are taken from and in the case of C60, C79, C101 and C174 data sets the $E_{\rm p}$ values are mixed between the ones from A118 and A102 data sets. Instead, by definition, the C51 data set is the only one sharing the $E_{\rm p}$ values of the A118 data set, which are inferred from a more refined spectral analysis (see discussion above). For this reason, the C51 data set and the A118 data for the Amati case, are expected to provide more refined and reliable cosmological constraints with respect to the other data sets (see Sec.\ 5).

In this paper we also use 11 BAO and 31 $H(z)$ measurements. The BAO data span the redshift range $0.0106 \leq z \leq 2.33$ while the $H(z)$ data cover the redshift range $0.07 \leq z \leq 1.965$. The BAO data are given in Table 1 of \cite{KhadkaRatra2021} and the $H(z)$ data are listed in Table 2 of \cite{Ryanetal2018}. We compare cosmological constraints obtained using each GRB data set with those obtained using the BAO + $H(z)$ data. Cosmological constraints obtained using A118 data are consistent with those obtained using the BAO + $H(z)$ data so we also analyse A118 GRB data in conjunction with the BAO + $H(z)$ data.

In this paper we choose not to utilize CMB data because our aim is to compare GRB constraints with the ones obtained from other data sets at intermediate redshifts. On the other hand, we do not consider SNIa or CMB data sets because of the existing tension between the measurements of $H_0$ from these two probes \cite{2019ApJ...876...85R,PlanckCollaboration2020}. For these reasons we employ $H(z)$ data, so far the only one providing cosmological-model-independent measurements of the Hubble rate. BAO data was also considered because measurements of the characteristic size of the BAO feature --- which relate to the known comoving BAO scale $r_d$ --- in the radial direction provide complementary estimates of Hubble rate as well.

In general, BAO data requires the knowledge of $r_d$ and often this is typically got by using the CMB to fix the baryon density parameter $\Omega_b h^2$. We here use \texttt{CLASS} to compute $r_d$ in each model, which is a function of the cold dark matter, baryonic and neutrino energy density parameters $\Omega_c h^2$, $\Omega_b h^2$, $\Omega_\nu h^2$, and $h=H_0/(100\,{\rm km/s/Mpc})$. We fix $\Omega_\nu=0.0014$ \cite{Caoetal2020} and treat $\Omega_c h^2$ and $\Omega_b h^2$ as free cosmological parameters to be constrained by the data we use. Therefore, we determine $r_d$ from the data we use in this paper and not from CMB data.

\section{Analysis methods}\label{sec:methods}

The $E_{\rm p}$--$E_{\rm iso}$ or Amati correlation \citep{AmatiDellaValle2013} relates the rest-frame peak energy $E_{\rm p}$ of the GRB photon energy spectrum and the isotropic energy $E_{\rm iso}$. Its functional form is 
\begin{equation}
\label{eq:XCDM}
    \log \left(\frac{E_{\rm iso}} {\rm erg}\right) = a + b \log \left(\frac{E_{\rm p}}{\rm keV}\right)\,,
\end{equation}
where $a$ and $b$ are free parameters to be determined from the data. A further free parameter of the correlation is the intrinsic dispersion $\sigma_{\rm ext}$ \citep{Dago2005}. $E_{\rm iso}$ and $E_{\rm p}$ are not observed quantities. They are derived quantities defined through
\begin{align}
\label{eq:XCDM1}
    E_{\rm iso} &= 4\pi D^2_L(z, p) S_{\rm bolo}(1+z)^{-1} ,\\
\label{eq:XCDM2}
    E_{\rm p} &= E_{\rm p}^{\rm obs} (1+z)  ,
\end{align}
where $S_{\rm bolo}$ is the measured bolometric fluence, computed in the standard rest-frame energy band $1$--$10^4$~keV, and $E_{\rm p}^{\rm obs}$ is the measured peak energy of the GRB spectrum. The luminosity distance $D_L(z,p)$ is a function of $z$ and the cosmological parameters $p$ of the model under consideration and is given by
\begin{equation}
\label{eq:DM}
  \frac{H_0\sqrt{\left|\Omega_{k0}\right|}D_L(z, p)}{(1+z)} = 
    \begin{cases}
    {\rm sinh}\left[g(z)\right] & \text{if}\ \Omega_{k0} > 0, \\
    \vspace{1mm}
    g(z) & \text{if}\ \Omega_{k0} = 0,\\
    \vspace{1mm}
    {\rm sin}\left[g(z)\right] & \text{if}\ \Omega_{k0} < 0.
    \end{cases}   
\end{equation}
Here
\begin{equation}
\label{eq:XCDM}
   g(z) = H_0\sqrt{\left|\Omega_{k0}\right|}\int^z_0 \frac{dz'}{H(z')},
\end{equation}
and expressions for $H(z)$ are given in Sec.\ 2 for the cosmological models we consider. GRB data alone are unable to constrain $H_0$ because of the degeneracy between $H_0$ and the correlation intercept parameter $a$. For GRB-only analyses we fix $H_0 = 70$ ${\rm km}\hspace{1mm}{\rm s}^{-1}{\rm Mpc}^{-1}$ but when using GRB measurements together with BAO and $H(z)$ data $H_0$ is allowed to be a free parameter.
\begin{table}
	\centering
	\caption{Summary of the non-zero flat prior parameter ranges.}
	\label{tab:prior}
	\begin{threeparttable}
	\begin{tabular}{l|c}
	\hline
	Parameter & Prior range \\
	\hline
	$\Omega_bh^2$ & $[0, 1]$ \\
	$\Omega_ch^2$ & $[0, 1]$ \\
    $\Omega_{m0}$ & $[0, 1]$ \\
    $\Omega_{k0}$ & $[-2, 2]$ \\
    $\omega_{X}$ & $[-5, 0.33]$ \\
    $\alpha$ & $[0, 10]$ \\
    $\sigma_{\rm ext}$ & $[0, 5]$ \\
    $a$ & $[0, 300]$ \\
    $b$ & $[0, 5]$ \\
    $q_0$ & $[0, 100]$ \\
    $q_1$ & $[0, 5]$ \\
	\hline
	\end{tabular}
    \end{threeparttable}
\end{table}
The GRB data likelihood function is \citep{Dago2005}
\begin{equation}
\label{eq:chi2}
    \ln({\rm LF}) = -\frac{1}{2}\sum^{N_A}_{i = 1} \left[\frac{\left(\log E^{\rm obs}_{{\rm iso},i} - \log E^{\rm th}_{{\rm iso},i}\right)^2}{s^2_{E,i}} + \ln(2\pi s^2_{E,i})\right],
\end{equation}
where $s^2_{E,i} = \sigma^2_{\log E_{{\rm iso},i}} + b^2 \sigma^2_{\log E_{{\rm p},i}} + \sigma_{\rm ext}^2$. Here, $\sigma_{\log E_{{\rm iso},i}}$ is the error in the measured value of $\log E_{{\rm iso},i}$ and $\sigma_{\log E_{{\rm p},i}}$ is the error in $\log E_{{\rm p},i}$. We maximize this likelihood function to determine best-fit values and errors of the free parameters.

The Combo correlation links the $\gamma$-ray observable $E_{\rm p}$ with the X-ray afterglow observables $\tau$, $\alpha$, and
\begin{equation}
\label{eq:L0Combo}
    L_0= 4\pi D^2_L(z, p) F_0\,,
\end{equation}
which is the luminosity of the plateau of the luminosity light curve in the $0.3$--$10$ keV rest-frame energy band.
The Combo-correlation is defined as
\begin{equation}
\label{eq:XCDM}
    \log \left(\frac{L_0}{\rm erg/s}\right) = q_0 + q_1 \log \left(\frac{E_{\rm p}}{\rm keV}\right) - \log\left(\frac{\tau /s}{|1+\alpha |}\right)\,,
\end{equation}
where $q_0$, $q_1$ and the intrinsic dispersion $\sigma_{\rm ext}$ are the free parameters to be determined.
The corresponding likelihood function is
\begin{equation}
\label{eq:chi2}
    \ln({\rm LF}) = -\frac{1}{2}\sum^{N}_{i = 1} \left[\frac{\left(\log L^{\rm obs}_{0,i} - \log L^{\rm th}_{0,i}\right)^2}{s^2_{L,i}} + \ln(2\pi s^2_{L,i})\right],
\end{equation}
where $s^2_{L,i} = \sigma^2_{\log L_{0,i}} + q_1^2 \sigma^2_{\log E_{{\rm p},i}} + \sigma^2_{\log \left({\tau}/{|1+\alpha |}\right)} + \sigma_{\rm ext}^2$. Here, $\sigma_{\log L_{0,i}}$ is the error in the measured value of $\log L_{0,i}$, $\sigma_{\log E_{{\rm p},i}}$ is the error in $\log E_{{\rm p},i}$ and $\sigma_{\log \left({\tau}/{|1+\alpha |}\right)}$ is the error in $\log\left({\tau}/{|1+\alpha |}\right)$.

To determine cosmological constraints from BAO + $H(z)$ data, we follow the method outlined by \cite{KhadkaRatra2021}.

For the model comparisons, we compute the Akaike and Bayes Information Criterion ($AIC$ and $BIC$) values,
\begin{align}
\label{eq:AIC}
    AIC =& -2\ln(LF_{\rm max}) + 2d,\\
\label{eq:BIC}
    BIC =& -2\ln(LF_{\rm max}) + d\ln{N}\,.
\end{align}
Here $d$ is the number of free parameters, $N$ the number of data points, and we define the degree of freedom $dof = N - d$. For $N \gtrsim 7.4$, as is the case for all data sets we consider, larger values of $d$ are penalized by $BIC$ more severely than by $AIC$. We also compute the differences $\Delta AIC$ and $\Delta BIC$ with respect to a reference model. Negative values of $\Delta AIC$ and $\Delta BIC$ indicate that the model under investigation performs better than the reference model. On the other hand, positive values show the opposite case. In particular, $\Delta AIC(BIC) \in [0, 2]$ provides weak evidence in favor of the reference model, $\Delta AIC(BIC) \in(2, 6]$ gives positive evidence against the given model, and $\Delta AIC(BIC)>6$ is strong evidence against the given model. 

The likelihood analysis for each data set and cosmological model is done using the MCMC method as implemented in the \textsc{MontePython} code \citep{Brinckmann2019}. Convergence of the MCMC chains for each parameter is determined with the Gelman-Rubin criterion $(R-1 < 0.05)$. For each free parameter we assume a top hat prior which is non-zero over the ranges given in Table \ref{tab:prior}.

\section{Results}
\textbf{\subsection{A118, A102, and A220 data constraints}}
\label{Amati}

Results for the A118, A102 and A220 data sets are given Tables 3 and 4. The unmarginalized best-fit parameter values are listed in Table 3 and the marginalized one-dimensional best-fit parameter values and limits are listed in Table 4. The corresponding plots of two-dimensional likelihood contours and one-dimensional likelihoods are shown in Figs.\ 1--3. Results for the A118, A102, and A220 data sets are plotted in magenta, blue, and green, respectively. The A118 results here are an update of those of \cite{KhadkaRatra2020c}, based on the slightly updated A118 sample now analyzed here using \textsc{MontePython} (instead of \textsc{emcee}).

The use of these GRB data to constrain cosmological parameters is based on the validity of the Amati $E_p-E_{\rm iso}$ correlation. We use these three data sets to also simultaneously determine the slope $b$ and the intercept $a$ of the Amati relation and test whether this relation is independent of the assumed cosmological model. For these three data sets, for all models, best-fit values of $a$ lie in the range $\sim 49-50$ and best-fit values of $b$ lie in the range $\sim 1.2 - 1.3$. For a given data set, the $a$ and $b$ values are almost constant, the same for all models, indicating that the Amati relation is the same in all models and that the GRBs in a given data set are standardized. However, A118 GRBs favor an Amati relation with a somewhat larger value for the intercept and a somewhat shallower slope than do the A102 GRBs.

The minimum value of the intrinsic dispersion $\sigma_{\rm ext}$, $\sim 0.39$, is obtained for the A118 data set and the maximum value, $\sim 0.52$, is obtained for the A102 data set, with the A220 data value in-between, $\sim 0.46$. These differences are large and indicate that the A118 data are of significantly better quality than the A102 data and so provide more reliable cosmological constraints than do the A102 and A220 data sets. As discussed above in Sec.\ 3, the A118 data are what should be used to determine current GRB data cosmological constraints, and the A102 and A220 data sets should not be used to constrain cosmological parameters. 

In an attempt to determine if the large intrinsic dispersion of the A220 data is the consequence of only a few GRBs, and if $\sigma_{\rm ext}$ can be decreased by removing a few unreliable GRBs, we repeated the analyses for a number of different subsets of the A220 data compilation. These subsets included: $(i)$ four bins in redshift space with approximately the same number of GRBs in each bin, from which we found the Amati relation parameters, $a$ and $b$, were independent of redshift within the error bars\footnote{We did not repeat this computation for the A118 data set, which would have fewer bursts in each redshift bin, resulting in greater uncertainties for the determined $a$ and $b$ parameters in each bin.}  and that $\sigma_{\rm ext}$ was large for the first three redshift bins but smaller, $\sim 0.33-0.34$ depending on cosmological model, for the highest $2.68 < z  \le 8.2$ bin; and, $(ii)$ dividing the A220  data into two parts at a number of different redshifts, from $0.48$ to $1.8$, from which we found the Amati relation parameters, $a$ and $b$, were again independent of redshift within the error bars, and that $\sigma_{\rm ext}$ was large for all the lower redshift data subsets (and increases as the cutoff $z$ was decreased) and smallest, $\sim 0.4$ depending on cosmological model, for the $z > 1.8$ data. We conclude from these analyses that the larger $\sigma_{\rm ext}$ value for the A220 data is caused more by lower redshift GRBs but we were not able to find a fair procedure to settle on which low redshift GRBs to discard from the A220 data set to try to improve it so that it could be used for cosmological purposes.\footnote{We note that the analyses of \cite{Demianskietal2017b, Demianskietal2019, Rezaeietal2020, Lussoetal2019, 2019MNRAS.486L..46A} are based on GRB data that include a significant number of A220 GRBs. That they find these data inconsistent with the standard flat $\Lambda$CDM model is probably more correctly viewed as a reflection of the inappropriateness of using the A220 GRBs to constrain cosmological parameters rather than an inadequacy of the standard flat $\Lambda$CDM model.} Our recommendation is that, for the purpose of constraining cosmological parameters it is only appropriate to use the A118 data set.

From Figs.\ 1--3, we see that for the A118 and A102 data sets a significant part of the probability favors currently accelerating cosmological expansion while for the A220 data set currently decelerating cosmological expansion is more favored. 

The A118 data largely provide a lower limit on the value of $\Omega_{m0}$, except in the non-flat $\phi$CDM model where they result in $\Omega_{m0} = 0.560^{+0.210}_{-0.250}$. Representing this case by the 2$\sigma$ lower limit, these lower limits range from $> 0.060$ for the non-flat $\phi$CDM model to $> 0.238$ for the flat $\phi$CDM model. In all six cosmological models the A118 $\Omega_{m0}$ values are consistent with those from the BAO + $H(z)$ data. For the A102 data, the value of $\Omega_{m0}$ ranges from $> 0.223$ to $> 0.272$. The minimum value is obtained in the flat XCDM model and the maximum value is obtained in the flat $\phi$CDM model. These values are consistent with those from the BAO + $H(z)$ data. For the A220 data, the value of $\Omega_{m0}$ ranges from $> 0.327$ to $> 0.481$. The minimum value is obtained in the flat XCDM model and the maximum value is obtained in the non-flat $\Lambda$CDM model. These values are mostly inconsistent with those from the BAO + $H(z)$ data.

From Table 4, for all three data sets, in the flat $\Lambda$CDM model, the value of $\Omega_{\Lambda}$ ranges from $< 0.545$ to $< 0.770$.\footnote{We derive chains for $\Omega_{\Lambda}$ in each computation using the equation $\Omega_{\Lambda}= 1-\Omega_{m0}-\Omega_{k0}$ (where for the spatially-flat $\Lambda$CDM model $\Omega_{k0}=0$). Then, from those chains, using the \textsc{python} package \textsc{getdist} \citep{Lewis_2019} we determine the best-fit values and uncertainties for $\Omega_{\Lambda}$. We also use this \textsc{python} package to plot all one-dimensional likelihoods and two-dimensional contours and to compute the best-fit values and uncertainties of the free parameters.} The minimum value is obtained using the A220 data and the maximum value is obtained using the A118 data. In the non-flat $\Lambda$CDM model, the value of $\Omega_{\Lambda}$ ranges from $< 0.910$ to $< 1.310$. The minimum value is obtained using the A118 data and the maximum value is obtained using the A102 data.

For these three data sets, for all three non-flat models, the value of $\Omega_{k0}$ ranges from $-0.136^{+0.796}_{-0.424}$ to $0.330^{+0.520}_{-0.360}$. Both minimum and maximum values are obtained in non-flat $\Lambda$CDM model using A220 and A118 data sets respectively. These values are mutually consistent with those from the BAO + $H(z)$ data.\footnote{In the non-flat $\phi$CDM model, the value of $\Omega_{\phi}(z, \alpha)$ is determined using the numerical solutions of the dynamical equations and its current value always lies in the range $0 \leq \Omega_{\phi}(0, \alpha) \leq 1$. In the non-flat $\phi$CDM model plots, this restriction on $\Omega_{\phi}(0,\alpha)$ can be seen in the $\Omega_{m0}-\Omega_{k0}$ sub-panel in the form of straight line boundaries.}

For all three data sets, for the flat and non-flat XCDM parametrizations, the value of the equation of state parameter ($\omega_X$) ranges from $< -0.143$ to $< 0.256$. The minimum value is obtained in the flat XCDM parametrization using the A118 data and the maximum value is obtained in the non-flat XCDM parametrization using the A220 data. These data sets provide very weak constraints on $\omega_X$. In the flat and non-flat $\phi$CDM models, all three data sets are unable to constrain the scalar field potential energy density parameter $\alpha$.

Only the A118 data set should be used to constrain cosmological parameters, and perhaps the most useful constraint it provides for this purpose is the (weak) lower limit on $\Omega_{m0}$.

In Table 3, for all three data sets, the values of $AIC$ and $BIC$ and their differences $\Delta AIC$ and $\Delta BIC$ with respect to the flat $\Lambda$CDM values, assumed as a reference, are listed. In almost the totality of models we have $\Delta AIC\lesssim2$, implying at most weak evidence for the flat $\Lambda$CDM model; only the non-flat XCDM and $\phi$CDM cases have $2\lesssim\Delta AIC\lesssim6$, positive evidence for the flat $\Lambda$CDM model. Moving to the $\Delta BIC$ values, now almost the totality of models exhibit positive evidence for the flat $\Lambda$CDM one, whereas the non-flat XCDM and $\phi$CDM cases indicate strong evidence for the flat $\Lambda$CDM one.

\begin{figure*}
\begin{multicols}{2}    
    \includegraphics[width=\linewidth]{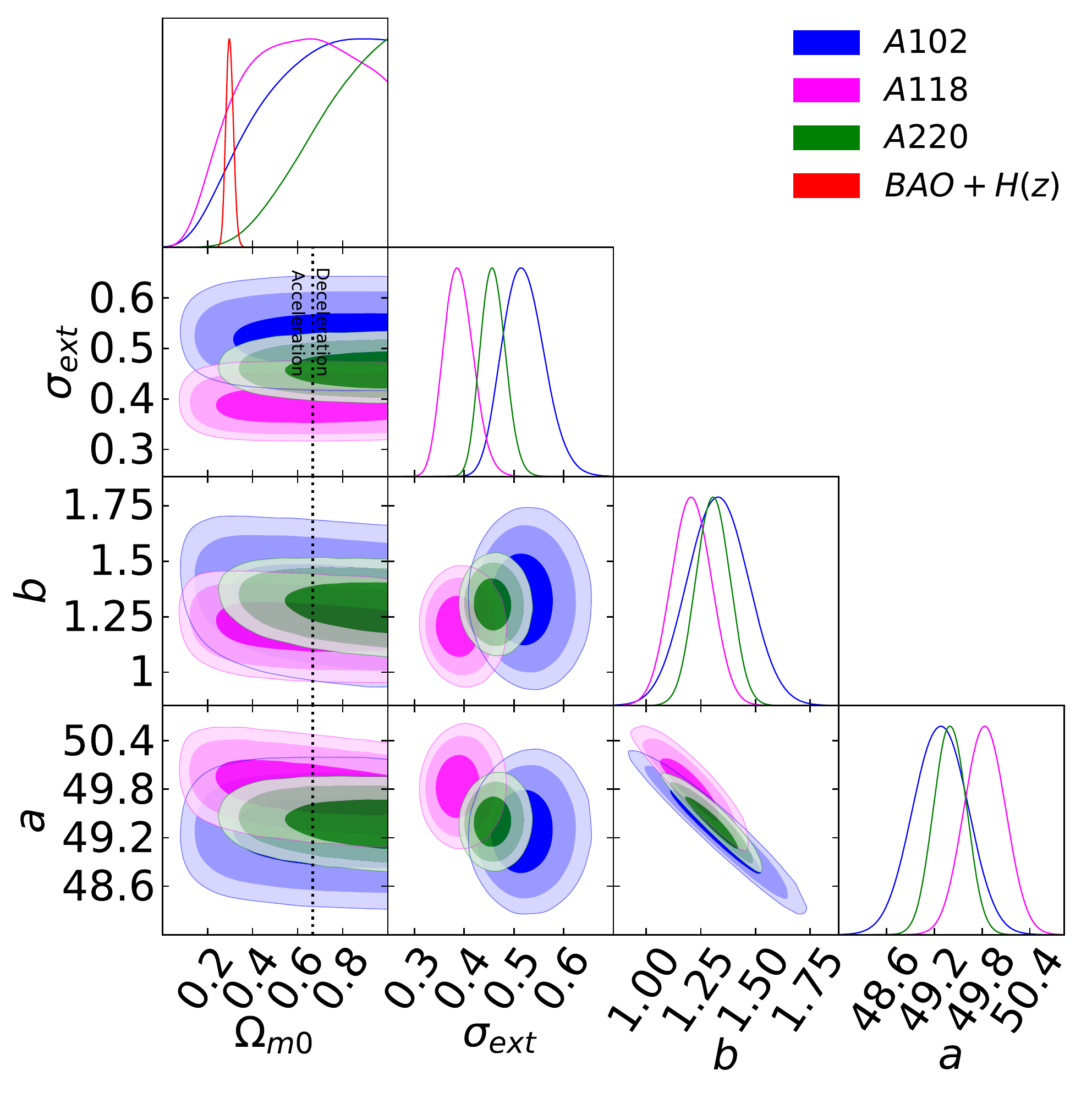}\par
    \includegraphics[width=\linewidth]{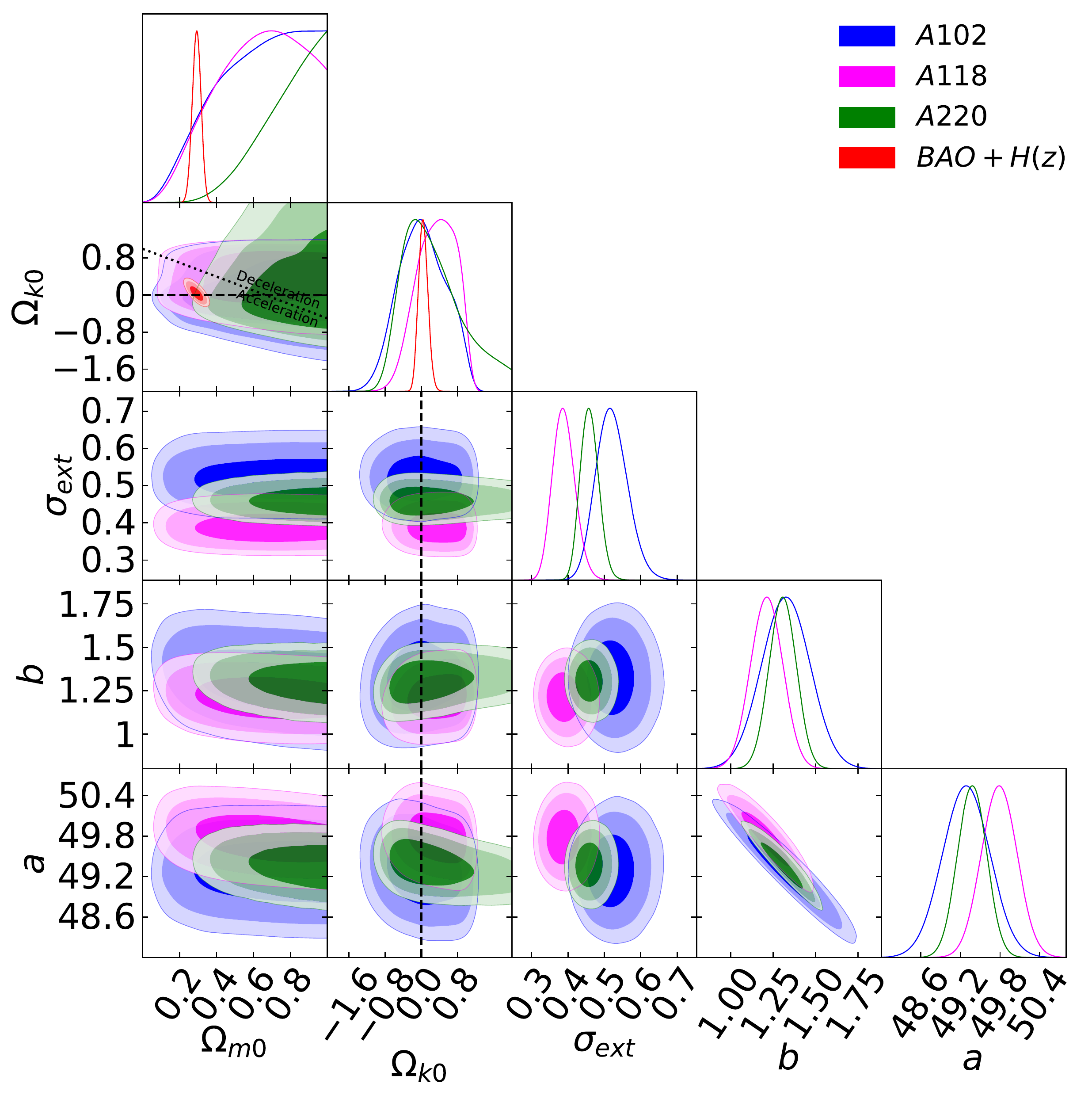}\par
\end{multicols}
\caption{One-dimensional likelihood distributions and two-dimensional contours at 1$\sigma$, 2$\sigma$, and 3$\sigma$ confidence levels using A118 (blue), A102 (magenta), A220 (green),  and BAO + $H(z)$ (red) data for all free parameters. Left panel shows the flat $\Lambda$CDM model. The black dotted vertical lines are the zero acceleration lines with currently accelerated cosmological expansion occurring to the left of the lines. Right panel shows the non-flat $\Lambda$CDM model. The black dotted sloping line in the $\Omega_{k0}-\Omega_{m0}$ subpanel is the zero acceleration line with currently accelerated cosmological expansion occurring to the lower left of the line. The black dashed horizontal or vertical line in the $\Omega_{k0}$ subpanels correspond to $\Omega_{k0} = 0$.}
\label{fig:13}
\end{figure*}

\begin{landscape}
\centering
\small\addtolength{\tabcolsep}{0.0pt}
\begin{threeparttable}
\caption{Unmarginalized one-dimensional best-fit parameters for Amati correlation GRB and BAO + $H(z)$ data sets. For each data set, $\Delta AIC$ and $\Delta BIC$ values are computed with respect to the $AIC$ and $BIC$  values of the flat  $\Lambda$CDM model.}\label{tab:1d_BFP2}
\setlength{\tabcolsep}{1.3mm}{
\begin{tabular}{lccccccccccccccccc}
\toprule
Model & Data set & $\Omega_{b}h^2$ & $\Omega_{c}h^2$& $\Omega_{\rm m0}$ & $\Omega_{\rm k0}$ & $\omega_{X}$ & $\alpha$ & $H_0$\tnote{a} & $\sigma_{\rm ext}$ & $a$ & $b$ & $dof$ & $-2\ln L_{\rm max}$ & $AIC$ & $BIC$ & $\Delta AIC$ & $\Delta BIC$\\
\hline
& A220 & - & -& 0.997 & - & - & - & - & 0.450 & 49.355 & 1.295 & 216 & 292.44 & 300.44 & 314.01 & - & -\\
Flat & A118 & - & -& 0.611 & - & - & - &- & 0.379 & 49.816 & 1.207 & 114 & 117.70 & 125.70 & 136.78 & - & -\\
$\Lambda$CDM & A102 & - & - & 0.992 & - & - & - &- & $0.503$ & 49.263 & 1.291 & 98 & 156.98 & 164.98 & 175.48 & - & -\\
&  BAO + $H(z)$ & 0.024 & 0.119 & 0.298 & - & - & - &69.119&-&-&-& 39 & 23.66&29.66&34.87 & - & -\\
& A118 + BAO + $H(z)$ & 0.024 & 0.118 & 0.296 & - & - & - &69.125&0.379&49.935&1.230& 154 & 142.12&154.12&172.57 & - & -\\
\hline
& A220 & - & -& 0.989 & 0.037 & - &-&- & 0.452 & 49.375 & 1.288 & 215 & 292.44 & 302.44 & 319.41 & 2.00 & 5.40\\
Non-flat & A118 & - & - & 0.939 & 0.098 & - & - &- & 0.374 & 49.661 & 1.212 & 113 & 116.76 & 126.76 & 140.61 & 1.06 & 3.83\\
$\Lambda$CDM & A102 &- & -& 0.960 & 0.035 & - & - &- & 0.505 & 49.289 & 1.284 & 97 & 156.98 & 166.98 & 180.10 & 2.00 & 4.62\\
& BAO + $H(z)$ & 0.025 & 0.114 & 0.294 & 0.021 & - & - &68.701&-&-&-&38&23.60&31.60&38.55 & 1.94 & 3.68\\
& A118 + BAO + $H(z)$ & 0.024 & 0.116 & 0.297 & 0.024 & - & - &68.537&0.379&49.946&1.228&153&142.08&156.08&177.61 & 1.94 & 5.04\\
\hline
& A220 &- & -& 0.114 & - & 0.118 &-&- & 0.451 & 49.360 & 1.268 & 215 & 292.18 & 302.18 & 319.15 & 1.74 & 5.14\\
Flat & A118 &- & -& 0.029 & - & $-$0.161 & - &- & 0.376 & 49.809 & 1.199 & 113 & 117.18 & 127.18 & 141.03 & 1.48 & 4.25\\
XCDM & A102 &- & -& 0.800 & - & 0.020 & - &- & 0.504 & 49.234 & 1.300 & 97 & 156.98 & 166.98 & 180.10 & 2.00 & 4.62\\
& BAO + $H(z)$ & 0.031 & 0.088 & 0.280 & - & $-$0.691 & - &65.036& - & - & -&38&19.66&27.66&34.61 & $-2.00$ & $-0.26$\\
& A118 + BAO + $H(z)$ & 0.031 & 0.086 & 0.279 & - & $-$0.684 & - &64.757& 0.374 & 49.980 & 1.216&153&137.94&151.94&173.47 & $-2.08$ & 0.90\\
\hline
& A220 &- & -& 0.639 & $-$0.160 & 0.013 &-&- & 0.456 & 49.401 & 1.260 & 214 & 292.19 & 304.19 & 324.55 & 3.75 & 10.54\\
Non-flat & A118 &- & -& 0.975 & 0.923 & $-$0.663 & - &- & 0.375 & 49.672 & 1.217 & 112 & 116.84 & 128.84 & 145.46 & 3.14 & 8.68\\
XCDM & A102 &- & -& 0.324 & $-$0.001 & 0.050 & - &- & 0.508 & 49.229 & 1.295 & 96 & 156.98 & 168.98 & 184.73 & 4.00 & 9.25\\
& BAO + $H(z)$ & 0.030 & 0.094 & 0.291 & $-$0.147 & $-$0.641 & - &65.204& - & - & -&37&18.34&28.34&37.03 & $-1.32$ & 2.16\\
& A118 + BAO + $H(z)$ & 0.028 & 0.010 & 0.295 & $-$0.137 & $-$0.677 & - &65.893& 0.385 & 49.952 & 1.216&152&136.72&152.72&177.32 & $-1.40$ & 4.75\\
\hline
& A220 &- & -& 0.995 & - & - &8.335&- & 0.452 & 49.360 & 1.293 & 215 & 292.44 & 302.44 & 319.41 & 2.00 & 5.40\\
Flat & A118 &- & -& 0.630 & - & - & 9.477 &- & 0.378 & 49.819 & 1.202 & 113 & 117.70 & 127.70 & 141.55 & 2.00 & 4.97\\
$\phi$CDM & A102 &- & -& 0.996 & - & - & 9.074 &- & 0.503 & 49.262 & 1.291 & 97 & 156.98 & 166.98 & 180.10 & 2.00 & 5.03\\
& BAO + $H(z)$ & 0.033 & 0.080 & 0.265 & - & - & 1.445 &65.272& - & - & -&38&19.56&27.56&34.51 & $-2.10$ & $-0.36$\\
& A118 + BAO + $H(z)$ & 0.036 & 0.068 & 0.252 & - & - & - &64.381&0.378&49.956&1.223& 153 & 137.72&151.72&173.25 & $-2.40$ & 0.68\\
\hline
& A220 &- & -& 0.999 & $-$0.085 & - & 7.277 &- & 0.452 & 49.364 & 1.290 & 214 & 292.44 & 304.44 & 324.80 & 4.00 & 10.79\\
Non-flat & A118 &- & -& 0.642 & 0.353 & - & 6.515 & - & 0.378 & 48.792 & 1.206 & 112 & 117.16 & 129.16 & 145.78 & 3.46 & 9.00\\
$\phi$CDM & A102 &- & -& 0.993 & $-$0.053 & - & 3.081 &- & 0.507 & 49.280 & 1.288 & 96 & 156.98 & 168.98 & 184.73 & 4.00 & 9.25\\
& BAO + $H(z)$ & 0.035 & 0.078 & 0.261 & $-$0.155 & - & 2.042 &65.720& - & - & -&37&18.16&28.16&36.85 & $-1.50$ & 1.98\\
& A118 + BAO + $H(z)$ & 0.038 & 0.072 & 0.252 & $-$0.150 & - & 2.323 &66.124& 0.383 & 49.920 & 1.221&152&136.58&152.58&177.18 & $-1.54$ & 4.61\\
\bottomrule
\end{tabular}}
\begin{tablenotes}[flushleft]
\item[a]${\rm km}\hspace{1mm}{\rm s}^{-1}{\rm Mpc}^{-1}$. $H_0$ is set to $70$ ${\rm km}\hspace{1mm}{\rm s}^{-1}{\rm Mpc}^{-1}$ for the GRB-only data analyses.
\end{tablenotes}
\end{threeparttable}
\end{landscape}

\begin{landscape}
\centering
\small\addtolength{\tabcolsep}{0.0pt}
\begin{threeparttable}
\caption{Marginalized one-dimensional best-fit parameters with 1$\sigma$ confidence intervals for Amati correlation GRB and BAO + $H(z)$ data sets. A 2$\sigma$ limit is given when only an upper or lower limit exists.}\label{tab:1d_BFP2}
\setlength{\tabcolsep}{1.3mm}{
\begin{tabular}{lccccccccccccc}
\toprule
Model & Data set & $\Omega_{b}h^2$ & $\Omega_{c}h^2$ & $\Omega_{\rm m0}$ & $\Omega_{\Lambda}$ & $\Omega_{\rm k0}$ & $\omega_{X}$ & $\alpha$ &$H_0$\tnote{a}& $\sigma_{\rm ext}$ & $a$ & $b$\\
\midrule
 & A220 &-&-& $> 0.455$ & $< 0.545$ & - & - & - & - & $0.459^{+0.023}_{-0.027}$ & $49.400^{-0.200}_{-0.200}$ & $1.306^{+0.077}_{-0.077}$\\
Flat & A118 &-&-& $> 0.230$ & $< 0.770$ & - & - & - &- & $0.390^{+0.026}_{-0.032}$ & $49.830^{-0.260}_{-0.260}$ & $1.207^{+0.091}_{-0.091}$\\
$\Lambda$CDM & A102 &-&-& $> 0.267$ & $< 0.733$ & - & - & - &- & $0.521^{+0.037}_{-0.046}$ & $49.280^{-0.340}_{-0.340}$ & $1.330^{+0.140}_{-0.140}$\\
& BAO + $H(z)$& $0.024^{+0.003}_{-0.003}$ & $0.119^{+0.008}_{-0.008}$ & $0.299^{+0.015}_{-0.017}$ & $0.700^{+0.017}_{-0.015}$ & - & - & - &$69.300^{+1.800}_{-1.800}$&-&-&-\\
&  A118 + BAO + $H(z)$ &$0.024^{+0.003}_{-0.003}$&$0.120^{+0.007}_{-0.008}$& $0.300^{+0.015}_{-0.017}$ & $0.700^{+0.017}_{-0.015}$ & - & - & - &$69.200^{+1.700}_{-1.700}$&$0.390^{+0.025}_{-0.030}$&$49.940^{+0.250}_{-0.250}$&$1.228^{+0.088}_{-0.088}$\\
\hline
& A220 &-&-& $> 0.481$ & $< 1.100$ & $-0.136^{+0.796}_{-0.424}$ & - &-&- & $0.460^{+0.023}_{-0.027}$ & $49.380^{+0.220}_{-0.220}$ & $1.306^{+0.079}_{-0.079}$\\
Non-flat & A118 &-&-& $> 0.261$ & $< 0.910$ & $0.330^{+0.520}_{-0.360}$ & - & - &- & $0.389^{+0.027}_{-0.033}$ & $49.790^{-0.260}_{-0.260}$ & $1.212^{+0.090}_{-0.090}$\\
$\Lambda$CDM & A102 &-&-& $> 0.247$ & $< 1.310$ & $0.020^{+0.510}_{-0.580}$ & - & - &- & $0.521^{+0.039}_{-0.047}$ & $49.290^{-0.340}_{-0.340}$ & $1.320^{+0.140}_{-0.140}$\\
& BAO + $H(z)$& $0.025^{+0.004}_{-0.004}$ & $0.113^{+0.019}_{-0.019}$ & $0.292^{+0.023}_{-0.023}$ & $0.667^{+0.093}_{+0.081}$ & $-0.014^{+0.075}_{-0.075}$ & - & - &$68.700^{+2.300}_{-2.300}$&-&-&-\\
& A118 + BAO + $H(z)$& $0.025^{+0.004}_{-0.005}$ & $0.114^{+0.018}_{-0.018}$ & $0.294^{+0.023}_{-0.023}$ & $0.669^{+0.088}_{+0.076}$ & $0.037^{+0.089}_{-0.100}$ & - & - &$68.700^{+2.200}_{-2.200}$&$0.389^{+0.250}_{-0.300}$&$49.950^{+0.240}_{-0.240}$&$1.227^{+0.086}_{-0.086}$\\
\hline
& A220 &-&-& > 0.327 & - & - & $< 0.163$ &-&- & $0.459^{+0.023}_{-0.027}$ & $49.43^{+0.210}_{-0.210}$ & $1.302^{+0.077}_{-0.077}$\\
Flat & A118 &-&-& $> 0.171$ & - & - & $< -0.143 $ & - &- & $0.390^{+0.026}_{-0.032}$ & $49.900^{-0.270}_{-0.310}$ & $1.201^{+0.091}_{-0.091}$\\
XCDM & A102 &-&-& $> 0.223$ & - & - & $< 0.100 $ & - &- & $0.521^{+0.037}_{-0.045}$ & $49.320^{-0.350}_{-0.350}$ & $1.320^{+0.140}_{-0.140}$\\
&BAO + $H(z)$ & $0.030^{+0.005}_{-0.005}$ & $0.093^{+0.019}_{-0.017}$ & $0.282^{+0.021}_{-0.021}$ & - & - & $-0.744^{+0.140}_{-0.097}$ & - &$65.800^{+2.200}_{-2.500}$& - & - & -\\
& A118 + BAO + $H(z)$ & $0.030^{+0.005}_{-0.005}$ & $0.092^{+0.019}_{-0.017}$ & $0.282^{+0.023}_{-0.020}$ & - & - & $-0.733^{+0.150}_{-0.095}$ & - &$65.600^{+2.200}_{-2.500}$& $0.389^{+0.025}_{-0.031}$ & $49.950^{0.240}_{-0.240}$ & $1.225^{+0.086}_{-0.086}$\\
\hline
& A220 &-&-& $> 0.335$ & - & $-0.089^{+0.499}_{-0.421}$ & $< 0.256$ &-&- & $0.460^{+0.023}_{-0.027}$ & $49.400^{+0.230}_{-0.230}$ & $1.297^{+0.081}_{-0.081}$\\
Non-flat & A118 &-&-& $> 0.179$ & - & $0.018^{+0.392}_{-0.398}$ & $< 0.000 $ & - &- & $0.390^{+0.026}_{-0.032}$ & $49.810^{-0.270}_{-0.270}$ & $1.213^{+0.091}_{-0.091}$\\
XCDM & A102 &-&-& --- & - & $0.008^{+0.522}_{-0.428}$ & $< 0.100 $ & - &- & $0.522^{+0.039}_{-0.046}$ & $49.280^{-0.350}_{-0.350}$ & $1.320^{+0.140}_{-0.140}$\\
& BAO + $H(z)$ & $0.029^{+0.005}_{-0.005}$ & $0.099^{+0.021}_{-0.021}$ & $0.293^{+0.027}_{-0.027}$ & - & $-0.120^{+0.130}_{-0.130}$ & $-0.693^{+0.130}_{-0.077}$ & - &$65.900^{+2.400}_{-2.400}$& - & - & -\\
& A118 + BAO + $H(z)$ & $0.029^{+0.005}_{-0.006}$ & $0.097^{+0.021}_{-0.021}$ & $0.291^{+0.026}_{-0.026}$ & - & $-0.110^{+0.120}_{-0.120}$ & $-0.694^{+0.140}_{-0.079}$ & - &$65.800^{+2.200}_{-2.500}$& $0.389^{+0.025}_{-0.030}$ & $49.940^{+0.240}_{-0.240}$ & $1.218^{+0.087}_{-0.087}$\\
\hline
& A220 &-&-& > 0.454 & - & - & - &---&- & $0.458^{+0.022}_{-0.026}$ & $49.40^{+0.200}_{-0.200}$ & $1.306^{+0.075}_{-0.075}$\\
Flat & A118 &-&-& $> 0.238$ & - & - & - & --- &- & $0.389^{+0.025}_{-0.030}$ & $49.830^{-0.240}_{-0.240}$ & $1.206^{+0.086}_{-0.086}$\\
$\phi$CDM & A102 &-&-& $> 0.272$ & - & - & - & --- &- & $0.519^{+0.034}_{-0.043}$ & $49.280^{-0.320}_{-0.320}$ & $1.330^{+0.130}_{-0.130}$\\
& BAO + $H(z)$ & $0.032^{+0.006}_{-0.003}$ & $0.081^{+0.017}_{-0.017}$ & $0.266^{+0.023}_{-0.023}$ & - & - & - & $1.530^{+0.620}_{-0.850}$ &$65.100^{+2.100}_{-2.100}$& - & - & -\\
& A118 + BAO + $H(z)$ & $0.033^{+0.007}_{-0.003}$ & $0.080^{+0.017}_{-0.020}$ & $0.266^{+0.024}_{-0.024}$ & - & - & - & $1.580^{+0.660}_{-0.890}$ &$65.000^{+2.200}_{-2.200}$& $0.389^{+0.026}_{-0.032}$ & $49.95^{+0.250}_{-0.250}$ & $1.225^{+0.088}_{-0.088}$\\

\hline
& A220 &-&-& $> 0.453$ & - & $-0.127^{+0.237}_{-0.263}$ & - &--- &- & $0.459^{+0.023}_{-0.026}$ & $49.420^{+0.200}_{-0.200}$ & $1.300^{+0.076}_{-0.076}$\\
Non-flat & A118 &-&-& $0.560^{+0.210}_{-0.250}$ & - & $0.049^{+0.291}_{-0.249}$ & - & --- &- & $0.389^{+0.026}_{-0.031}$ & $49.840^{-0.240}_{-0.240}$ & $1.210^{+0.087}_{-0.087}$\\
$\phi$CDM & A102 &-&-& $> 0.271$ & - & $-0.081^{+0.304}_{-0.336}$ & - & --- &- & $0.521^{+0.038}_{-0.046}$ & $49.300^{-0.330}_{-0.330}$ & $1.320^{+0.130}_{-0.130}$\\
& BAO + $H(z)$ & $0.032^{+0.006}_{-0.004}$ & $0.085^{+0.017}_{-0.021}$ & $0.271^{+0.024}_{-0.028}$ & - & $-0.080^{+0.100}_{-0.100}$ & - & $1.660^{+0.670}_{-0.830}$ &$65.500^{+2.500}_{-2.500}$& - & - & -\\
& A118 + BAO + $H(z)$ & $0.032^{+0.007}_{-0.003}$ & $0.084^{+0.018}_{-0.022}$ & $0.271^{+0.025}_{-0.028}$ & - & $-0.080^{+0.100}_{-0.100}$ & - & $1.710^{+0.700}_{-0.850}$ &$65.400^{+2.200}_{-2.200}$& $0.389^{+0.025}_{-0.031}$ & $49.95^{+0.240}_{-0.240}$ & $1.220^{+0.086}_{-0.086}$\\
\bottomrule
\end{tabular}}
\begin{tablenotes}[flushleft]
\item[a]${\rm km}\hspace{1mm}{\rm s}^{-1}{\rm Mpc}^{-1}$. $H_0$ is set to $70$ ${\rm km}\hspace{1mm}{\rm s}^{-1}{\rm Mpc}^{-1}$ for GRB-only data analyses.
\end{tablenotes}
\end{threeparttable}
\end{landscape}

\begin{figure*}
\begin{multicols}{2}    
    \includegraphics[width=\linewidth]{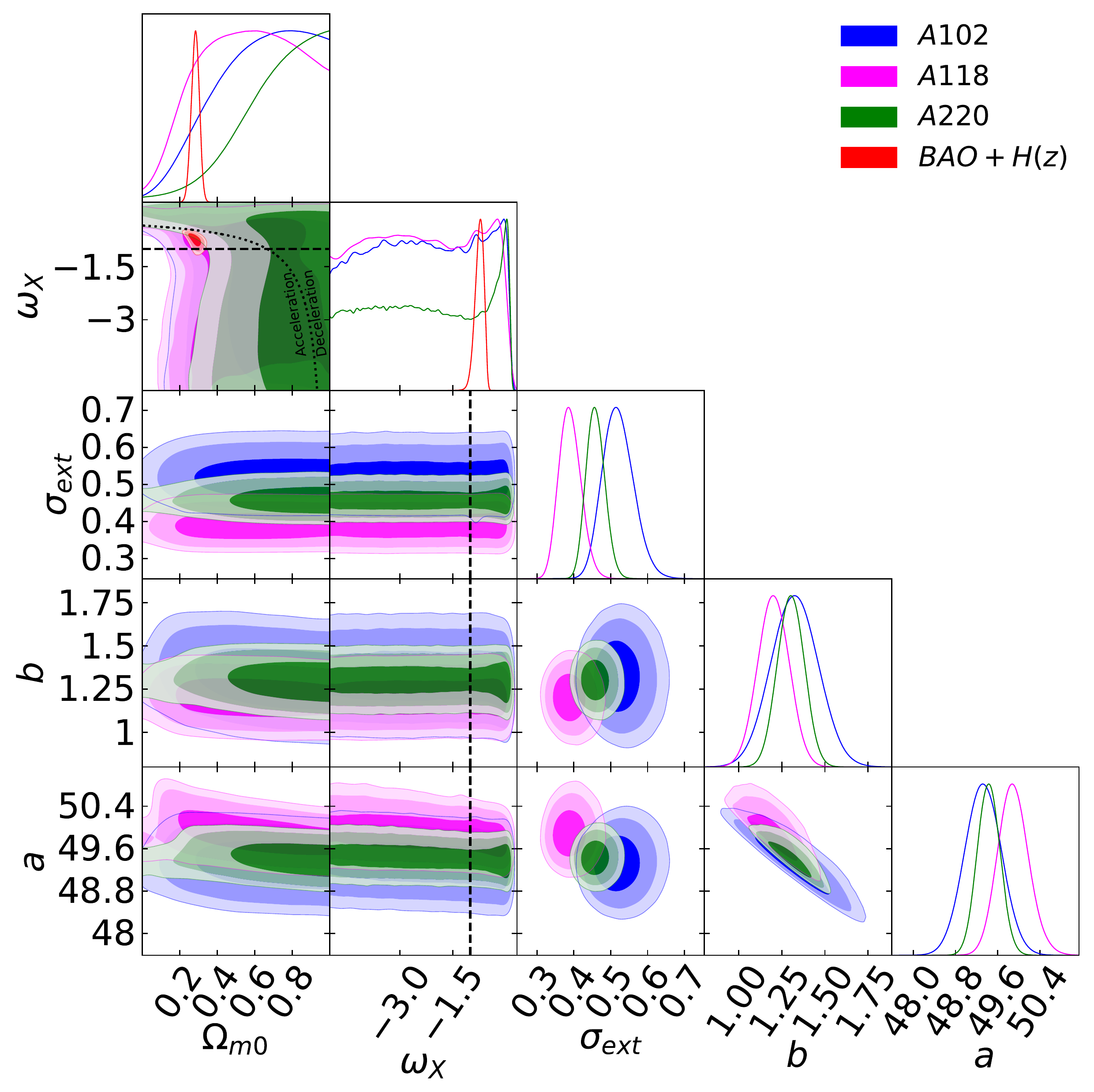}\par
    \includegraphics[width=\linewidth]{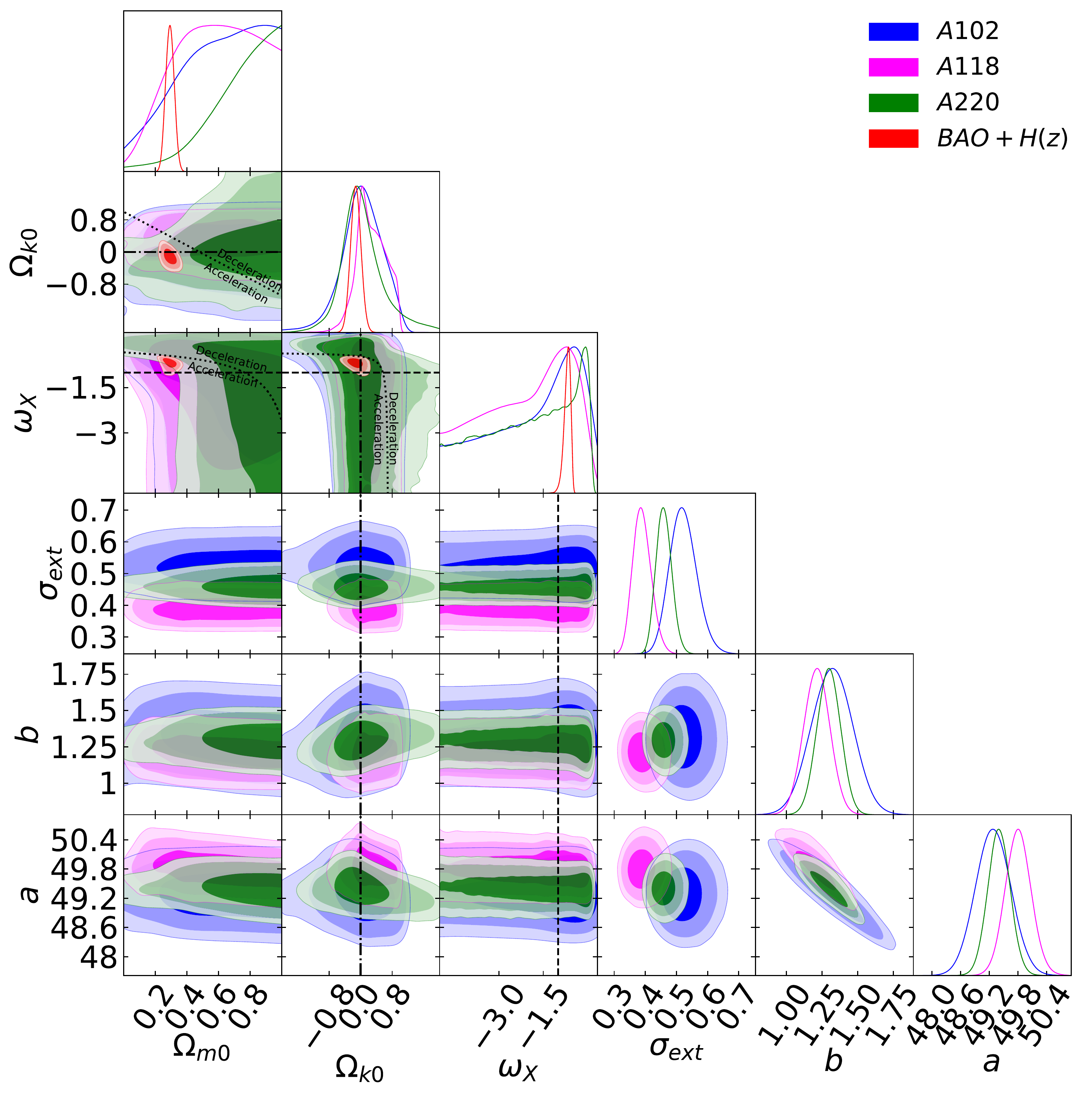}\par
\end{multicols}
\caption{One-dimensional likelihood distributions and two-dimensional contours at 1$\sigma$, 2$\sigma$, and 3$\sigma$ confidence levels using A118 (blue), A102 (magenta), A220 (green),  and BAO + $H(z)$ (red) data for all free parameters. Left panel shows the flat XCDM parametrization. The black dotted curved line in the $\omega_X-\Omega_{m0}$ subpanel is the zero acceleration line with currently accelerated cosmological expansion occurring below the line and the black dashed straight lines correspond to the $\omega_X = -1$ $\Lambda$CDM model. Right panel shows the non-flat XCDM parametrization. The black dotted lines in the $\Omega_{k0}-\Omega_{m0}$, $\omega_X-\Omega_{m0}$, and $\omega_X-\Omega_{k0}$ subpanels are the zero acceleration lines with currently accelerated cosmological expansion occurring below the lines. Each of the three lines is computed with the third parameter set to the BAO + $H(z)$ data best-fit value of Table 3. The black dashed straight lines correspond to the $\omega_x = -1$ $\Lambda$CDM model. The black dotted-dashed straight lines correspond to $\Omega_{k0} = 0$.}
\label{fig:13}
\end{figure*}

\begin{figure*}
\begin{multicols}{2}    
    \includegraphics[width=\linewidth]{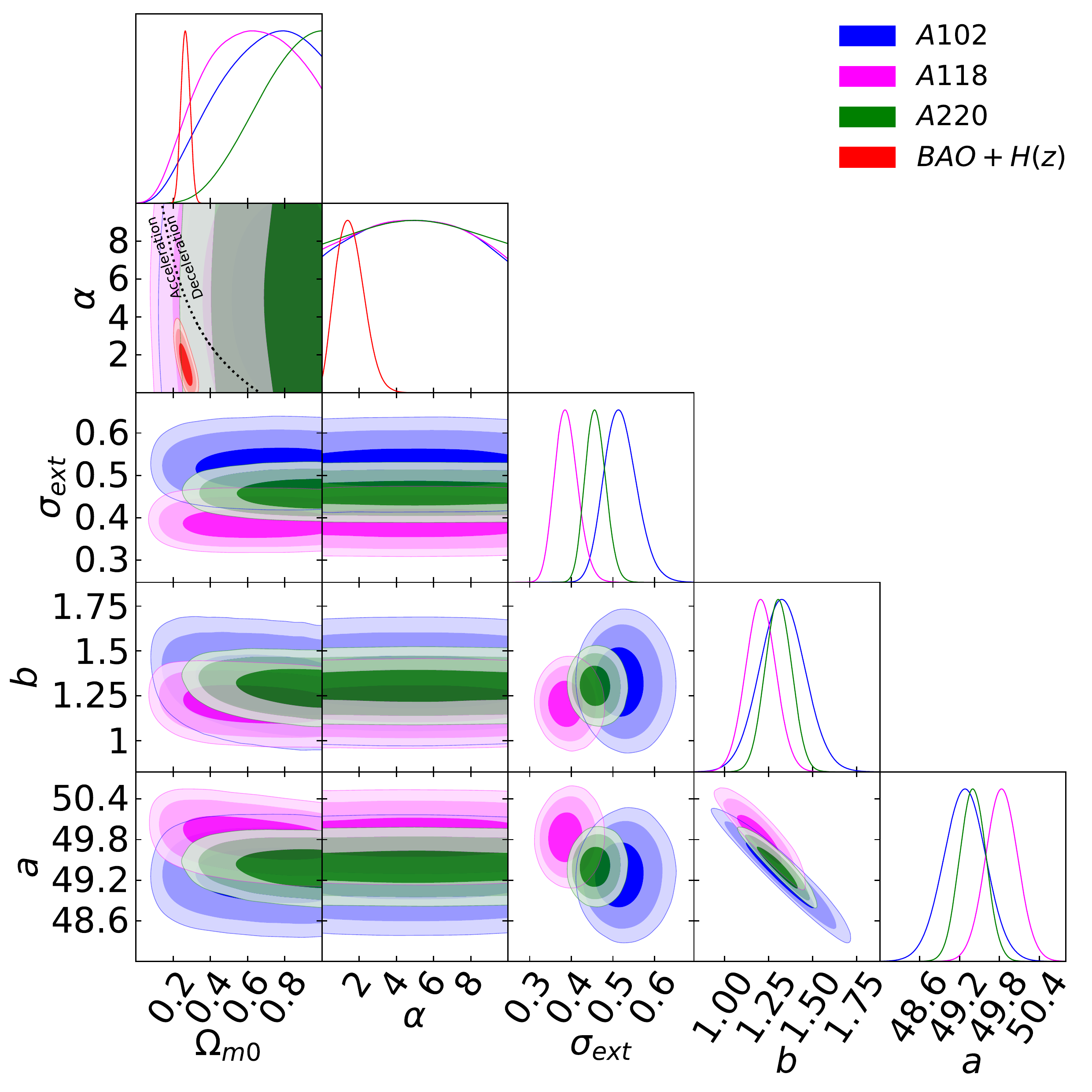}\par
    \includegraphics[width=\linewidth]{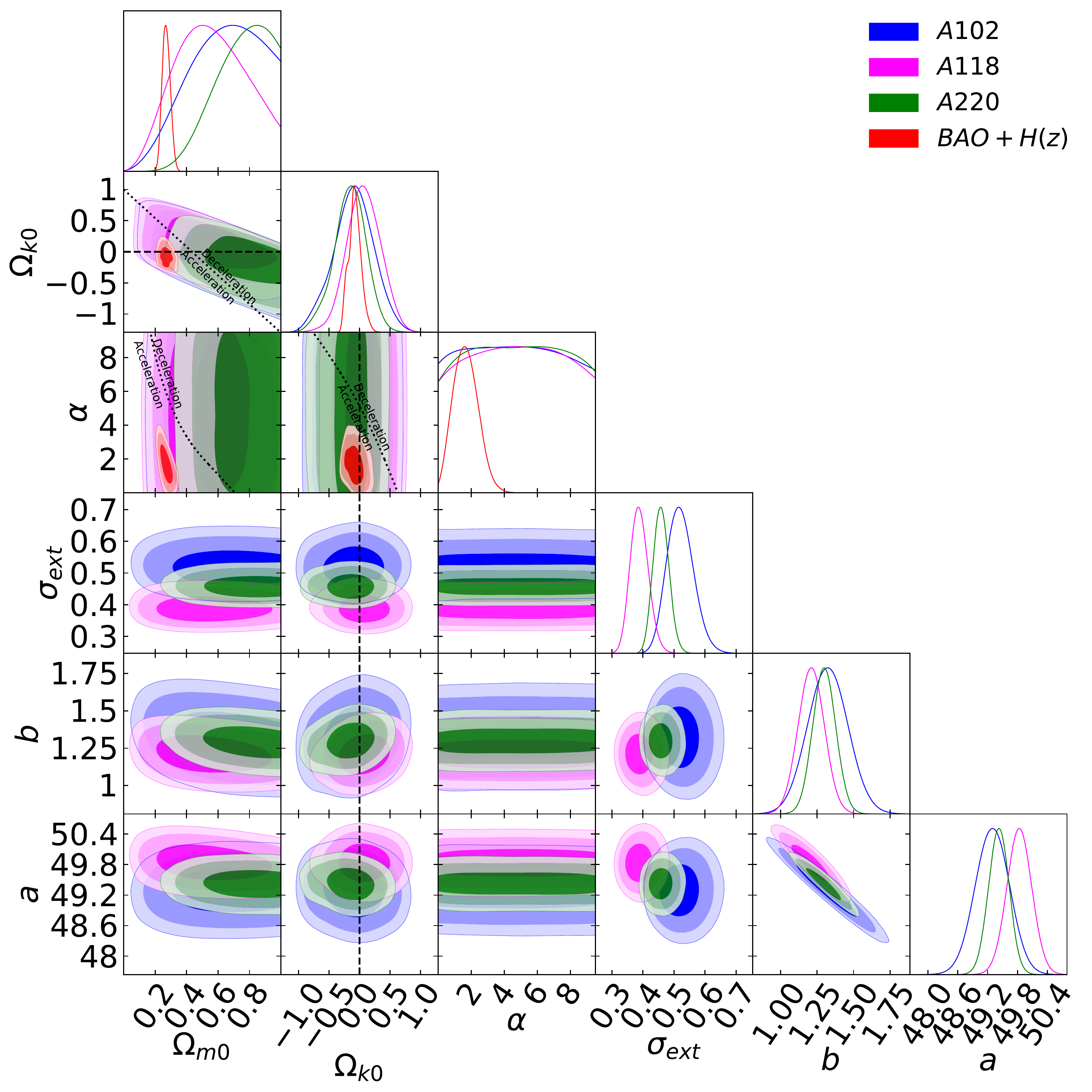}\par
\end{multicols}
\caption{One-dimensional likelihood distributions and two-dimensional contours at 1$\sigma$, 2$\sigma$, and 3$\sigma$ confidence levels using A118 (blue), A102 (magenta), A220 (green),  and BAO + $H(z)$ (red) data for all free parameters. The $\alpha = 0$ axes correspond to the $\Lambda$CDM model. Left panel shows the flat $\phi$CDM model. The black dotted curved line in the $\alpha - \Omega_{m0}$ subpanel is the zero acceleration line with currently accelerated cosmological expansion occurring to the left of the line. Right panel shows the non-flat $\phi$CDM model. The black dotted lines in the $\Omega_{k0}-\Omega_{m0}$, $\alpha-\Omega_{m0}$, and $\alpha-\Omega_{k0}$ subpanels are the zero acceleration lines with currently accelerated cosmological expansion occurring below the lines. Each of the three lines is computed with the third parameter set to the BAO + $H(z)$ data best-fit value of Table 3. The black dashed straight lines correspond to $\Omega_{k0} = 0$.}
\label{fig:13}
\end{figure*}

\subsection{BAO + $H(z)$ and A118 + BAO + $H(z)$ data constraints}
\label{Amati_BAO}

\begin{figure*}
\begin{multicols}{2}    
    \includegraphics[width=\linewidth]{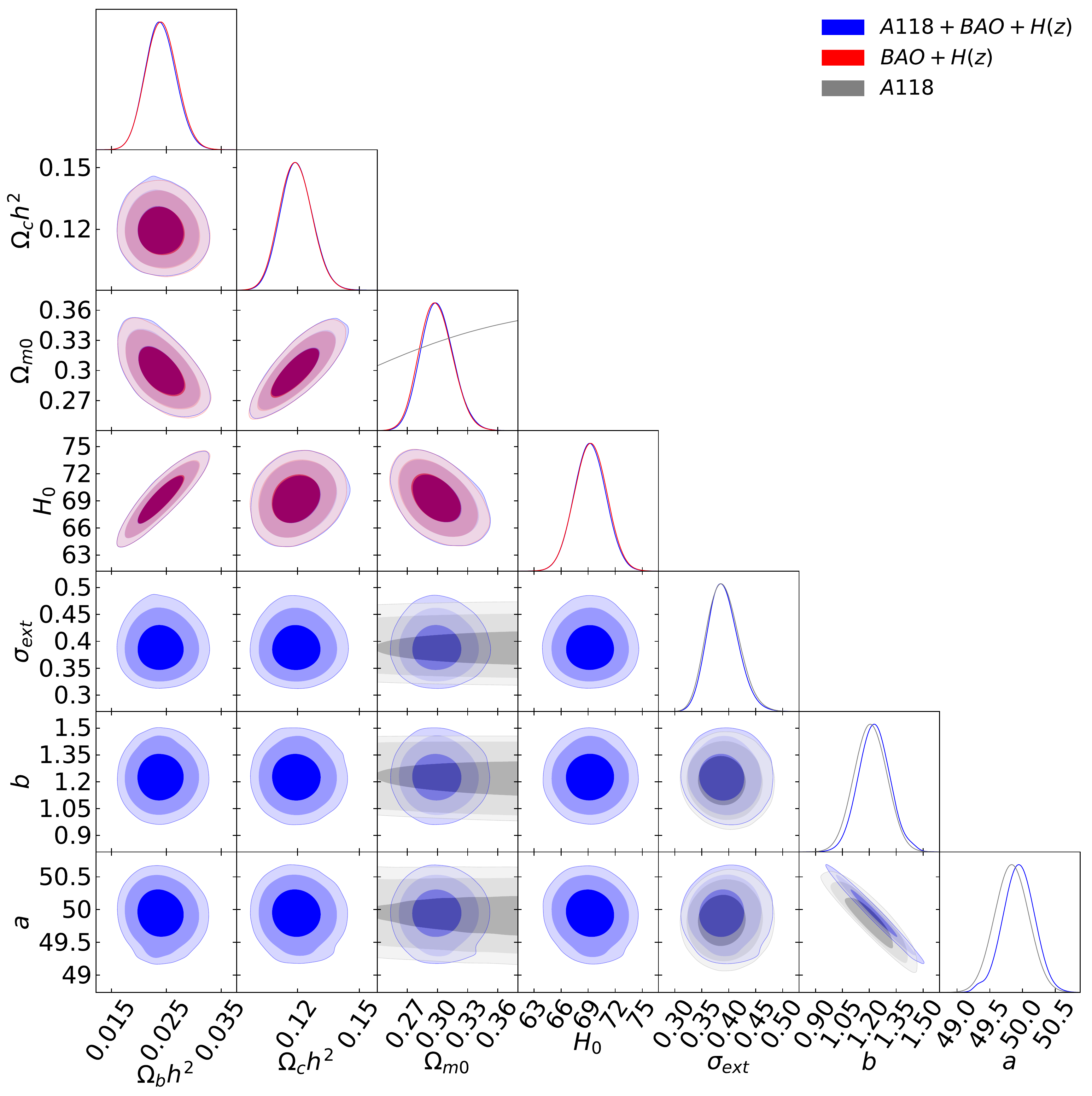}\par
    \includegraphics[width=\linewidth]{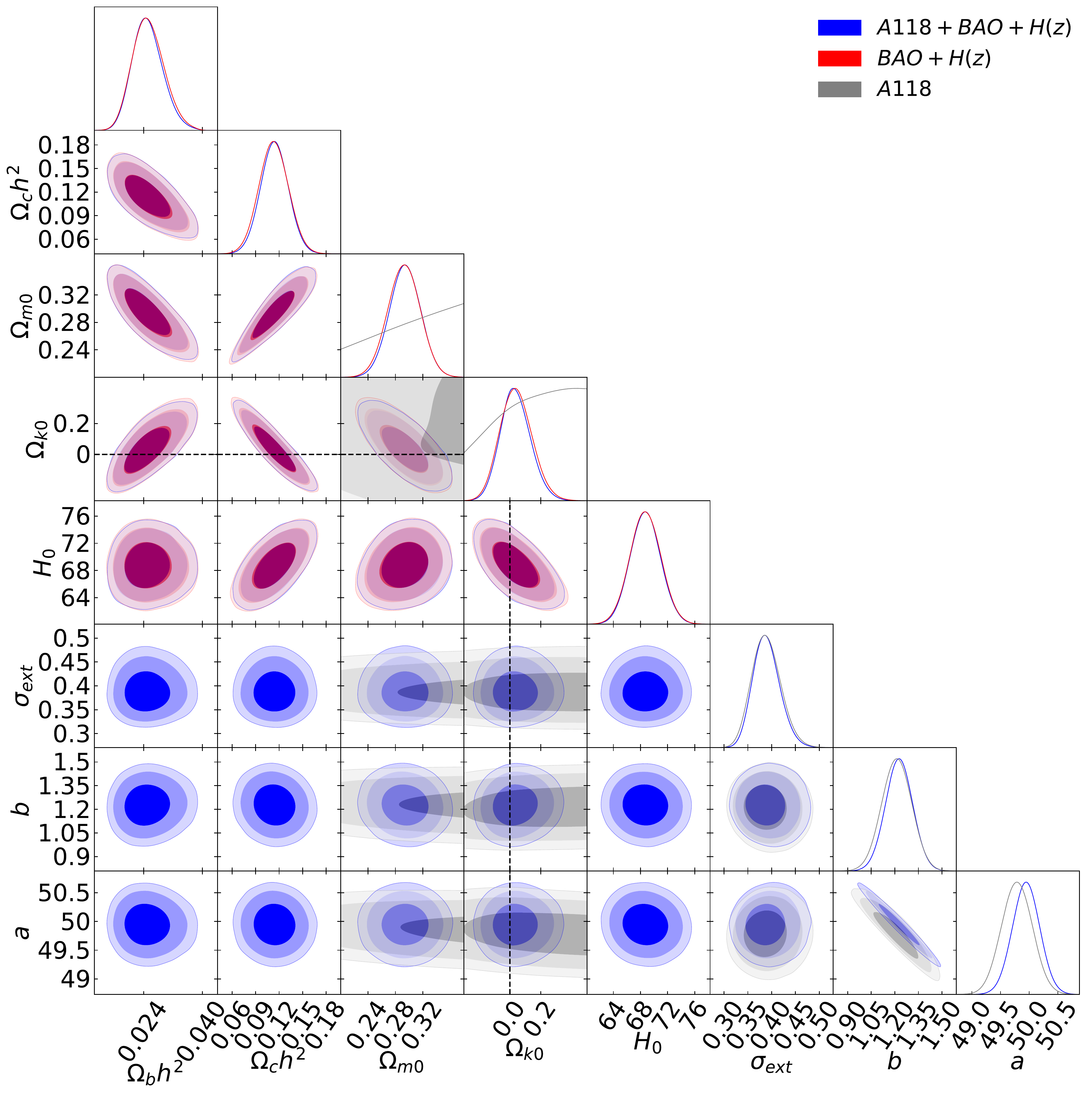}\par
\end{multicols}
\caption{One-dimensional likelihood distributions and two-dimensional contours at 1$\sigma$, 2$\sigma$, and 3$\sigma$ confidence levels using A118 (gray), BAO + $H(z)$ (red), and A118 + BAO + $H(z)$ (blue) data for all free parameters. Left panel shows the flat $\Lambda$CDM model and right panel shows the non-flat $\Lambda$CDM model. The black dashed straight lines in the right panel correspond to $\Omega_{k0} = 0$.}
\label{fig:13}
\end{figure*}

\begin{figure*}
\begin{multicols}{2}    
    \includegraphics[width=\linewidth]{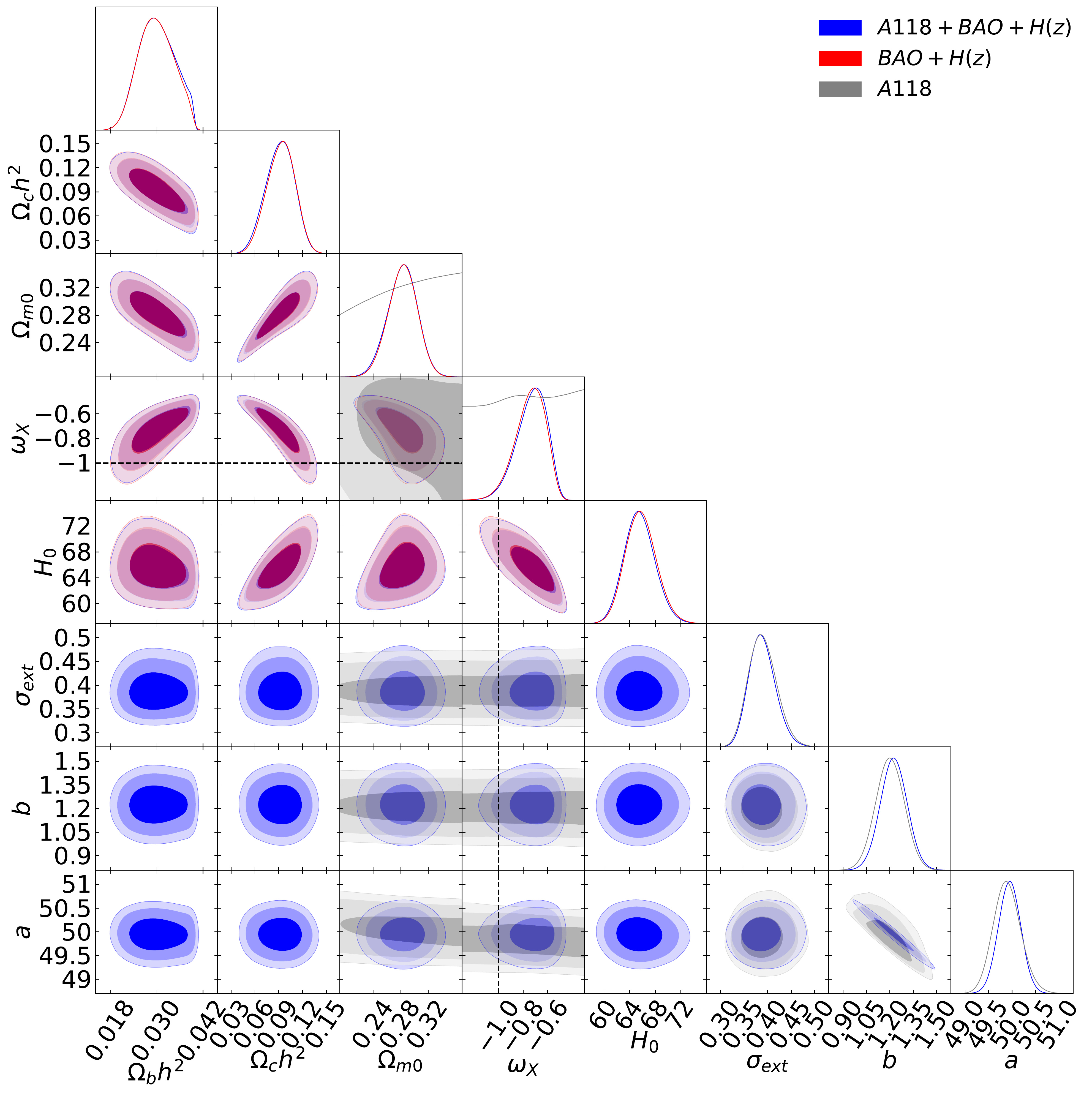}\par
    \includegraphics[width=\linewidth]{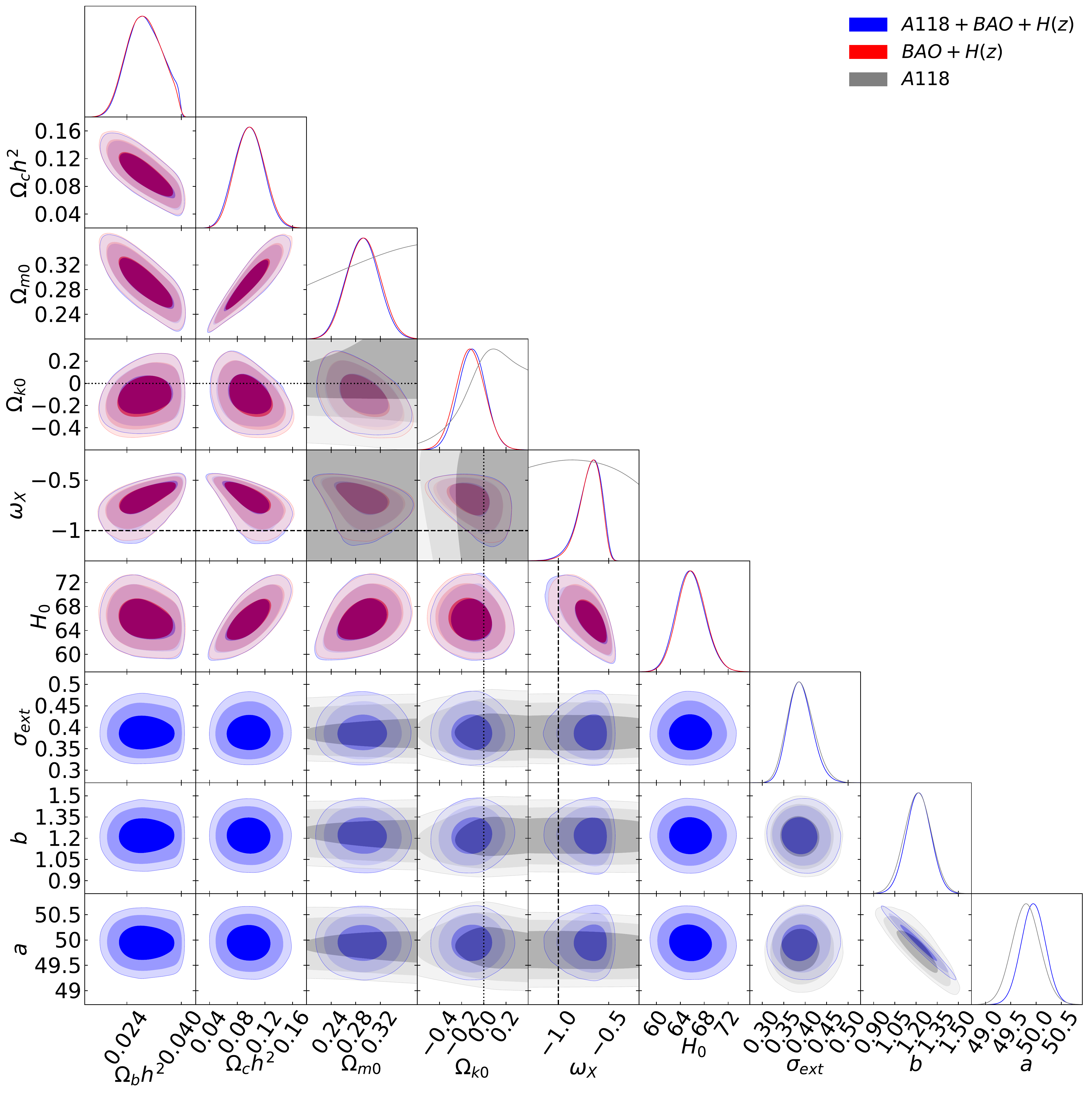}\par
\end{multicols}
\caption{One-dimensional likelihood distributions and two-dimensional contours at 1$\sigma$, 2$\sigma$, and 3$\sigma$ confidence levels using A118 (gray), BAO + $H(z)$ (red), and A118 + BAO + $H(z)$ (blue) data for all free parameters. Left panel shows the flat XCDM parametrization. Right panel shows the non-flat XCDM parametrization. The black dashed straight lines in both panels correspond to the $\omega_x = -1$ $\Lambda$CDM models. The black dotted straight lines in the $\Omega_{k0}$ subpanels in the right panel correspond to $\Omega_{k0} = 0$.}
\label{fig:13}
\end{figure*}

\begin{figure*}
\begin{multicols}{2}    
    \includegraphics[width=\linewidth]{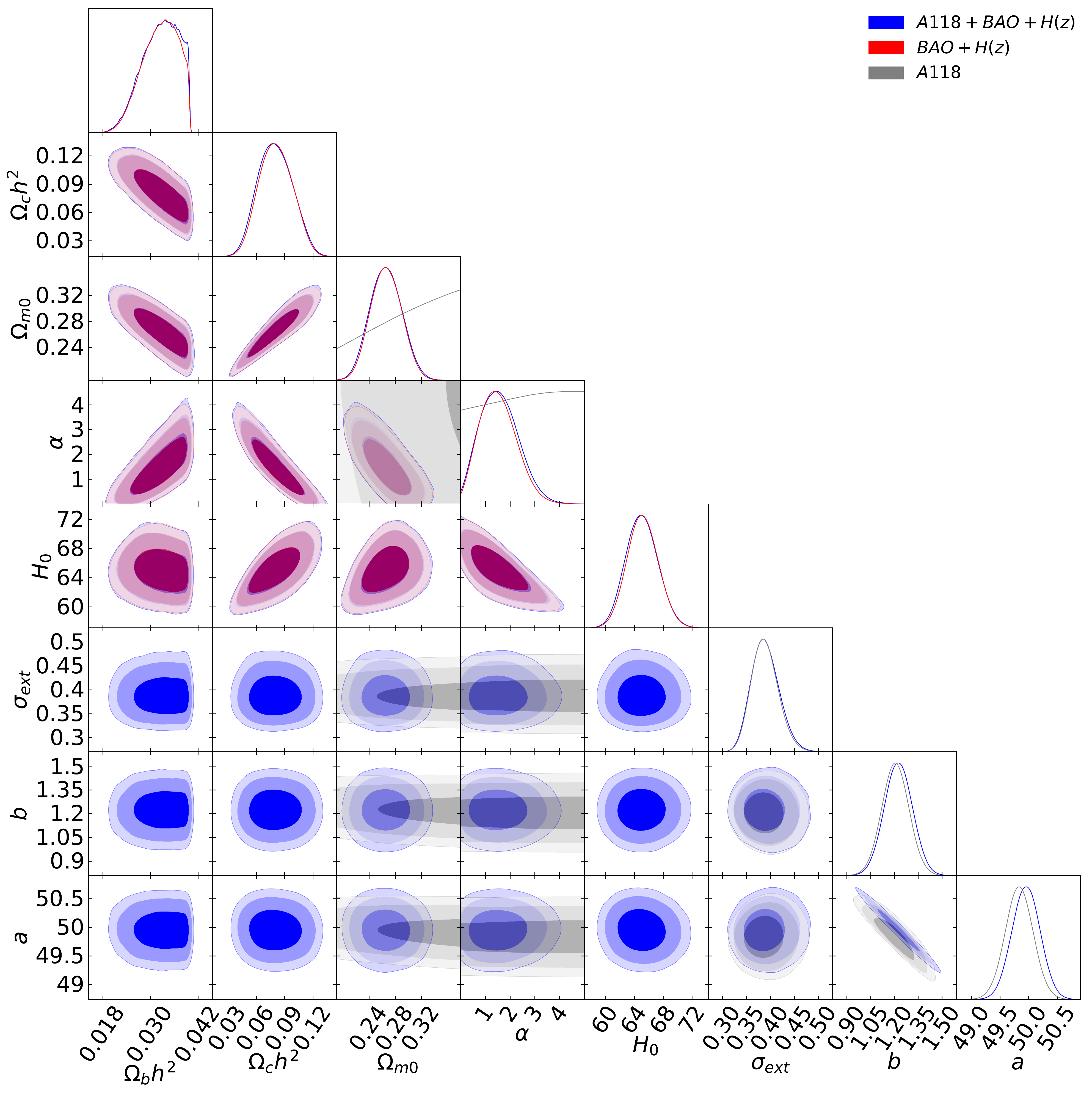}\par
    \includegraphics[width=\linewidth]{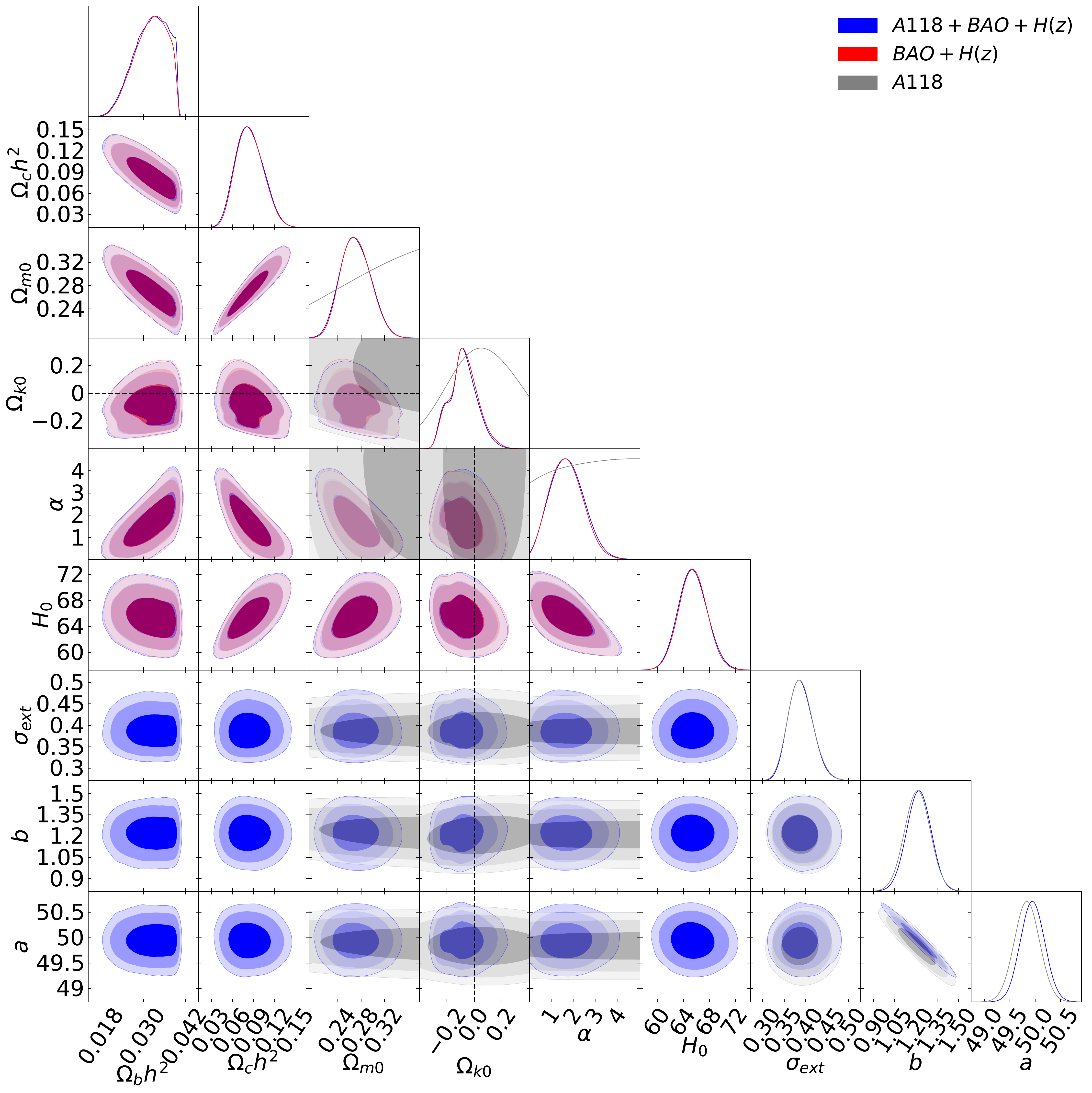}\par
\end{multicols}
\caption{One-dimensional likelihood distributions and two-dimensional contours at 1$\sigma$, 2$\sigma$, and 3$\sigma$ confidence levels using A118 (gray), BAO + $H(z)$ (red), and A118 + BAO + $H(z)$ (blue) data for all free parameters. Left panel shows the flat $\phi$CDM model and right panel shows the non-flat $\phi$CDM model. The $\alpha = 0$ axes correspond to the $\Lambda$CDM models. The black dashed straight lines in the $\Omega_{k0}$ subpanels in the right panel correspond to $\Omega_{k0} = 0$.}
\label{fig:13}
\end{figure*}

Results for the BAO + $H(z)$ data are listed in Tables 3 and 4 and shown in Figs.\ 1--6. These results are from \cite{KhadkaRatra2021}. We use them here for comparing to the GRB results we derive here, to examine consistency between the GRB data constraints and constraints from the better-established BAO + $H(z)$ cosmological probes.

From Figs.\ 1--3, we see that the BAO + $H(z)$ data results are not inconsistent with the GRB A118 data cosmological constraints --- and that both of these data sets favor currently accelerated cosmological expansion [the BAO + $H(z)$ more so than the A118 GRB] --- so it is reasonable to derive cosmological constraints from joint analyses of A118 + BAO + $H(z)$ data. On the other hand,  the BAO + $H(z)$ data constraints are somewhat inconsistent with those from the GRB A102 data and especially with those from the A220 data (that more favor currently decelerated cosmological expansion). This is likely related to the higher intrinsic dispersion of the A102 and A220 data sets, as discussed above, and so it is not appropriate to perform joint analyses of the A102 or A220 data with BAO + $H(z)$ data for the purpose of constraining cosmological parameters.   

Results for the A118 + BAO + $H(z)$ data are listed in Tables 3 and 4. One-dimensional likelihoods and two-dimensional constraint contours for these data are shown in blue in Figs.\ 4--6, where the A118 data results are shown in grey and the BAO + $H(z)$ data results are shown in red. These results are an update of those of \cite{KhadkaRatra2020c}; here we have slightly updated the BAO and GRB data sets used there, now use $\Omega_b h^2$ and $\Omega_c h^2$ as free parameters in the analyses here (compared to using $\Omega_{m0}$ as the free parameter there), and now use \textsc{MontePython} (instead of \textsc{emcee}) for the analyses here. These updates lead to only small changes compared to the results shown in \cite{KhadkaRatra2020c} so our discussion on them is brief.

The figures show that the A118 GRB data constraints are significantly broader than those that follow from the BAO +$H(z)$ data. Consequently, a joint analysis of the A118 + BAO +$H(z)$ data combination will result in only marginally more restrictive constraints on cosmological parameters compared to the constraints from the BAO + $H(z)$ data. 

From Table 4, for the A118 + BAO + $H(z)$ data, the value of $\Omega_{b}h^2$ ranges from $0.024^{+0.003}_{-0.003}$ to $0.033^{+0.007}_{-0.003}$. The minimum value is obtained in the flat $\Lambda$CDM model and the maximum value is obtained in the flat $\phi$CDM model. The value of $\Omega_{c}h^2$ ranges from $0.080^{+0.017}_{-0.020}$ to $0.120^{+0.007}_{+0.008}$. The minimum value is obtained in the flat $\phi$CDM model while the maximum value is obtained in the flat $\Lambda$CDM model. The value of $\Omega_{m0}$ ranges from $0.266^{+0.024}_{-0.024}$ to $0.300^{+0.015}_{+0.017}$. The minimum value is obtained in the flat $\phi$CDM model and the maximum value is obtained in the flat $\Lambda$CDM model. As expected, these values are almost identical to the BAO + $H(z)$ results.

From Table 4, in the flat $\Lambda$CDM model, the value of $\Omega_{\Lambda}$ is determined to be $0.700^{+0.017}_{-0.015}$. In the non-flat $\Lambda$CDM model, the value of $\Omega_{\Lambda}$ is determined to be $0.669^{+0.088}_{-0.076}$.

In case of the A118 + BAO + $H(z)$ analyses $H_0$ is a free parameter to be determined from the data. The value of $H_0$ ranges from $65.0 \pm 2.2$ to $69.2 \pm 1.7$ km s${}^{-1}$ Mpc${}^{-1}$. The minimum value is obtained in the flat $\phi$CDM model and the maximum value is obtained in the flat $\Lambda$CDM model.

For this joint analysis, for the three non-flat models, the value of $\Omega_{k0}$ ranges from $-0.110^{+0.120}_{-0.120}$ to $0.037^{+0.089}_{-0.100}$. The minimum value is obtained in the non-flat XCDM parametrization and the maximum value is obtained in the non-flat $\Lambda$CDM model. These data are consistent with flat spatial hypersurfaces but do not rule out a little spatial curvature.

From Table 4, for the flat and non-flat XCDM parametrization, the values of $\omega_X$ are determined to be $-0.733^{+0.150}_{-0.095}$ and $-0.694^{+0.140}_{-0.079}$ respectively. In the flat and non-flat $\phi$CDM model, the values of scalar field potential energy density parameter $\alpha$ are determined to be $1.580^{+0.660}_{-0.890}$ and $1.710^{+0.700}_{-0.850}$ respectively. In all four cases, dynamical dark energy is favored at 1.8$\sigma$ to 3.9$\sigma$ statistical significance over the cosmological constant.

From Table 3, all the $\Delta AIC$ values for the two data sets are positive or negative, but with |$\Delta AIC|\lesssim2$, indicating the models are almost indistinguishable from the $\Lambda$CDM one. Moving to the $\Delta BIC$ values, we see that for all data sets the non-flat $\Lambda$CDM, XCDM, and $\phi$CDM cases exhibit positive evidence for the flat $\Lambda$CDM model.

\textbf{\subsection{C60, C51, C79, C101, and C174 data constraints}}
\label{combo}

\begin{landscape}
\centering
\small\addtolength{\tabcolsep}{0.0pt}
\begin{threeparttable}
\centering
\small
\caption{Unmarginalized one-dimensional best-fit parameters for Combo correlation GRB data sets. For each data set, $\Delta AIC$ and $\Delta BIC$ values are computed with respect to the $AIC$ and $BIC$  values of the flat  $\Lambda$CDM model.}\label{tab:1d_BFP3}
\setlength{\tabcolsep}{1.3mm}{
\begin{tabular}{lcccccccccccccccccc}
\toprule
Model & Data set & $\Omega_{\rm m0}$ & $\Omega_{\rm k0}$ & $\omega_{X}$ & $\alpha$  & $\sigma_{\rm ext}$ & $q_0$ & $q_1$ & $dof$ & $-2\ln L_{\rm max}$ & $AIC$ & $BIC$ & $\Delta AIC$ & $\Delta BIC$\\
\hline
& C101 & 0.998 & - & - & - & 0.373 & 49.908 & 0.631 & 97 & 111.18 & 119.18 & 129.64 & - & -\\
Flat & C51 & 0.861 & - & - & - & 0.372 & 50.405 & 0.489 & 47 & 51.64 & 59.64 & 67.37 & - & -\\
$\Lambda$CDM& C60 & 0.999 & - & - & - & 0.338 & 49.965 & 0.629 & 56 & 49.68 & 57.68 & 66.05 & - & -\\
& C174  & 0.998 & - & - & - & 0.356 & 49.831 & 0.658 & 170 & 194.14 & 202.14 & 214.78 & - & -\\
& C79  & 0.996 & - & - & - & 0.383 & 50.128 & 0.572 & 75 & 85.70 & 93.70 & 103.18 & - & -\\
\hline
& C101 & 0.997 & $-$0.322 & - &- & 0.375 & 49.929 & 0.628 & 96 & 111.00 & 121.00 & 134.08 & 1.82 & 4.44\\
Non-flat & C51 & 0.966 & 0.384  & - &- & 0.371 & 50.368 & 0.488 & 46 & 51.56 & 61.56 & 71.22 & 1.92 & 3.85\\
$\Lambda$CDM& C60 & 0.992 & $-$0.291 & - &- & 0.337 & 49.986 & 0.626 & 55 & 49.58 & 59.58 & 70.05 & 1.90 & 4.00\\
& C174 & 0.998 & $-$0.529 & - &- & 0.354 & 49.901 & 0.635 & 169 & 193.18 & 203.18 & 218.98 & 1.04 & 2.92\\
& C79 & 0.987 & $-$0.356 & - &- & 0.386 & 50.154 & 0.568 & 74 & 85.44 & 95.44 & 107.29 & 1.74 & 4.11\\
\hline
& C101 & 0.120 & - & 0.142 &-& 0.369 & 49.883 & 0.607 & 96 & 110.06 & 120.06 & 133.14 & 0.88 & 3.50\\
Flat & C51 & 0.428 & - & $-$4.470 & - & 0.369 & 50.679 & 0.493 & 46 & 51.60 & 61.60 & 71.26 & 1.96 & 5.21\\
XCDM & C60 & 0.251 & - & 0.139 &-& 0.338 & 49.964 & 0.605 & 55 & 49.38 & 59.38 & 69.85 & 1.70 & 3.80\\
& C174 & 0.081 & - & 0.141 & - & 0.351 & 49.803 & 0.636 & 169 & 191.90 & 201.90 & 217.70 & $-0.24$ & 2.92\\
& C79 & 0.152 & - & 0.128 & - & 0.383 & 50.082 & 0.563 & 74 & 85.24 & 95.24 & 107.09 & 1.54 & 3.91\\
\hline
& C101 & 0.187 & $-$0.773 & 0.095 & - & 0.372 & 49.824 & 0.585 & 95 & 109.44 & 121.44 & 137.13 & 2.26 & 7.49\\
Non-flat & C51 & 0.988 & 0.378 & $-$0.717 &- & 0.372 & 50.334 & 0.500 & 45 & 51.56 & 63.56 & 75.15 & 5.88 & 7.78\\
XCDM& C60 & 0.440 & $-$0.295 & 0.040 &- & 0.335 & 49.927 & 0.613 & 54 & 49.36 & 61.36 & 73.92 & 3.68 & 7.87\\
& C174 & 0.129 & $-$0.747 & 0.088 &- & 0.351 & 49.794 & 0.595 & 168 & 190.38 & 202.38 & 221.33 & 0.24 & 6.55\\
& C79 & 0.482 & $-$0.503 & 0.095 &- & 0.381 & 50.099 & 0.527 & 73 & 85.14 & 97.14 & 111.36 & 3.44 & 8.36\\

\hline
& C101 & 0.999 & - & - &6.357 & 0.373 & 49.883 & 0.642 & 96 & 111.20 & 121.20 & 134.28 & 2.02 & 4.64\\
Flat & C51 & 0.834 & - & - & 1.891 & 0.370 & 50.408 & 0.491 & 46 & 51.64 & 61.64 & 71.30 & 2.00 & 3.93\\
$\phi$CDM & C60 & 0.994 & - & - & 0.711 & 0.339 & 49.953 & 0.633 & 55 & 49.68 & 59.68 & 70.15 & 2.00 & 4.10\\
& C174 & 0.998 & - & - & 1.495 & 0.352 & 49.844 & 0.653 & 168 & 194.14 & 204.14 & 219.94 & 2.00 & 5.16\\
& C79 & 0.994 & - & - & 6.188 & 0.387 & 50.142 & 0.568 & 74 & 85.70 & 95.70 & 107.55 & 2.00 & 4.37\\

\hline
& C101 & 0.999 & $-$0.389 & - & 5.399 & 0.369 & 49.944 & 0.621 & 95 & 110.98 & 122.98 & 138.67 & 3.80 & 9.03\\
Non-flat & C51 & 0.835 & 0.122 & - & 0.411 & 0.370 & 50.409 & 0.491 & 45 & 51.60 & 63.60 & 75.19 & 3.96 & 7.82\\
$\phi$CDM & C60 & 0.997 & $-$0.330 & - & 1.104 & 0.337 & 50.000 & 0.619 & 54 & 49.58 & 61.58 & 74.15 & 4.00 & 8.10\\
& C174 & 0.996 & $-$0.560 & - & 0.183 & 0.358 & 49.913 & 0.632 & 168 & 193.20 & 205.20 & 224.15 & 3.06 & 9.37\\
& C79 & 0.992 & $-$0.492 & - & 3.215 & 0.387 & 50.159 & 0.567 & 73 & 85.44 & 97.44 & 111.66 & 3.74 & 8.48\\
\bottomrule
\end{tabular}}
\begin{tablenotes}[flushleft]
\item[a]${\rm km}\hspace{1mm}{\rm s}^{-1}{\rm Mpc}^{-1}$. $H_0$ is set to $70$ ${\rm km}\hspace{1mm}{\rm s}^{-1}{\rm Mpc}^{-1}$ for the GRB-only data analyses.
\end{tablenotes}
\end{threeparttable}
\end{landscape}

\begin{landscape}
\centering
\small\addtolength{\tabcolsep}{5.0pt}
\begin{threeparttable}
\centering
\caption{Marginalized one-dimensional best-fit parameters with 1$\sigma$ confidence intervals for Combo correlation GRB data sets. A 2$\sigma$ limit is given when only an upper or lower limit exists.}\label{tab:1d_BFP3}
\setlength{\tabcolsep}{1.5mm}{
\begin{tabular}{lcccccccccccccccccc}
\toprule
Model & Data set & $\Omega_{\rm m0}$ & $\Omega_{\Lambda}$ & $\Omega_{\rm k0}$ & $\omega_{X}$ & $\alpha$ & $\sigma_{\rm ext}$ & $q_0$ & $q_1$ \\
\hline
Flat $\Lambda$CDM & C101 & $> 0.450$ & $< 0.550$ & - & - & - & $0.388^{+0.031}_{-0.039}$ & $49.930^{-0.290}_{-0.290}$ & $0.650^{+0.110}_{-0.110}$\\
& C51 & $> 0.193$ & $< 0.807$ & - & - & - & $0.395^{+0.040}_{-0.055}$ & $50.490^{-0.390}_{-0.390}$ & $0.500^{+0.140}_{-0.140}$\\
& C60 & $> 0.327$ & $< 0.673$ & - & - & - & $0.358^{+0.032}_{-0.043}$ & $49.990^{-0.280}_{-0.280}$ & $0.660^{+0.110}_{-0.110}$\\
& C174 & $> 0.579$ & $< 0.421$ & - & - & - & $0.363^{+0.024}_{-0.028}$ & $49.860^{-0.210}_{-0.210}$ & $0.668^{+0.078}_{-0.078}$\\
& C79 & $> 0.332$ & $< 0.668$ & - & - & - & $0.402^{+0.032}_{-0.042}$ & $50.170^{-0.270}_{-0.270}$ & $0.590^{+0.100}_{-0.100}$\\
\hline
Non-flat $\Lambda$CDM & C101 & $> 0.395$ & $< 1.300$ & $-0.068^{+0.668}_{-0.712}$ & - &-& $0.389^{+0.033}_{-0.039}$ & $49.930^{+0.290}_{-0.290}$ & $0.650^{+0.110}_{-0.110}$\\
& C51 & $> 0.190$ & $< 1.200$ & $-0.053^{+1.203}_{-0.447}$ & - &-& $0.397^{+0.043}_{-0.057}$ & $50.450^{+0.390}_{-0.390}$ & $0.500^{+0.140}_{-0.140}$\\
& C60 & $> 0.307$ & $< 1.400$ & $-0.233^{+0.873}_{-0.507}$ & - &-& $0.360^{+0.032}_{-0.044}$ & $49.990^{+0.300}_{-0.300}$ & $0.650^{+0.120}_{-0.120}$\\
& C174 & $> 0.593$ & $< 1.300$ & $-0.200^{+0.290}_{-0.63}$ & - &-& $0.363^{+0.024}_{-0.028}$ & $49.890^{+0.220}_{-0.220}$ & $0.655^{+0.080}_{-0.080}$\\
& C79 & $> 0.330$ & $< 1.500$ & $-0.226^{+0.786}_{-0.624}$ & - &-& $0.403^{+0.033}_{-0.042}$ & $50.170^{+0.280}_{-0.280}$ & $0.590^{+0.100}_{-0.100}$\\
\hline
Flat XCDM & C101 & > 0.248 & - & - & $< 0.059$ &-& $0.388^{+0.031}_{-0.039}$ & $49.970^{+0.300}_{-0.300}$ & $0.650^{+0.110}_{-0.110}$\\
& C51 & > 0.149 & - & - & $< -0.150$ &-& $0.395^{+0.040}_{-0.055}$ & $50.560^{+0.420}_{-0.420}$ & $0.500^{+0.140}_{-0.140}$\\
& C60 & > 0.209 & - & - & $< 0.010$ &-& $0.358^{+0.032}_{-0.043}$ & $50.040^{+0.300}_{-0.200}$ & $0.650^{+0.110}_{-0.110}$\\
& C174 & > 0.280 & - & - & $< 0.097$ &-& $0.363^{+0.024}_{-0.028}$ & $49.880^{+0.210}_{-0.210}$ & $0.665^{+0.079}_{-0.079}$\\
& C79 & > 0.243 & - & - & $< -0.036$ &-& $0.402^{+0.033}_{-0.042}$ & $50.220^{+0.280}_{-0.280}$ & $0.590^{+0.100}_{-0.100}$\\
\hline
Non-flat XCDM & C101 & $> 0.206$ & - & $-0.217^{+0.617}_{-0.683}$ & $< 0.200$ &-& $0.388^{+0.030}_{-0.037}$ & $49.970^{+0.310}_{-0.310}$ & $0.630^{+0.110}_{-0.110}$\\
& C51 & $> 0.190$ & - & $-0.045^{+0.975}_{-0.395}$ & $< 0.000$ &-& $0.396^{+0.038}_{-0.052}$ & $50.460^{+0.390}_{-0.390}$ & $0.500^{+0.130}_{-0.130}$\\
& C60 & $> 0.260$ & - & $-0.166^{+0.686}_{-0.514}$ & $< 0.100$ &-& $0.360^{+0.032}_{-0.043}$ & $50.010^{+0.320}_{-0.320}$ & $0.640^{+0.120}_{-0.120}$\\
& C174 & --- & - & $-0.424^{+0.414}_{-0.646}$ & --- &-& $0.361^{+0.024}_{-0.028}$ & $49.880^{+0.230}_{-0.250}$ & $0.625^{+0.085}_{-0.085}$\\
& C79 & $> 0.230$ & - & $-0.190^{+0.630}_{-0.590}$ & < 0.100 &-& $0.404^{+0.033}_{-0.042}$ & $50.190^{+0.300}_{-0.300}$ & $0.580^{+0.110}_{-0.110}$\\

\hline
Flat $\phi$CDM & C101 & > 0.463 & - & - & - &---& $0.387^{+0.029}_{-0.036}$ & $49.930^{+0.270}_{-0.270}$ & $0.650^{+0.100}_{-0.100}$\\
& C51 & > 0.201 & - & - & - &---& $0.392^{+0.037}_{-0.052}$ & $50.490^{+0.370}_{-0.370}$ & $0.500^{+0.130}_{-0.130}$\\
& C60 & > 0.336 & - & - & - &---& $0.357^{+0.032}_{-0.043}$ & $49.990^{+0.280}_{-0.280}$ & $0.660^{+0.110}_{-0.110}$\\
& C174 & > 0.582 & - & - & - &---& $0.363^{+0.023}_{-0.027}$ & $49.860^{+0.200}_{-0.200}$ & $0.668^{+0.076}_{-0.076}$\\
& C79 & > 0.338 & - & - & - &---& $0.401^{+0.032}_{-0.041}$ & $50.170^{+0.260}_{-0.260}$ & $0.590^{+0.100}_{-0.100}$\\
\hline
Non-flat $\phi$CDM & C101 & $> 0.467$ & - & $-0.178^{+0.288}_{-0.312}$ & - &--- & $0.387^{+0.031}_{-0.037}$ & $49.960^{+0.270}_{-0.270}$ & $0.640^{+0.100}_{-0.100}$\\
& C51 & $0.572^{+0.288}_{-0.222}$ & - & $-0.035^{+0.155}_{-0.485}$ & - &--- & $0.393^{+0.040}_{-0.053}$ & $50.490^{+0.370}_{-0.370}$ & $0.500^{+0.130}_{-0.130}$\\
& C60 & $> 0.349$ & - & $-0.141^{+0.311}_{-0.319}$ & - &--- & $0.357^{+0.034}_{-0.044}$ & $50.020^{+0.280}_{-0.280}$ & $0.650^{+0.110}_{-0.110}$\\
& C174 & $> 0.587$ & - & $-0.309^{+0.329}_{-0.301}$ & - &--- & $0.362^{+0.023}_{-0.027}$ & $49.900^{+0.200}_{-0.200}$ & $0.654^{+0.076}_{-0.076}$\\
& C79 & $> 0.354$ & - & $-0.155^{+0.315}_{-0.325}$ & - &--- & $0.401^{+0.034}_{-0.042}$ & $50.190^{+0.260}_{-0.260}$ & $0.590^{+0.100}_{-0.100}$\\
\bottomrule
\end{tabular}}
\begin{tablenotes}[flushleft]
\item[] $H_0$ is set to $70$ ${\rm km}\hspace{1mm}{\rm s}^{-1}{\rm Mpc}^{-1}$ for these GRB-only data analyses.
\end{tablenotes}
\end{threeparttable}
\end{landscape}

Results for C60, C51, C79, C101, and C174 data sets are given Tables 5 and 6. The unmarginalized best-fit parameter values are listed in Table 5 and marginalized one-dimensional best-fit parameter values and limits are listed in Table 6. The corresponding two-dimensional likelihood contours and one-dimensional likelihoods are plotted in Figs.\ 7--12 where the C60, C51, C79, C101, and C174 data set results are shown in indigo, blue, peru/orange, green, and olive colors respectively.

\begin{figure*}
\begin{multicols}{2}    
    \includegraphics[width=\linewidth]{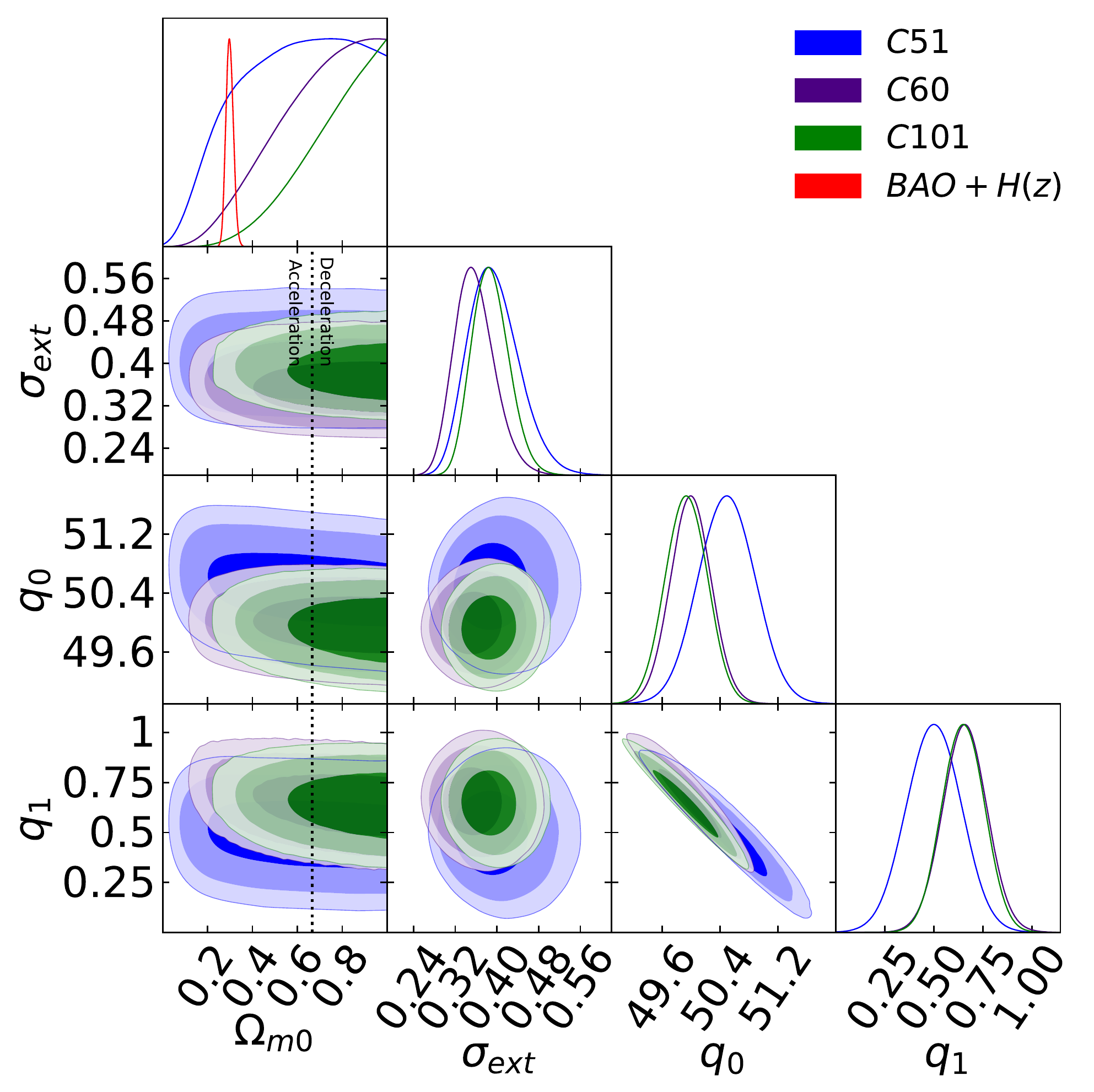}\par
    \includegraphics[width=\linewidth]{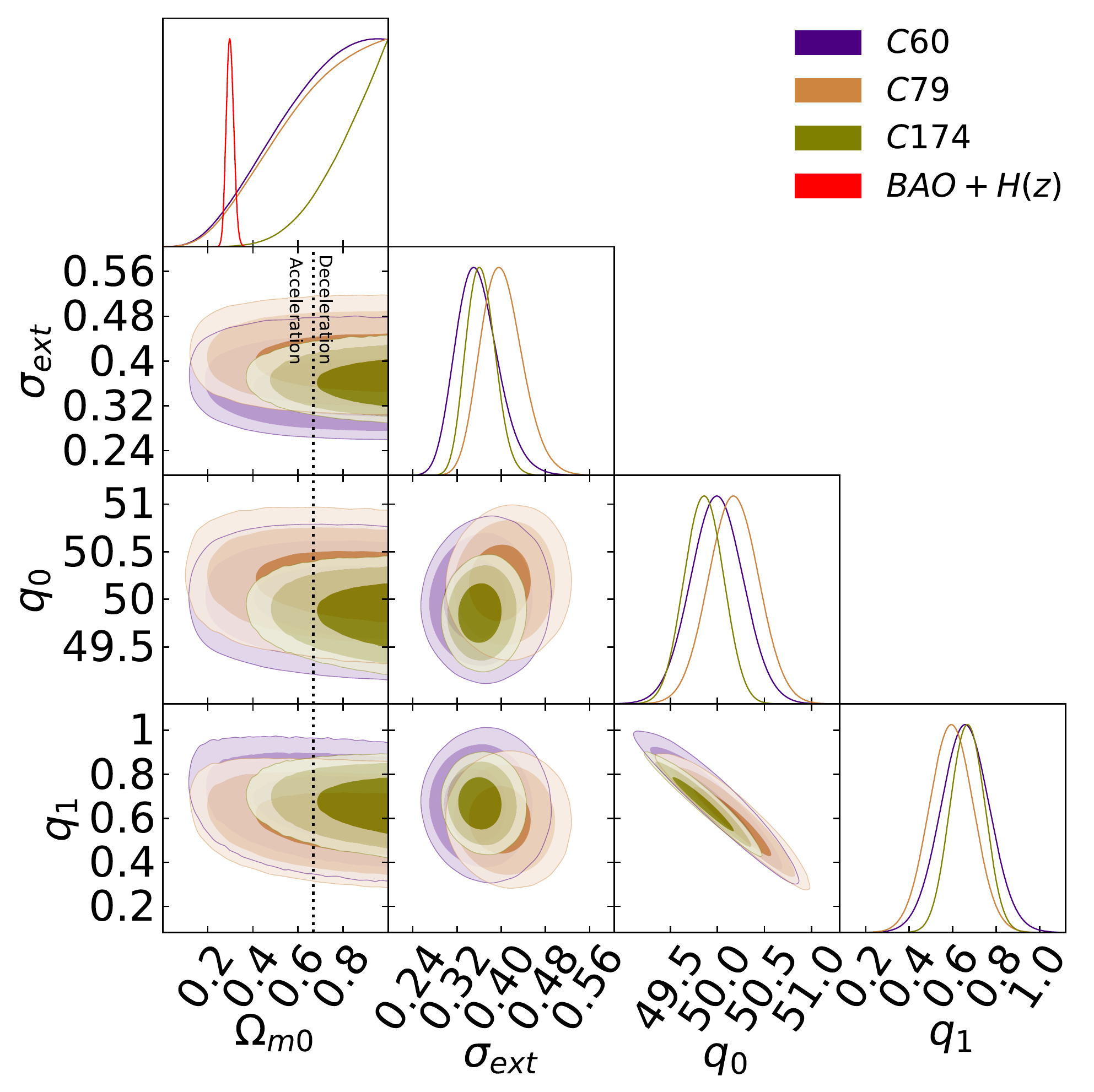}\par
\end{multicols}
\caption{One-dimensional likelihood distributions and two-dimensional contours at 1$\sigma$, 2$\sigma$, and 3$\sigma$ confidence levels for all free parameters in the flat $\Lambda$CDM model. Left panel shows the plots for the C51 (blue), C60 (indigo), C101 (green), and BAO + $H(z)$ (red) data sets. Right panel shows the plots for the C60 (indigo), C79 (peru), C174 (olive) and BAO + $H(z)$ (red) data sets. The black dotted vertical lines are the zero acceleration lines with currently accelerated cosmological expansion occurring to the left of the lines.}
\label{fig:13}
\end{figure*}

\begin{figure*}
\begin{multicols}{2}    
    \includegraphics[width=\linewidth]{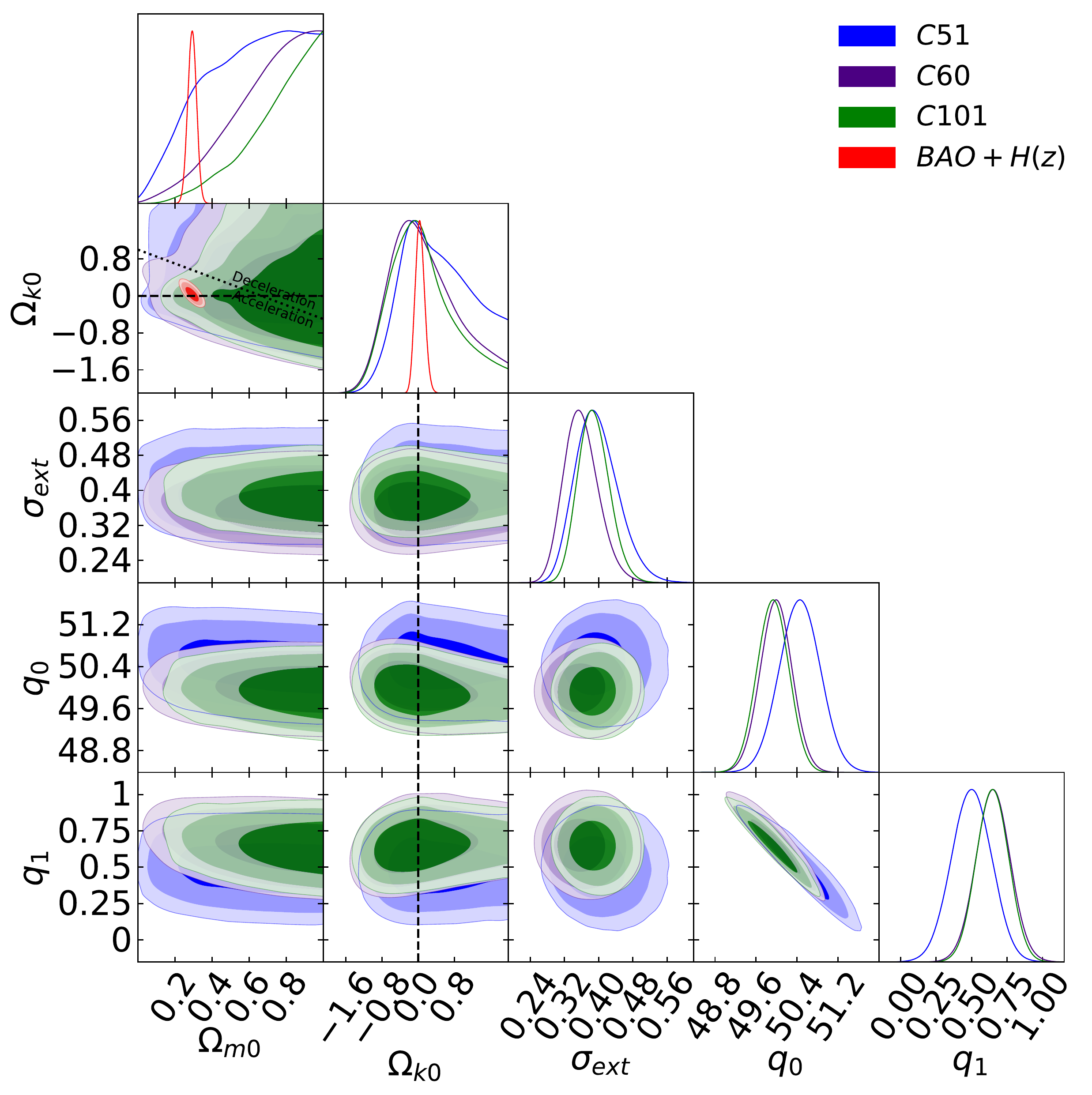}\par
    \includegraphics[width=\linewidth]{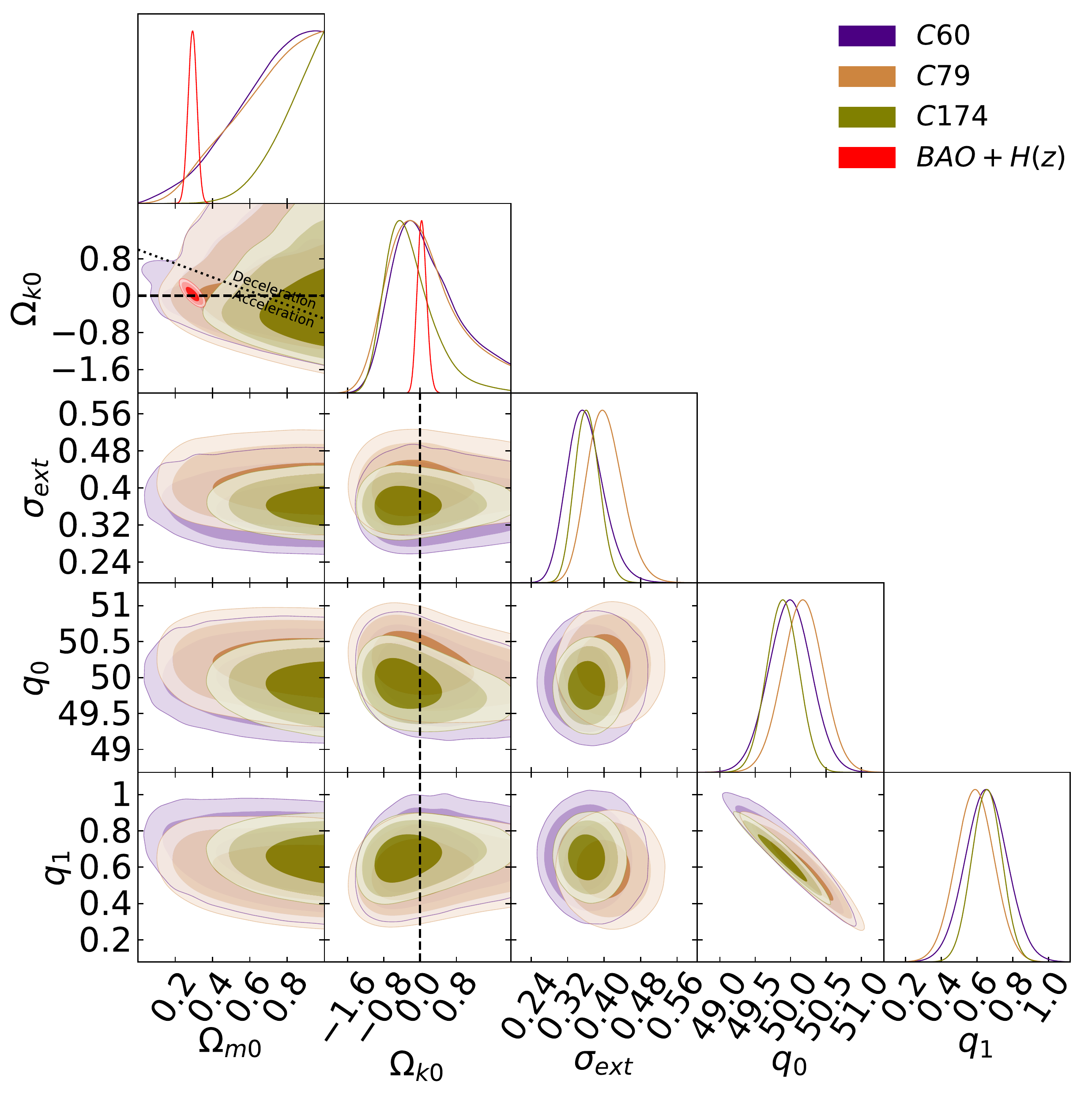}\par
\end{multicols}
\caption{One-dimensional likelihood distributions and two-dimensional contours at 1$\sigma$, 2$\sigma$, and 3$\sigma$ confidence levels for all free parameters in the non-flat $\Lambda$CDM model. Left panel shows the plots for the C51 (blue), C60 (indigo), C101 (green), and BAO + $H(z)$ (red) data sets. Right panel shows the plots for the C60 (indigo), C79 (peru), C174 (olive) and BAO + $H(z)$ (red) data sets. The black dotted sloping line in the $\Omega_{k0}-\Omega_{m0}$ subpanels is the zero acceleration line with currently accelerated cosmological expansion occurring to the lower left of the line. The black dashed horizontal or vertical line in the $\Omega_{k0}$ subpanels correspond to $\Omega_{k0} = 0$.}
\label{fig:13}
\end{figure*}

\begin{figure*}
\begin{multicols}{2}    
    \includegraphics[width=\linewidth]{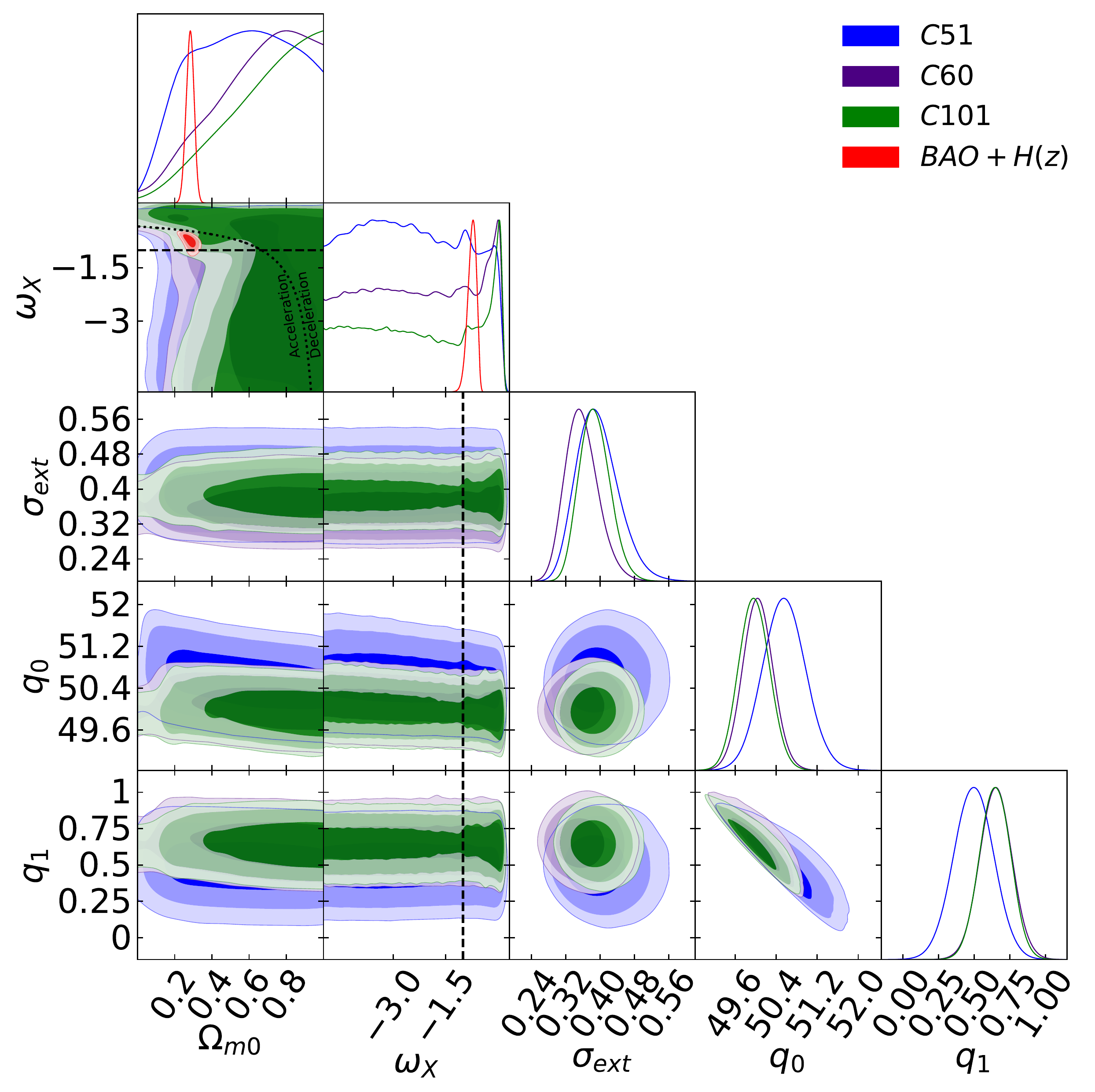}\par
    \includegraphics[width=\linewidth]{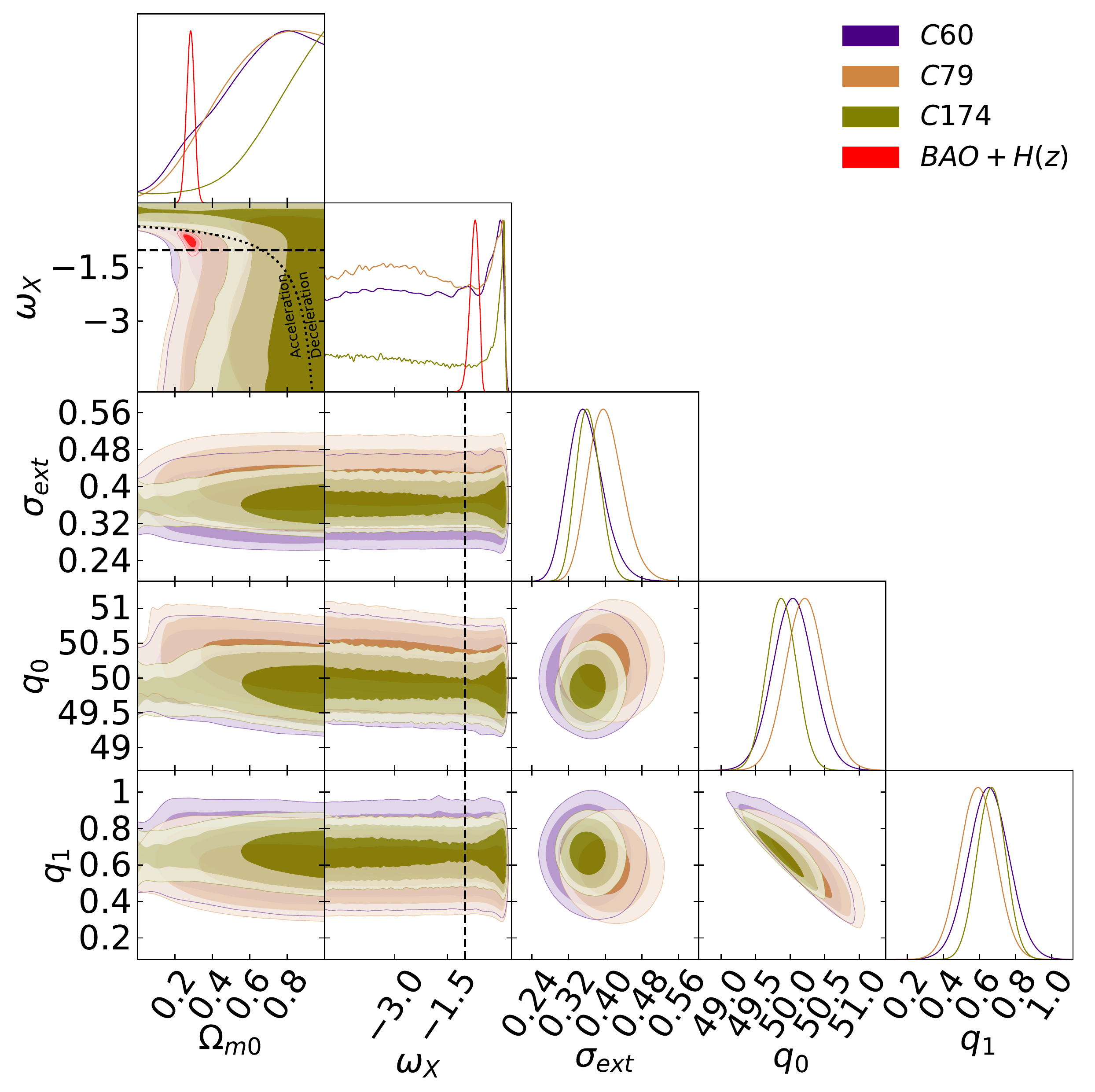}\par
\end{multicols}
\caption{One-dimensional likelihood distributions and two-dimensional contours at 1$\sigma$, 2$\sigma$, and 3$\sigma$ confidence levels for all free parameters in the flat XCDM model. Left panel shows the plots for the C51 (blue), C60 (indigo), C101 (green), and BAO + $H(z)$ (red) data sets. Right panel shows the plots for the C60 (indigo), C79 (peru), C174 (olive) and BAO + $H(z)$ (red) data sets. The black dotted curved line in the $\omega_X-\Omega_{m0}$ subpanels is the zero acceleration line with currently accelerated cosmological expansion occurring below the line and the black dashed straight lines correspond to the $\omega_X = -1$ $\Lambda$CDM model.}
\label{fig:13}
\end{figure*}

\begin{figure*}
\begin{multicols}{2}    
    \includegraphics[width=\linewidth]{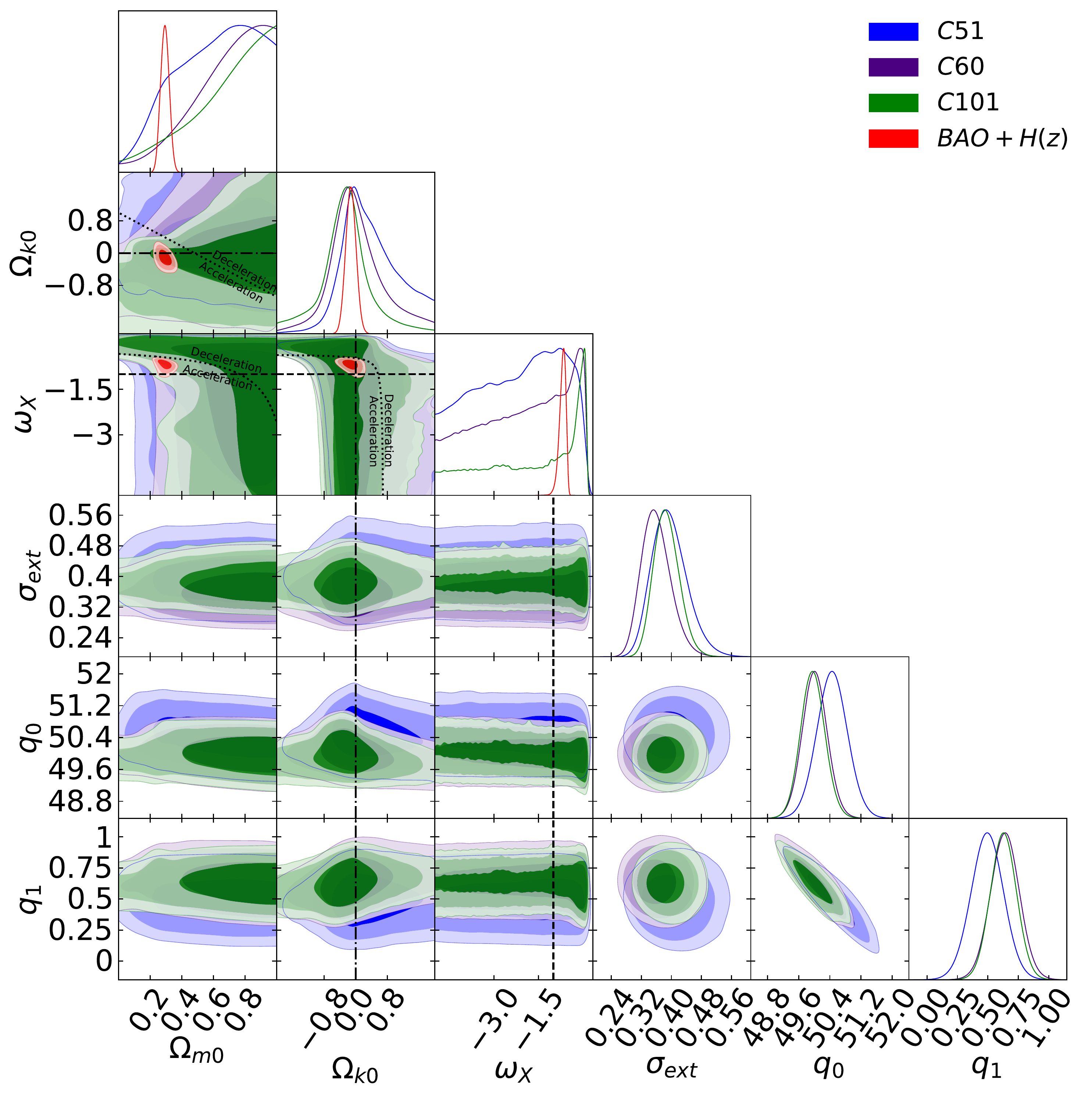}\par
    \includegraphics[width=\linewidth]{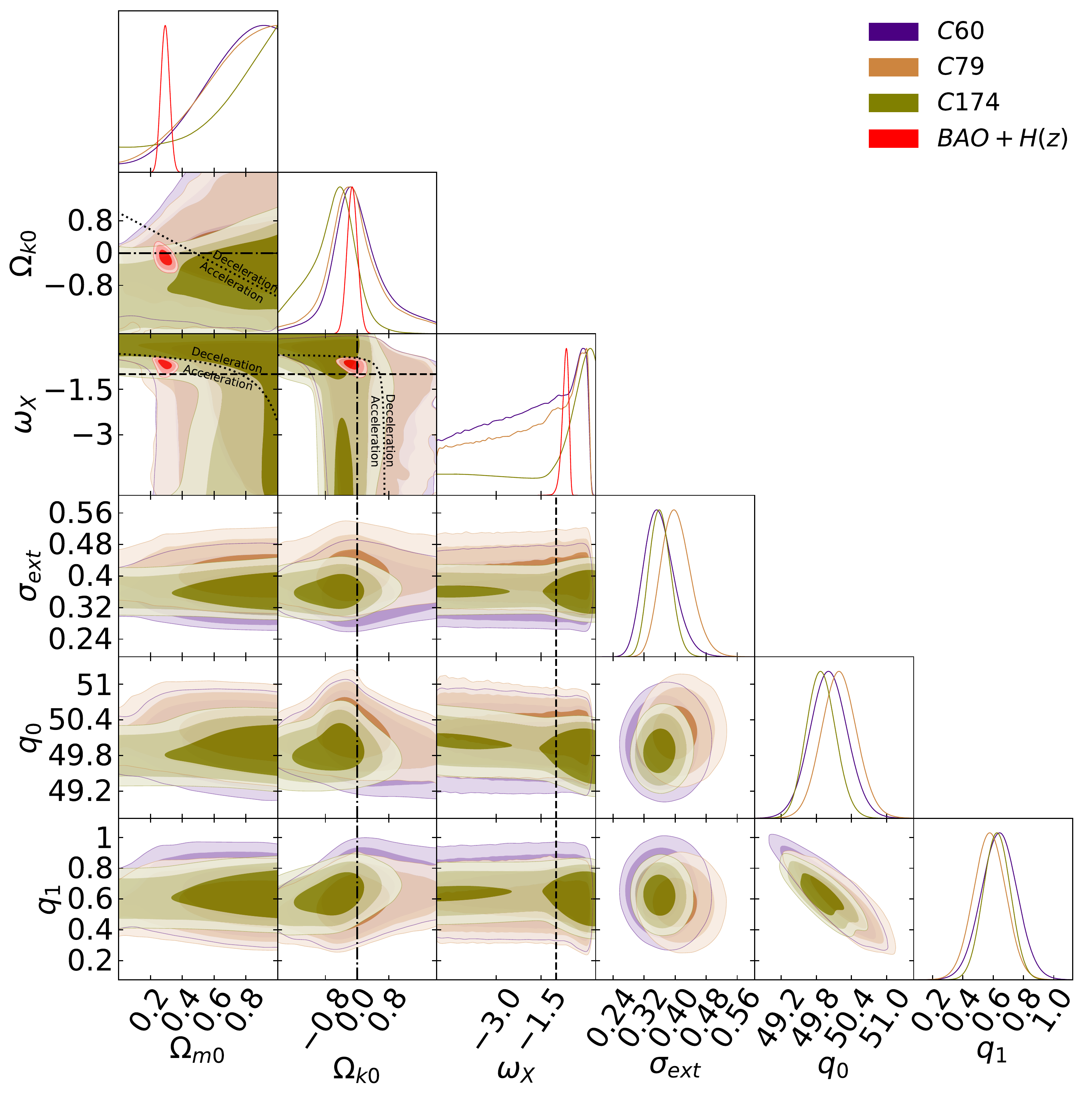}\par
\end{multicols}
\caption{One-dimensional likelihood distributions and two-dimensional contours at 1$\sigma$, 2$\sigma$, and 3$\sigma$ confidence levels for all free parameters in the non-flat XCDM model. Left panel shows the plots for the C51 (blue), C60 (indigo), C101 (green), and BAO + $H(z)$ (red) data sets. Right panel shows the plots for the C60 (indigo), C79 (peru), C174 (olive) and BAO + $H(z)$ (red) data sets. The black dotted lines in the $\Omega_{k0}-\Omega_{m0}$, $\omega_X-\Omega_{m0}$, and $\omega_X-\Omega_{k0}$ subpanels are the zero acceleration lines with currently accelerated cosmological expansion occurring below the lines. Each of the three lines is computed with the third parameter set to the BAO + $H(z)$ data best-fit value of Table 3. The black dashed straight lines correspond to the $\omega_x = -1$ $\Lambda$CDM model. The black dotted-dashed straight lines correspond to $\Omega_{k0} = 0$.}
\label{fig:13}
\end{figure*}

\begin{figure*}
\begin{multicols}{2}    
    \includegraphics[width=\linewidth]{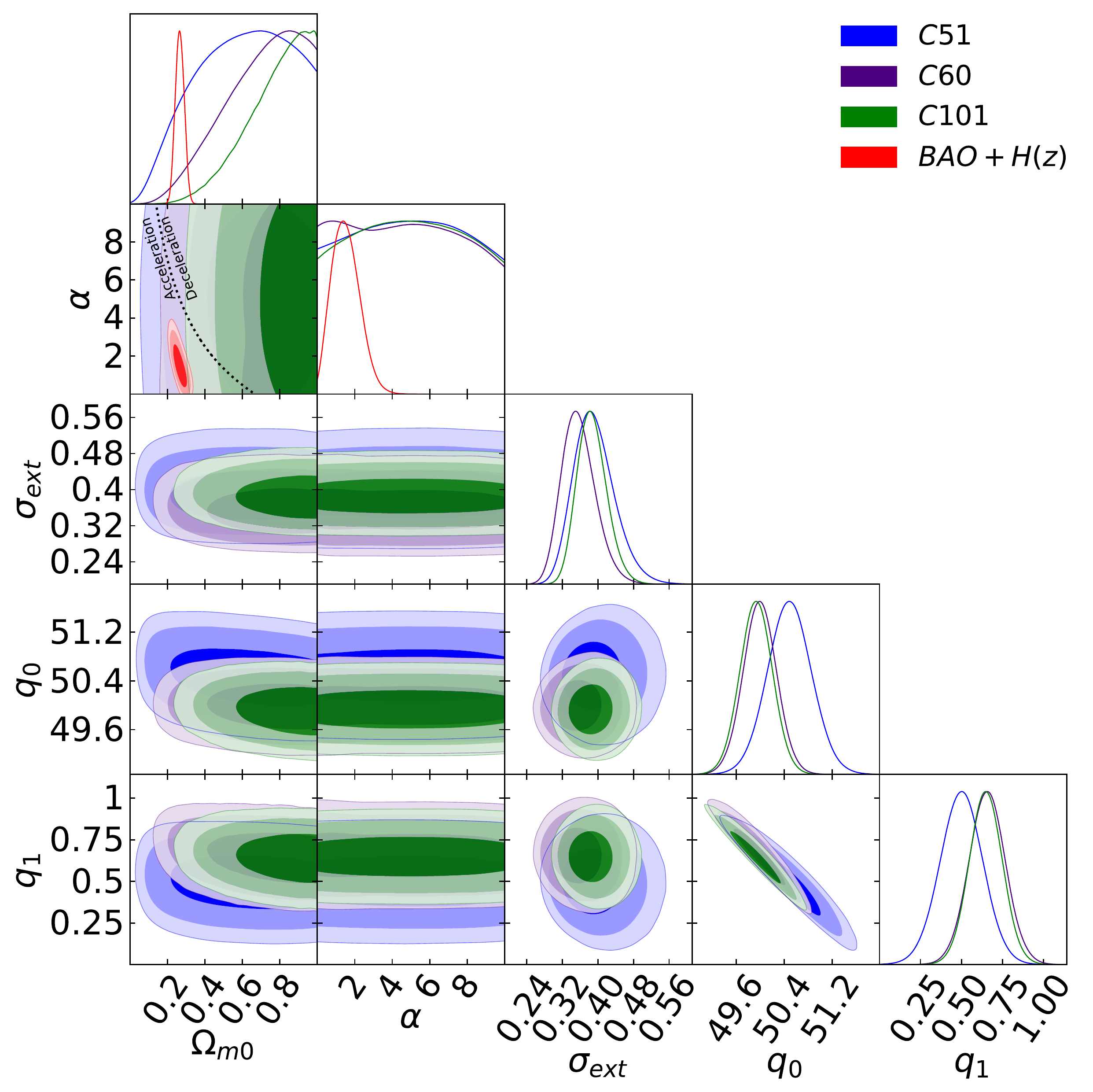}\par
    \includegraphics[width=\linewidth]{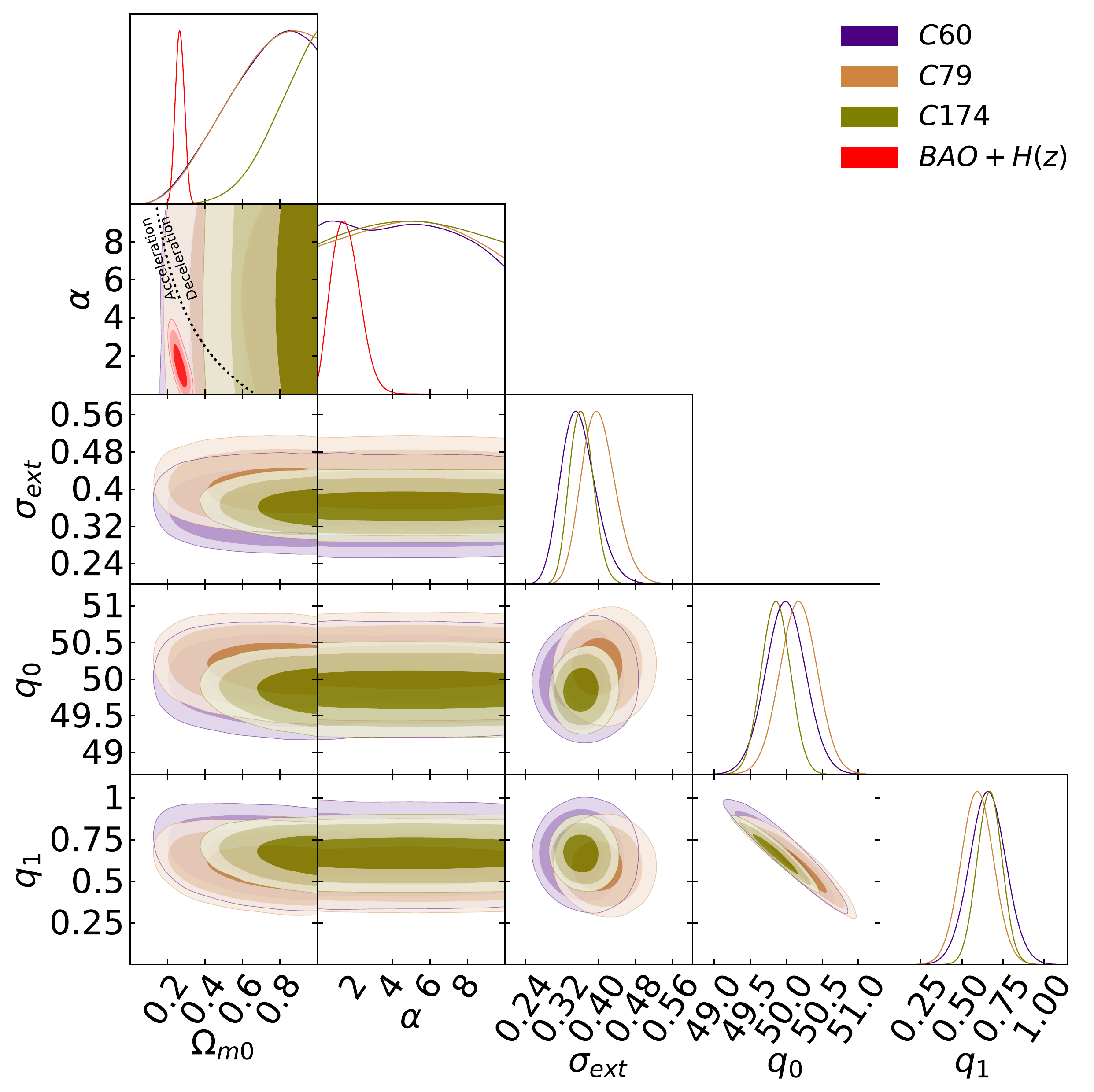}\par
\end{multicols}
\caption{One-dimensional likelihood distributions and two-dimensional contours at 1$\sigma$, 2$\sigma$, and 3$\sigma$ confidence levels for all free parameters in the flat $\phi$CDM model. Left panel shows the plots for the C51 (blue), C60 (indigo), C101 (green), and BAO + $H(z)$ (red) data sets. Right panel shows the plots for the C60 (indigo), C79 (peru), C174 (olive) and BAO + $H(z)$ (red) data sets. The $\alpha = 0$ axes correspond to the $\Lambda$CDM model. The black dotted curved line in the $\alpha - \Omega_{m0}$ subpanels is the zero acceleration line with currently accelerated cosmological expansion occurring to the left of the line.}
\label{fig:13}
\end{figure*}

\begin{figure*}
\begin{multicols}{2}    
    \includegraphics[width=\linewidth]{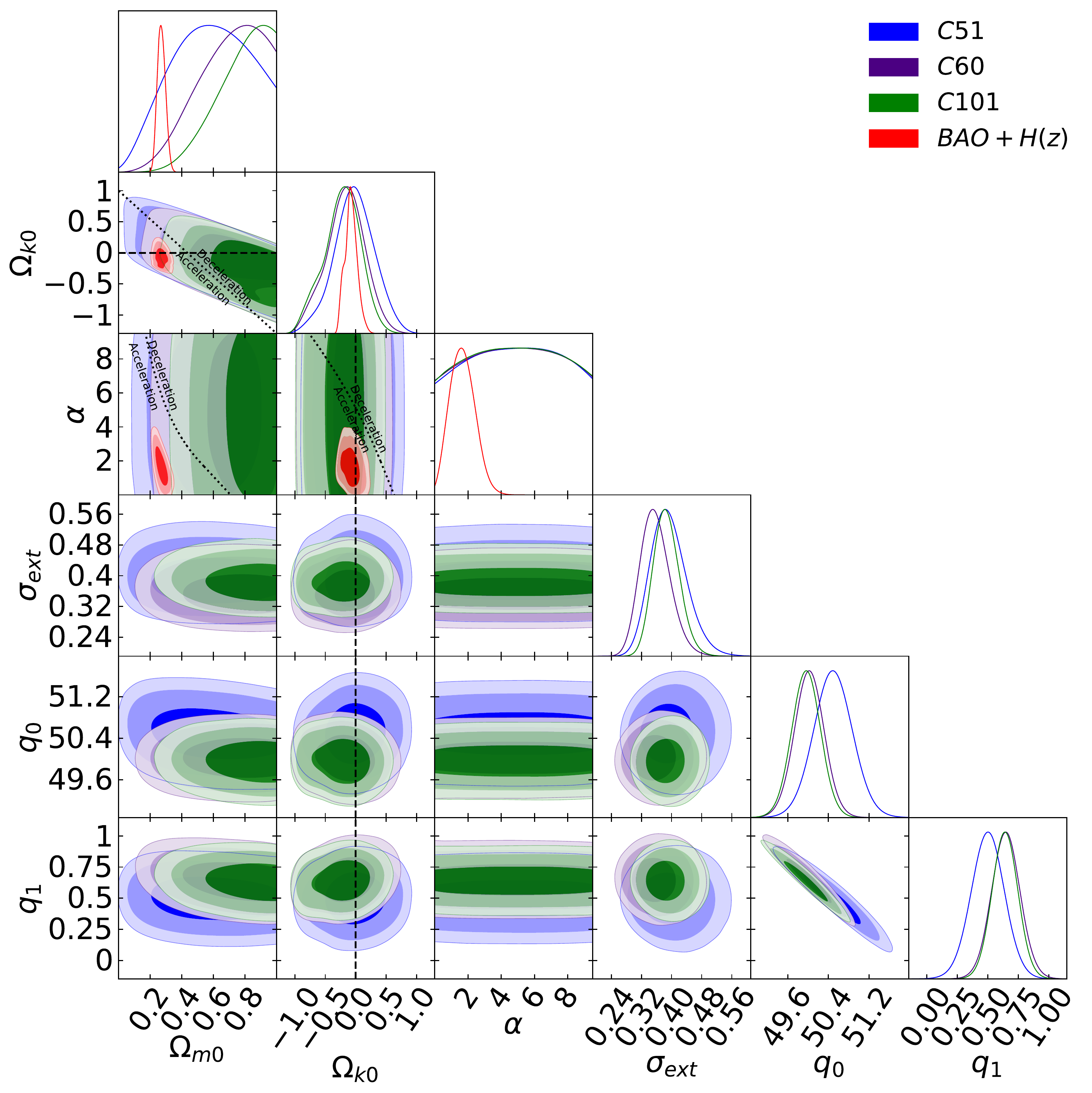}\par
    \includegraphics[width=\linewidth]{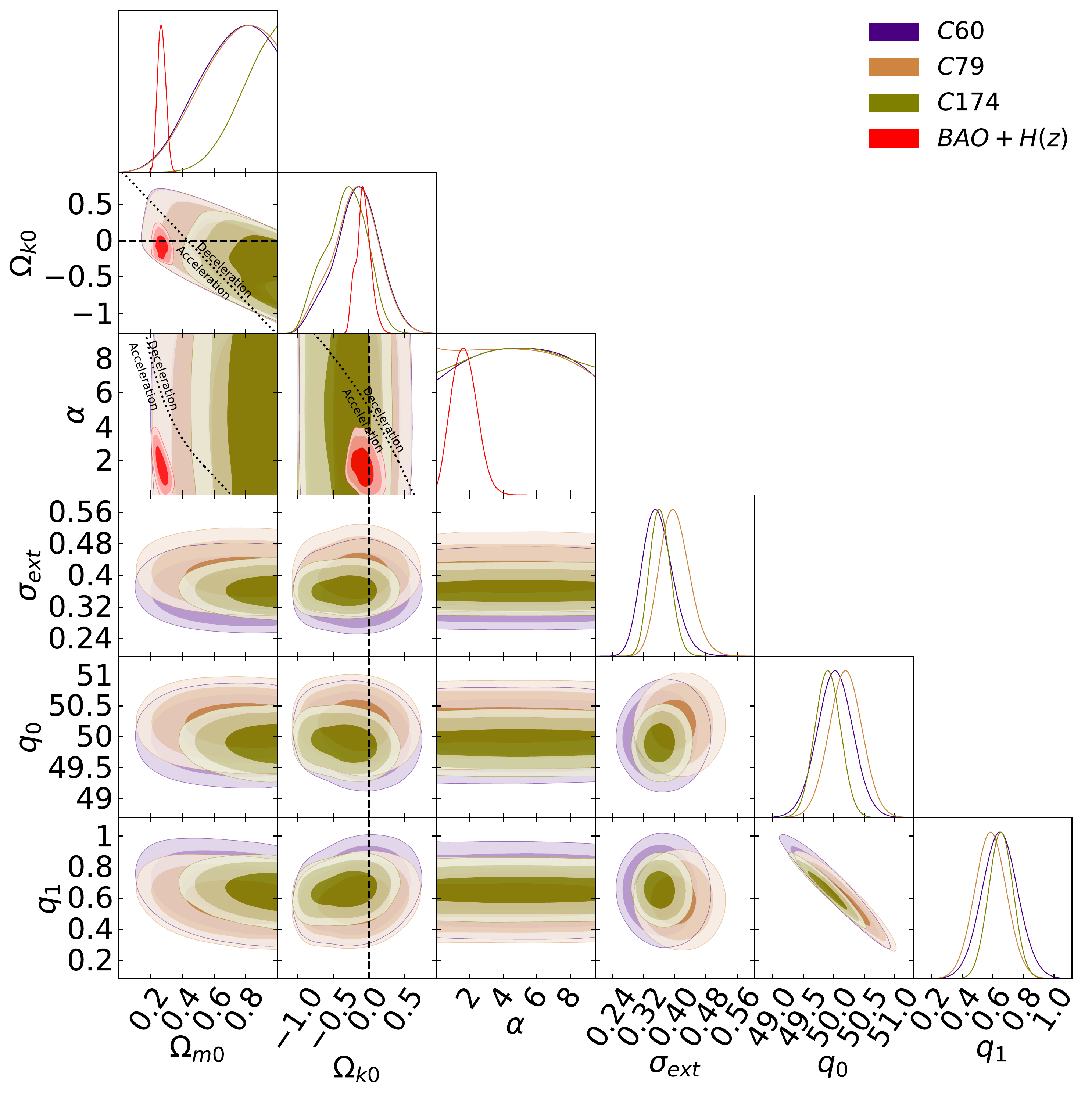}\par
\end{multicols}
\caption{One-dimensional likelihood distributions and two-dimensional contours at 1$\sigma$, 2$\sigma$, and 3$\sigma$ confidence levels for all free parameters in the non-flat $\phi$CDM model. Left panel shows the plots for the C51 (blue), C60 (indigo), C101 (green), and BAO + $H(z)$ (red) data sets. Right panel shows the plots for the C60 (indigo), C79 (peru), C174 (olive) and BAO + $H(z)$ (red) data sets. The $\alpha = 0$ axes correspond to the $\Lambda$CDM model. The black dotted lines in the $\Omega_{k0}-\Omega_{m0}$, $\alpha-\Omega_{m0}$, and $\alpha-\Omega_{k0}$ subpanels are the zero acceleration lines with currently accelerated cosmological expansion occurring below the lines. Each of the three lines is computed with the third parameter set to the BAO + $H(z)$ data best-fit value of Table 3. The black dashed straight lines correspond to $\Omega_{k0} = 0$.}
\label{fig:13}
\end{figure*}

The use of Combo GRBs to constrain cosmological parameters is based on the validity of the Combo correlation. For these five data sets, for all models, values of the Combo correlation intercept $q_0$ lie in the range $\sim 49-51$ and values of the Combo correlation slope $q_1$ lie in the range $\sim 0.5 - 0.7$. For a given data set the measured $q_0$ and $q_1$ values --- and so the Combo relation --- are independent of cosmological model, indicating that these GRBs are standardized. However, the $q_0$ and $q_1$ values are somewhat different for each data set.

The minimum value of the intrinsic dispersion $\sigma_{\rm ext}$, $\sim 0.36$, is obtained for the C60 data set and the maximum value of $\sigma_{\rm ext}$, $\sim 0.40$, is obtained for the C79 data set. As discussed in Sec.\ 3 above, the ``lower" quality C101 data set includes the 101 GRBs common to the A220 and C174 compilations, while the ``higher" quality C51 data set includes the 51 GRBs common to the A118 and C174 compilations. There is however no difference between the $\sigma_{\rm ext}$ values for C51, $\sim 0.39-0.40$ depending on model, and for C101, $\sim 0.39$. This indicates that unlike for the Amati correlation, for the Combo correlation $\sigma_{\rm ext}$ is not a good probe of the quality of the GRB data for cosmological purposes, thus rendering the Combo correlation less useful for cosmological purposes.

Both the C101 compilation with $\sigma_{\rm ext} \sim 0.39$ and the C174 compilation with the even lower $\sigma_{\rm ext} \sim 0.36$ result in cosmological constraints that mostly favor currently decelerating cosmological expansion and that are mostly inconsistent with the BAO + $H(z)$ constraints, except for the flat and non-flat XCDM parametrizations for the C101 case and the flat XCDM parametrization for the C174 case. On the other hand the C51 data with  $\sigma_{\rm ext} \sim 0.39-0.40$, the C60 data with  $\sigma_{\rm ext} \sim 0.36$, and the C79 data with  $\sigma_{\rm ext} \sim 0.40$, result in constraints that more favor currently accelerating cosmological expansion and that are mostly consistent with the BAO + $H(z)$ constraints, except for the flat and non-flat $\phi$CDM models for the C60 and C79 compilations. These findings also support the conclusion that the value of $\sigma_{\rm ext}$ for current Combo correlation data does not properly reflect the cosmological quality of these data, again casting doubt on the validity of the Combo correlation for cosmological purposes. Given that it is inappropriate to use current Combo data to constrain cosmological parameters, the following discussion of the resulting constraints from these data is brief.   

For Combo data sets, from Table 6, the value of $\Omega_{m0}$ ranges from $0.572^{+0.288}_{-0.222}$ to $> 0.593$. The minimum value is obtained in the non-flat $\phi$CDM model using the C51 data and the maximum value is obtained in the non-flat $\Lambda$CDM model using the C174 data. $\Omega_{m0}$ values obtained using C60, C51, and C79 data sets are mostly consistent with those obtained using the BAO + $H(z)$ data while $\Omega_{m0}$ values from the C101 and C174 data sets are inconsistent with those obtained using the BAO + $H(z)$ data. 

From Table 6, for all Combo data sets, in the flat $\Lambda$CDM model, the value of $\Omega_{\Lambda}$ ranges from $< 0.421$ to $< 0.807$. The minimum value is obtained using the C174 data and the maximum value is obtained using the C51 data. In the non-flat $\Lambda$CDM model, the value of $\Omega_{\Lambda}$ ranges from $< 1.200$ to $< 1.500$. The minimum value results from the C51 data and the maximum value is from the C79 data.

From Table 6, for all three non-flat models, the value of $\Omega_{k0}$ ranges from $-0.424^{+0.414}_{-0.646}$ to $-0.035^{+0.155}_{-0.485}$. The minimum value is in the non-flat XCDM parametrization from the C174 data while the maximum value is in the non-flat $\phi$CDM model using the C51 data set. These values are consistent with those obtained using BAO + $H(z)$ data.

For all Combo data sets, for the flat and non-flat XCDM parametrization, the value of the equation of state parameter ($\omega_X$) ranges from $< -0.150$ to $< 0.200$. the minimum value is obtained in the flat XCDM parametrization using the C51 data and the maximum value is obtained in the non-flat XCDM parametrization using the C101 data. In the flat and non-flat $\phi$CDM model, these data sets are unable to constrain the scalar field potential energy density parameter $\alpha$.

From Table 5, for the C60, C51, C79, C101, and C174 data sets, from the $\Delta AIC$ values, we infer that the most favored model is the flat $\Lambda$CDM one (for the C174 data set, the flat XCDM model provides an $AIC$ value slightly smaller than the $\Lambda$CDM one but consistent with it).
Moving to the $\Delta BIC$ values, for all the Combo data sets, the non-flat XCDM and $\phi$CDM cases provide strong evidence for the flat $\Lambda$CDM model.

\section{Conclusions}

In this work we focused on two widely-used GRB correlations, i.e. $E_{\rm p}-E_{\rm iso}$ and Combo, and examined whether current GRB data can be used to reliably constrain cosmological model parameters. We considered eight different GRB data sets: three for the $E_{\rm p}-E_{\rm iso}$ case (dubbed A118, A102 and A220), and five for the Combo correlation (dubbed C51, C60, C79, C101 and C174).

The A118 sample is composed of $118$ long GRBs, $93$ from \cite{2016A&A...585A..68W} and $25$ \textit{Fermi}-GBM/LAT bursts from \cite{2019ApJ...887...13F}.  A number of these GRBs have the best constrained spectral properties, obtained from more refined fits performed by using multiple component models instead of just the Band model. The A102 sample is an additional data set of $102$ long GRBs taken from \cite{Demianskietal2017a} and \cite{2019MNRAS.486L..46A}, not considered by \cite{2019ApJ...887...13F}. The updated spectral parameters of the A102 bursts, obtained by us from referenced papers and from Gamma-ray Coordination Network circulars, have been inferred from simple Band model fits. The A220 sample is the union of the A118 and A102 samples.

The C60 sample is the original Combo data set \citep{Izzo2015} composed of $60$ long GRBs, whereas the C174 sample is a newer  Combo data set composed of $174$ GRBs \citep{2021ApJ...908..181M}. We have updated both these data sets, by using updated $E_{\rm p}$ values from the A118 and A102 data sets (that come from mixed methodologies involving both the simple Band model and multiple component models). To perform comparative studies of Amati and Combo correlations, from C174 we have extracted: the C101 sample of $101$ Combo GRBs common between the A220 and C174 samples; the C51 sample of $51$ Combo GRBs common between the A118 and C174 samples; and, the C79 sample, which is the union of the C60 and C51 samples.

We analyzed these eight GRB data sets in six different cosmological models using the MCMC procedure implemented in \textsc{MontePython} and simultaneously determined (best-fit values of and limits on) uncalibrated GRB correlation parameters and cosmological model parameters.

For the $E_{\rm p}-E_{\rm iso}$ correlation case, the A118 sample has a significantly lower intrinsic dispersion than the A102 sample, indicating that it is the favored compilation. This is not inconsistent with the fact that the simpler Band model was used in the determination of spectral parameters for the bursts in the A102 sample. On the other hand, there is almost no difference in the intrinsic dispersion of the five Combo data sets. Given that some of the Combo data sets consist of lower-quality GRB measurements (with spectral parameters determined using the simpler Band model), this result indicates that for current Combo data sets the intrinsic dispersion value is insenstive to the quality of the burst data.

Additionally, we compared cosmological constraints determined from the GRB data sets to those from better-established BAO + $H(z)$ data. We found that the A118 and C51 constraints, and less so the A102, C60, and C79 constraints, are consistent with those from the BAO + $H(z)$ data, while the A220, C101, and C174 cosmological constraints are inconsistent with the BAO + $H(z)$ ones. 

Our first main conclusion is that only the $E_{\rm p}-E_{\rm iso}$ correlation A118 GRB sample is reliable enough to be used to constrain cosmological parameters. This includes only a little more than half of the available $E_{\rm p}-E_{\rm iso}$ bursts. We emphasize that, besides future GRB data, there is still plenty of old GRB data (publicly available or not) that can be reanalyzed using better spectral models (as done in \cite{2019ApJ...887...13F}) and could then be candidates for inclusion in a future update of the reliable $E_{\rm p}-E_{\rm iso}$ correlation GRB data set.
    
Our second main conclusion is that cosmological constraints from the uncalibrated A118 sample are quite weak and also consistent with those from better established probes such as BAO, $H(z)$, SNIa, and CMB anisotropy data. The consistency means that it is justified to derive cosmological constraints from a joint analyses of A118 and BAO + $H(z)$ data, as we have done here, but given the weakness of the A118 constraints adding the A118 data to the mix does not much tighten the BAO + $H(z)$ constraints.

Our third main conclusion is that current uncalibrated Combo correlation GRB data sets are not reliable enough to be used to constrain cosmological parameters.

Our analyses here resolve the previous inconsistencies between cosmological constraints found from different GRB data sets. They indicate that only the A118 GRB data set provides reliable cosmological constraints, that these are broad, consistent with those from better-established cosmological probes, as well as consistent with the standard flat $\Lambda$CDM model, but also consistent with dynamical dark energy models and non-spatially-flat models.

It appropriate to stress again that the our results indicate that the Amati correlation represents a reliable instrument for the standardization of the A118 GRB data set.
However, we are not yet in the position to discriminate among possible cosmologies with GRB data only, since our analysis method does not address the circularity problem, being unable to produce distance GRB moduli independently from any cosmological model.

Looking forward, since GRBs probe a largely unexplored part of cosmological redshift space, it is a worthwhile task to acquire more and better-quality burst data that might provide valuable cosmological constraints on the cosmological models used in this paper and test alternative ones \citep[see. e.g.,][]{2018PhRvD..98j3520L}.

\begin{appendix}
\section{The GRB data sets}
\label{GRBdata}
\LTcapwidth=\linewidth
\begin{longtable}{lccc}
\caption{A118 long GRB sample. For each source, columns list: GRB name, redshift, $E_{\rm p}$, and $S_{\rm bolo}$. The upper part of the table, above the horizontal dividing line, lists the $25$ bursts from \cite{2019ApJ...887...13F}, whereas the lower part lists the updated $93$ bursts from \cite{2016A&A...585A..68W}. Errors are $1\sigma$ confidence level.} \label{tab:A118}\\
\hline\hline
GRB &   $z$   & $E_\mathrm{p}$  & $S_\mathrm{bolo}$ \\
         &        &  (keV)                        &  (10$^{-5}$ erg/cm$^2$)  \\
\hline
\endfirsthead
\caption{continued.}\\
\hline\hline
GRB &   $z$   & $E_\mathrm{p}$  & $S_\mathrm{bolo}$ \\
         &        &  (keV)                        &  (10$^{-5}$ erg/cm$^2$)  \\
\hline
\endhead
\hline
\endfoot
080916C & 4.35  & 6953.87 $\pm$ 1188.77           & 10.40 $\pm$ 0.24 \\
090323  & 3.57          & 2060.09 $\pm$ 138.07    & 15.76 $\pm$ 0.39     \\
090328  & 0.736         & 1221.71 $\pm$ 81.87        & 7.99 $\pm$ 0.20  \\
090424  & 0.544          & 236.91 $\pm$ 4.55     & 5.72 $\pm$ 0.09     \\
090902B & 1.822         & 2146.57 $\pm$  21.71    & 39.05 $\pm$ 0.22  \\
090926A & 2.1062        & 868.63 $\pm$ 13.85    & 17.90 $\pm$ 0.13  \\
091003A & 0.8969        & 857.81 $\pm$ 33.08    & 4.43 $\pm$ 0.08   \\
091127  & 0.49          & 60.32 $\pm$ 1.93     & 2.25 $\pm$ 0.04    \\
\hline
\end{longtable}

\LTcapwidth=\linewidth
\begin{longtable}{lccc}
\caption{A102 long GRB sample. For each source, columns list: GRB name, redshift, $E_{\rm p}$, and $S_{\rm bolo}$. Errors are $1\sigma$ confidence level.} \label{tab:A102}\\
\hline\hline
GRB &   $z$   & $E_\mathrm{p}$  & $S_\mathrm{bolo}$ \\
         &        &  (keV)                        &  (10$^{-5}$ erg/cm$^2$)  \\
\hline
\endfirsthead
\caption{continued.}\\
\hline\hline
GRB &   $z$   & $E_\mathrm{p}$  & $S_\mathrm{bolo}$ \\
         &        &  (keV)                        &  (10$^{-5}$ erg/cm$^2$)  \\
\hline
\endhead
\hline
\endfoot
180325A &	2.24800  &   993.888 $\pm$ 146.16  & 2.143  $\pm$  0.338\\
171222A &	2.409 &  99.4746 $\pm$ 9.30657 & 0.324512 $\pm$ 0.0328681\\
171010A &	0.3285 & 204.589 $\pm$ 1.3285 & 90.6418 $\pm$ 0.535392\\
170705A &	2.01 & 301. $\pm$ 30.1 & 1.65024 $\pm$ 0.0603516\\
170607A &	0.557 & 239.778 $\pm$ 20.241 & 1.11395 $\pm$ 0.0380721\\
170113A &	1.968 & 217.554 $\pm$ 99.7248 & 0.089964 $\pm$ 0.018213\\
161129A &	0.645 & 324.065 $\pm$ 34.545 & 0.904616 $\pm$ 0.044168\\
161117A &	1.549 & 173.434 $\pm$ 6.78034 & 4.81252 $\pm$ 0.120027\\
\hline
\end{longtable}

\LTcapwidth=\linewidth
\begin{longtable}{lccccc}
\caption{C174 and C60 Combo GRBs and their Combo relation parameters (reproduced from \cite{2021ApJ...908..181M} with their kind permission, except for some $E_{\rm p}$ values we have updated here). The first $60$ sources are the C60 sample. The C60 sample and the following sources in boldface form the C174 sample. For each source, columns list: GRB name, redshift, $\log E_{\rm p}$, $\log F_0$, $\log \tau$, and $\alpha$. Errors are $1\sigma$ confidence level.} \label{tab:C17460}\\
\hline\hline
GRB      &  $z$    &  $\log \left(E_{\rm p}/{\rm keV}\right)$  &  $\log \left(F_0/{\rm erg}/{\rm cm}^2{\rm s}\right)$  &  $\log \left(\tau/{\rm s}\right)$     &  $\alpha$ \\
\hline
\endfirsthead
\caption{continued.}\\
\hline\hline
GRB      &  $z$    &  $\log \left(E_{\rm p}/{\rm keV}\right)$  &  $\log \left(F_0/{\rm erg}/{\rm cm}^2{\rm s}\right)$  &  $\log \left(\tau/{\rm s}\right)$     &  $\alpha$ \\
\hline
\endhead
\hline
\endfoot
	050401		&	$	2.8983	$	&	$	2.67	^	{	+	0.13	}	_	{	-0.11	}	$	&	$	-9.57	^	{	+	0.02	}	_	{	-0.02	}	$	&	$	2.90	^	{	+	0.06	}	_	{	-0.07	}	$	&	$	-1.75	\pm	0.09	$	\\
	050922C		&	$	2.1995	$	&	$	2.62	^	{	+	0.13	}	_	{	-0.10	}	$	&	$	-8.79	^	{	+	0.05	}	_	{	-0.06	}	$	&	$	1.64	^	{	+	0.07	}	_	{	-0.09	}	$	&	$	-1.30	\pm	0.03	$	\\
	051109A		&	$	2.346	$	&	$	2.73	^	{	+	0.13	}	_	{	-0.19	}	$	&	$	-9.86	^	{	+	0.07	}	_	{	-0.09	}	$	&	$	2.97	^	{	+	0.08	}	_	{	-0.10	}	$	&	$	-1.36	\pm	0.03	$	\\
	060115		&	$	3.53	$	&	$	2.47	^	{	+	0.05	}	_	{	-0.05	}	$	&	$	-11.15	^	{	+	0.06	}	_	{	-0.06	}	$	&	$	3.99	^	{	+	0.13	}	_	{	-0.19	}	$	&	$	-2.64	\pm	0.44	$	\\
	060418		&	$	1.489	$	&	$	2.76	^	{	+	0.10	}	_	{	-0.12	}	$	&	$	-9.22	^	{	+	0.07	}	_	{	-0.08	}	$	&	$	2.49	^	{	+	0.08	}	_	{	-0.10	}	$	&	$	-1.64	\pm	0.06	$	\\
	060526		&	$	3.2213	$	&	$	2.02	^	{	+	0.08	}	_	{	-0.10	}	$	&	$	-11.31	^	{	+	0.04	}	_	{	-0.04	}	$	&	$	4.53	^	{	+	0.14	}	_	{	-0.21	}	$	&	$	-5.23	\pm	1.40	$	\\
	060707		&	$	3.424	$	&	$	2.44	^	{	+	0.13	}	_	{	-0.08	}	$	&	$	-10.60	^	{	+	0.14	}	_	{	-0.21	}	$	&	$	2.87	^	{	+	0.19	}	_	{	-0.36	}	$	&	$	-1.07	\pm	0.05	$	\\
	060729		&	$	0.54	$	&	$	1.89	^	{	+	0.12	}	_	{	-0.16	}	$	&	$	-10.54	^	{	+	0.01	}	_	{	-0.01	}	$	&	$	4.80	^	{	+	0.02	}	_	{	-0.02	}	$	&	$	-1.76	\pm	0.03	$	\\
\hline
\end{longtable}

\LTcapwidth=\linewidth
\begin{longtable}{lccccc}
\caption{ The C101 sample. For each source, columns list: GRB name, redshift, $\log E_{\rm p}$, $\log F_0$, $\log \tau$, and $\alpha$. Errors are $1\sigma$ confidence level.} \label{tab:C101}\\
\hline\hline
GRB      &  $z$    &  $\log \left(E_{\rm p}/{\rm keV}\right)$  &  $\log \left(F_0/{\rm erg}/{\rm cm}^2{\rm s}\right)$  &  $\log \left(\tau/{\rm s}\right)$     &  $\alpha$ \\
\hline
\endfirsthead
\caption{continued.}\\
\hline\hline
GRB      &  $z$    &  $\log \left(E_{\rm p}/{\rm keV}\right)$  &  $\log \left(F_0/{\rm erg}/{\rm cm}^2{\rm s}\right)$  &  $\log \left(\tau/{\rm s}\right)$     &  $\alpha$ \\
\hline
\endhead
\hline
\endfoot
050401		&	$	2.8983	$	&	$	2.67	^	{	+	0.13	}	_	{	-0.11	}	$	&	$	-9.57	^	{	+	0.02	}	_	{	-0.02	}	$	&	$	2.90	^	{	+	0.06	}	_	{	-0.07	}	$	&	$	-1.75	\pm	0.09	$	\\

	050922C		&	$	2.1995	$	&	$	2.62	^	{	+	0.13	}	_	{	-0.10	}	$	&	$	-8.79	^	{	+	0.05	}	_	{	-0.06	}	$	&	$	1.64	^	{	+	0.07	}	_	{	-0.09	}	$	&	$	-1.30	\pm	0.03	$	\\

	051109A		&	$	2.346	$	&	$	2.73	^	{	+	0.13	}	_	{	-0.19	}	$	&	$	-9.86	^	{	+	0.07	}	_	{	-0.09	}	$	&	$	2.97	^	{	+	0.08	}	_	{	-0.10	}	$	&	$	-1.36	\pm	0.03	$	\\

	060115		&	$	3.53	$	&	$	2.47	^	{	+	0.05	}	_	{	-0.05	}	$	&	$	-11.15	^	{	+	0.06	}	_	{	-0.06	}	$	&	$	3.99	^	{	+	0.13	}	_	{	-0.19	}	$	&	$	-2.64	\pm	0.44	$	\\

	060418		&	$	1.489	$	&	$	2.76	^	{	+	0.10	}	_	{	-0.12	}	$	&	$	-9.22	^	{	+	0.07	}	_	{	-0.08	}	$	&	$	2.49	^	{	+	0.08	}	_	{	-0.10	}	$	&	$	-1.64	\pm	0.06	$	\\

	060526		&	$	3.2213	$	&	$	2.02	^	{	+	0.08	}	_	{	-0.10	}	$	&	$	-11.31	^	{	+	0.04	}	_	{	-0.04	}	$	&	$	4.53	^	{	+	0.14	}	_	{	-0.21	}	$	&	$	-5.23	\pm	1.40	$	\\

	060707		&	$	3.424	$	&	$	2.44	^	{	+	0.13	}	_	{	-0.08	}	$	&	$	-10.60	^	{	+	0.14	}	_	{	-0.21	}	$	&	$	2.87	^	{	+	0.19	}	_	{	-0.36	}	$	&	$	-1.07	\pm	0.05	$	\\

	060729		&	$	0.54	$	&	$	1.89	^	{	+	0.12	}	_	{	-0.16	}	$	&	$	-10.54	^	{	+	0.01	}	_	{	-0.01	}	$	&	$	4.80	^	{	+	0.02	}	_	{	-0.02	}	$	&	$	-1.76	\pm	0.03	$	\\
\hline
\end{longtable}

\LTcapwidth=\linewidth
\begin{longtable}{lccccc}
\caption{ The C51 sample. For each source, columns list: GRB name, redshift, $\log E_{\rm p}$, $\log F_0$, $\log \tau$, and $\alpha$. Errors are $1\sigma$ confidence level.} \label{tab:C51}\\
\hline\hline
GRB      &  $z$    &  $\log \left(E_{\rm p}/{\rm keV}\right)$  &  $\log \left(F_0/{\rm erg}/{\rm cm}^2{\rm s}\right)$  &  $\log \left(\tau/{\rm s}\right)$     &  $\alpha$ \\
\hline
\endfirsthead
\caption{continued.}\\
\hline\hline
GRB      &  z    &  $\log \left(E_{\rm p}/{\rm keV}\right)$  &  $\log \left(F_0/{\rm erg}/{\rm cm}^2{\rm s}\right)$  &  $\log \left(\tau/{\rm s}\right)$     &  $\alpha$ \\
\hline
\endhead
\hline
\endfoot
050401		&	$	2.8983	$	&	$	2.67	^	{	+	0.13	}	_	{	-0.11	}	$	&	$	-9.57	^	{	+	0.02	}	_	{	-0.02	}	$	&	$	2.90	^	{	+	0.06	}	_	{	-0.07	}	$	&	$	-1.75	\pm	0.09	$	\\

	050922C		&	$	2.1995	$	&	$	2.62	^	{	+	0.13	}	_	{	-0.10	}	$	&	$	-8.79	^	{	+	0.05	}	_	{	-0.06	}	$	&	$	1.64	^	{	+	0.07	}	_	{	-0.09	}	$	&	$	-1.30	\pm	0.03	$	\\

	060115		&	$	3.53	$	&	$	2.47	^	{	+	0.05	}	_	{	-0.05	}	$	&	$	-11.15	^	{	+	0.06	}	_	{	-0.06	}	$	&	$	3.99	^	{	+	0.13	}	_	{	-0.19	}	$	&	$	-2.64	\pm	0.44	$	\\

	060418		&	$	1.489	$	&	$	2.76	^	{	+	0.10	}	_	{	-0.12	}	$	&	$	-9.22	^	{	+	0.07	}	_	{	-0.08	}	$	&	$	2.49	^	{	+	0.08	}	_	{	-0.10	}	$	&	$	-1.64	\pm	0.06	$	\\

	060526		&	$	3.2213	$	&	$	2.02	^	{	+	0.08	}	_	{	-0.10	}	$	&	$	-11.31	^	{	+	0.04	}	_	{	-0.04	}	$	&	$	4.53	^	{	+	0.14	}	_	{	-0.21	}	$	&	$	-5.23	\pm	1.40	$	\\

	060707		&	$	3.424	$	&	$	2.44	^	{	+	0.13	}	_	{	-0.08	}	$	&	$	-10.60	^	{	+	0.14	}	_	{	-0.21	}	$	&	$	2.87	^	{	+	0.19	}	_	{	-0.36	}	$	&	$	-1.07	\pm	0.05	$	\\

	060908		&	$	1.8836	$	&	$	2.74	^	{	+	0.07	}	_	{	-0.09	}	$	&	$	-8.50	^	{	+	0.07	}	_	{	-0.08	}	$	&	$	1.65	^	{	+	0.10	}	_	{	-0.13	}	$	&	$	-1.37	\pm	0.05	$	\\

	060927		&	$	5.467	$	&	$	2.44	^	{	+	0.10	}	_	{	-0.14	}	$	&	$	-10.80	^	{	+	0.06	}	_	{	-0.06	}	$	&	$	2.51	^	{	+	0.11	}	_	{	-0.15	}	$	&	$	-1.55	\pm	0.17	$	\\
\hline
\end{longtable}

\LTcapwidth=\linewidth
\begin{longtable}{lccccc}
\caption{ The C79 sample. For each source, columns list: GRB name, redshift, $\log E_{\rm p}$, $\log F_0$, $\log \tau$, and $\alpha$. Errors are $1\sigma$ confidence level.} \label{tab:C79}\\
\hline\hline
GRB      &  $z$    &  $\log \left(E_{\rm p}/{\rm keV}\right)$  &  $\log \left(F_0/{\rm erg}/{\rm cm}^2{\rm s}\right)$  &  $\log \left(\tau/{\rm s}\right)$     &  $\alpha$ \\
\hline
\endfirsthead
\caption{continued.}\\
\hline\hline
GRB      &  $z$    &  $\log \left(E_{\rm p}/{\rm keV}\right)$  &  $\log \left(F_0/{\rm erg}/{\rm cm}^2{\rm s}\right)$  &  $\log \left(\tau/{\rm s}\right)$     &  $\alpha$ \\
\hline
\endhead
\hline
\endfoot
050401		&	$	2.8983	$	&	$	2.67	^	{	+	0.13	}	_	{	-0.11	}	$	&	$	-9.57	^	{	+	0.02	}	_	{	-0.02	}	$	&	$	2.90	^	{	+	0.06	}	_	{	-0.07	}	$	&	$	-1.75	\pm	0.09	$	\\

	050922C		&	$	2.1995	$	&	$	2.62	^	{	+	0.13	}	_	{	-0.10	}	$	&	$	-8.79	^	{	+	0.05	}	_	{	-0.06	}	$	&	$	1.64	^	{	+	0.07	}	_	{	-0.09	}	$	&	$	-1.30	\pm	0.03	$	\\

	051109A		&	$	2.346	$	&	$	2.73	^	{	+	0.13	}	_	{	-0.19	}	$	&	$	-9.86	^	{	+	0.07	}	_	{	-0.09	}	$	&	$	2.97	^	{	+	0.08	}	_	{	-0.10	}	$	&	$	-1.36	\pm	0.03	$	\\

	060115		&	$	3.53	$	&	$	2.47	^	{	+	0.05	}	_	{	-0.05	}	$	&	$	-11.15	^	{	+	0.06	}	_	{	-0.06	}	$	&	$	3.99	^	{	+	0.13	}	_	{	-0.19	}	$	&	$	-2.64	\pm	0.44	$	\\

	060418		&	$	1.489	$	&	$	2.76	^	{	+	0.10	}	_	{	-0.12	}	$	&	$	-9.22	^	{	+	0.07	}	_	{	-0.08	}	$	&	$	2.49	^	{	+	0.08	}	_	{	-0.10	}	$	&	$	-1.64	\pm	0.06	$	\\

	060526		&	$	3.2213	$	&	$	2.02	^	{	+	0.08	}	_	{	-0.10	}	$	&	$	-11.31	^	{	+	0.04	}	_	{	-0.04	}	$	&	$	4.53	^	{	+	0.14	}	_	{	-0.21	}	$	&	$	-5.23	\pm	1.40	$	\\

	060707		&	$	3.424	$	&	$	2.44	^	{	+	0.13	}	_	{	-0.08	}	$	&	$	-10.60	^	{	+	0.14	}	_	{	-0.21	}	$	&	$	2.87	^	{	+	0.19	}	_	{	-0.36	}	$	&	$	-1.07	\pm	0.05	$	\\

	060729		&	$	0.54	$	&	$	1.89	^	{	+	0.12	}	_	{	-0.16	}	$	&	$	-10.54	^	{	+	0.01	}	_	{	-0.01	}	$	&	$	4.80	^	{	+	0.02	}	_	{	-0.02	}	$	&	$	-1.76	\pm	0.03	$	\\
\hline
\end{longtable}
\end{appendix}

\acknowledgments

We acknowledge support provided by the Ministry of Education and Science of the Republic of Kazakhstan, grant IRN AP08052311, INFN--Frascati National Laboratories, Iniziative Specifiche MOONLIGHT2, and DOE grant DE-SC0011840. Part of the computing for this project was performed on the Beocat Research Cluster at Kansas State University.


\bibliographystyle{unsrt}
\bibliography{biblio}



\end{document}